\documentclass{aa}  

\usepackage{booktabs}
\usepackage{graphicx}
\usepackage{txfonts}
\usepackage{physics}
\usepackage{xcolor}
\usepackage{multirow}
\usepackage{arydshln}
\usepackage{capt-of}

\newcommand*{\un}[2]{#1\,\mathrm{#2}}

\begin{document} 

\title{Weak S-type asteroids compared to C-type explain the observed size distribution of the main belt}

\author{
M. Vávra\inst{1,2}\and
M. Brož\inst{1}
}

\institute{
Charles University, Faculty of Mathematics and Physics, Institute of Astronomy, V Holešovičkách 2, CZ-18000 Praha~8, Czech Republic
\and
Astronomical Institute of the Czech Academy of Sciences, Fričova 298, CZ-25165 Ondřejov, Czech Republic;
\email{mvavra@asu.cas.cz}
}

\date{Received - / Accepted -}

\abstract{
The main belt, the region between the orbits of Mars and Jupiter,
is home to more than 1 million asteroids.
These asteroids form orbital groups,
(i.e., asteroid families formed by collisions)
and also spectral groups (taxonomies) with different chemical compositions,
in particular carbonaceous (C-types) and silicate (S-types).
In this paper, we extend the existing main-belt collisional model
by finding the appropriate strength-versus-size dependence
(also known as the scaling~law)
for these two groups.
We used color indices and geometric albedos of 56 and 72 
spectroscopically confirmed
C- and S-types (control samples), along with statistical methods on $1\,065\,034$ asteroids,
to assign C-, S-, or other types.
This allowed us to construct observed
size-frequency distributions (SFDs)
for several subpopulations
constrained by either semimajor axis (inner, middle, outer)
or taxonomy (C, S, other). 
Then we used a Monte Carlo collisional model
to compute the long-term collisional evolution (4.5 billion years)
and derive synthetic SFDs.
Our best-fit scaling laws indicate that
S-types must be weaker below approximately $0.2\,\mathrm{km}$
than C-types
to explain the deficiency of asteroids in the inner part of the main belt
near (and below) the observational limit.
This may correspond to differences in chemical composition or material porosity. 
Future research will focus on the scaling laws of asteroids
with rare or "extreme" taxonomies (e.g., V, M).
}

\keywords{Solar System --
asteroids --
collisions --
Monte-Carlo models
}

\maketitle

\section{Introduction}

As the main belt (MB) represents a huge sample of more than 1 million asteroids
\citep{Tedesco_2002AJ....123.2070T,Gladman_2009Icar..202..104G},
we need statistical descriptions of this sample.
The appropriate description is the
cumulative size distribution (hereafter SFD),
i.e., the total number of asteroids $N({\ge}D)$
greater than or equal to the size $D$.
The characteristics of this distribution are affected by
a series of processes, including
collisions,
fragmentation,
ejection,
perturbations by resonances,
and the Yarkovsky effect,
i.e., a drift of the semimajor axis due to thermal radiation forces
\citep{O'Brien_2005Icar..178..179O,Bottke_2006AREPS..34..157B,Vokrouhlicky_2015aste.book..509V,Granvik_2016Natur.530..303G}.

One of the first efforts
to explain the observed SFD was made by \citet{Anders_1965Icar....4..399A}, who
proposed that the initial (primordial) differential SFD was Gaussian. 
Later, this was disproved by \citet{Davis_1979aste.book..528D,Davis_1985Icar...62...30D}.
The first analytical model was provided by \citet{Dohnanyi_1969JGR....74.2531D}, who showed that if the asteroid destruction rate is equal to their production rate, i.e., they are in collisional equilibrium \citep{O'Brien_2003Icar..164..334O}, 
then the cumulative SFD of these asteroids is described by a power law with slope $-2.5$. 
However, he did not use any realistic scaling~laws.
The first realistic scaling~laws were used by \citet{Davis_1979aste.book..528D,Davis_1985Icar...62...30D}.
They emphasized that precise collisional models should include other effects,
for example, an orbital decay of asteroids over time. 
Later, \citet{Campo_1994P&SS...42.1079C} showed that
scaling~laws must be size-dependent to match the observations.
An explicit form of the scaling~law for the MB was found by 
\citet{O'Brien_2003Icar..164..334O,O'Brien_2005Icar..178..179O}.

In the following years, the collisional evolution of the MB was studied by several authors, focusing primarily on the characteristics of the evolved cumulative SFD.
For example,    
\citet{Campo_1994P&SS...42.1079C} found a cutoff at small sizes;
\citet{Bottke_2005Icar..175..111B} showed that the cumulative SFD at large sizes is determined by early evolution of the MB from $4.5$ to $3.9\,\mathrm{Gyr}$ ago;
\citet{Cibulkova_2014Icar..241..358C} found that most asteroids appear monolithic rather than porous; and
\citet{Bottke_2015aste.book..701B} noticed that the cumulative SFD is "wavy" for asteroids smaller than $\un{10}{km}$.

A substantial contribution is the work of \citet{Morbidelli_2009Icar..204..558M}, who published a Monte Carlo code \texttt{Boulder}
for simulations of collisional evolution
using a particle-in-a-box approach
\citep{Rubinstein2016-ju}.
The code includes results from smoothed-particle hydrodynamic (SPH) simulations
\citep{Benz_1999Icar..142....5B,Cossins_2010PhDT.......301C,Jutzi_2015P&SS..107....3J},
which approximate a projectile and a target by thousands to millions of particles,
tracking the relevant physical quantities.
\citet{Morbidelli_2009Icar..204..558M} themselves studied the primordial SFD of planetesimals.

Collisions are necessary to explain a number of other observations.
They can, for example, explain the formation of comets \citep{Bottke_2022DPS....5430403B},
the origin of dust in the Solar System \citep{Landgraf_2002AJ....123.2857L,Nesvorny_2006Icar..183..296N},
the presence of nonindigenous material on Vesta, Bennu, and Almahata Sitta \citep{Bottke_2020DPS....5240202B},
asteroid shapes \citep{Marchis_2021A&A...653A..57M,Broz_2022A&A...657A..76B},
and cratering on asteroids \citep{Bottke_2020AJ....160...14B}.

Last but not the least, collisions are crucial for the formation of asteroid families \citep{Michel_2001Sci...294.1696M,Zappala_1990AJ....100.2030Z,Nesvorny_2002Icar..157..155N,Vokrouhlicky_2006Icar..182...92V}. 
Collisional models can also
help identify families \citep{Nesvorny_2015aste.book..297N},
estimate their age \citep{Nesvorny_2006Icar..183..296N,Vokrouhlicky_2010AJ....139.2148V},
describe their evolution \citep{Vernazza_2018A&A...618A.154V},
and study material strength \citep{Marschall_2022AJ....164..167M}.

In this paper, we continue these studies by extending the collisional model of the MB. 
In particular, we split the MB into C- and S-types,
since these two spectral (taxonomic) groups are the most common 
\citep{Gradie_1982Sci...216.1405G,Tholen_1984PhDT.........3T}.
Our goal is to find explicit forms of their scaling laws. 
To do so, we need to set up proper initial conditions and scaling laws
for both populations and simulate their long-term collisional evolution.
The resulting synthetic cumulative SFDs are then compared to the corresponding observed SFDs.

\section{Identifying C- and S-types in catalogs} \label{Algorithm for an assignemt of C– and S–types}

The characteristics of reflectance spectrum determine the taxonomy of an asteroid \citep{Bus_2002Icar..158..146B,DeMeo_2009Icar..202..160D}.
These characteristics can also be described by magnitudes measured in several passbands -- for example, $g,r,i,z$ as measured by the Sloan Digital Sky Survey (SDSS) \citep{Blanton_2017AJ....154...28B}.
Their mutual subtraction yields distance-independent color indices.
The indices are further calibrated by subtracting the values corresponding to the solar analog \citep{Ivezic_2001AJ....122.2749I}.
For us, the most relevant color indices 
are $i-z$ and the first principal component (denoted $a^*$), obtained via principal component analysis by \citet{Ivezic_2001AJ....122.2749I}.
The two most populous taxonomies, i.e., 
C- and S-types
are distinctive in $a^*$, $i-z$, and the geometrical albedo $pV$
\citep{Zellner_1979aste.book..783Z,Ivezic_2001AJ....122.2749I,Delbo_2004PhDT.......371D,Parker_2008Icar..198..138P}.

\subsection{Control sample of C- and S-types}

To define our control samples,
we first retrieved 328 C-type and 491 S-type asteroids identified by \citet{Bus_2002Icar..158..146B,Bus_2002Icar..158..106B} during a spectroscopic survey of the MB. Their $a^*$ and $i-z$ values were obtained from the SDSS catalog \citep{Blanton_2017AJ....154...28B}, and their albedos from the AKARI \citep{Yamauchi_2011PASP..123..852Y} catalog or, if unavailable, the Wide-field Infrared Survey Explorer (WISE; \citealt{Masiero_2011ApJ...741...68M}) catalog.

\begin{figure}
    \centering
    \includegraphics[width=7cm]{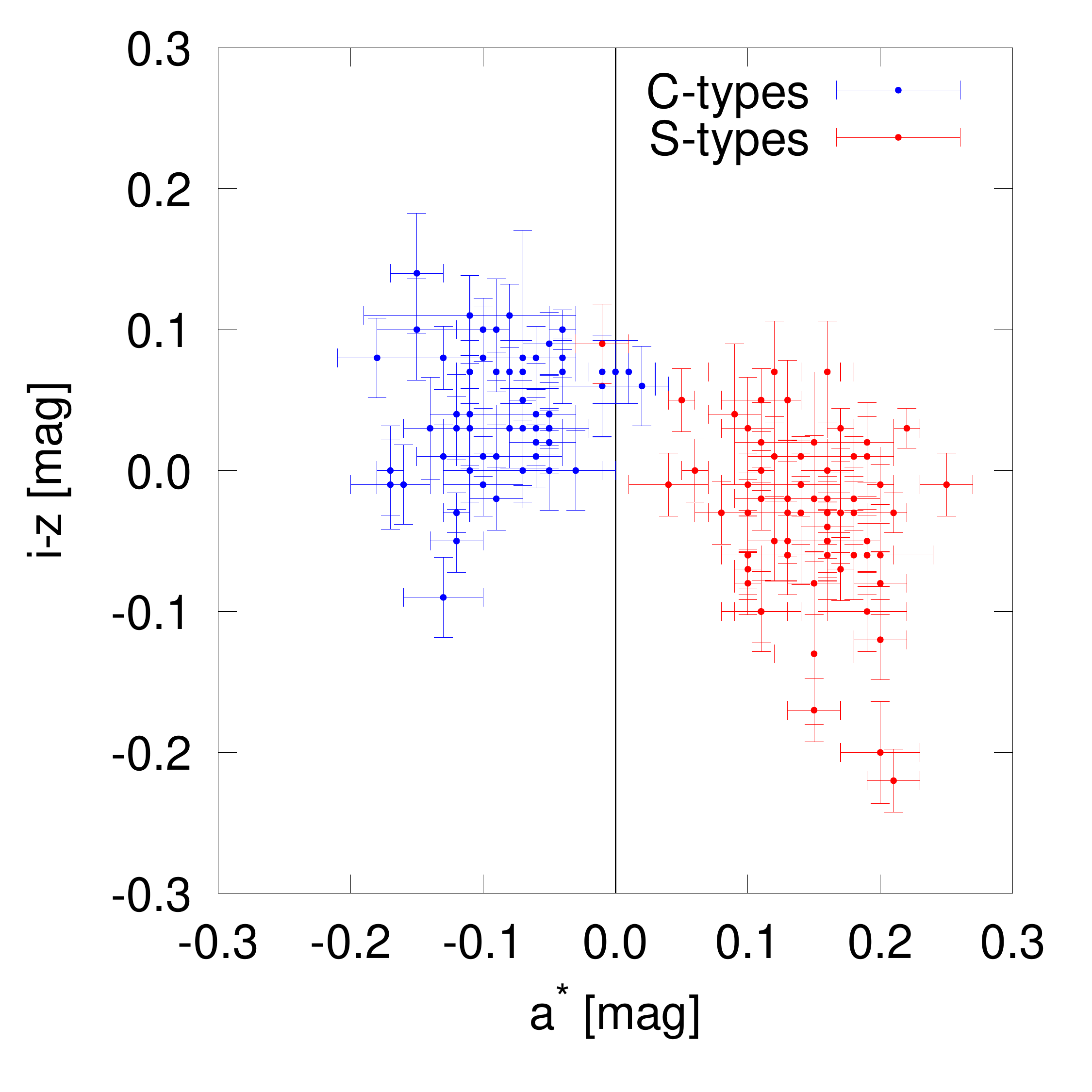}
    \caption{
    Scatter plot of the C- (blue) and S-type (red) control samples in the $i-z$ vs. $a^*$ plane, along with their uncertainties $\sigma_{i-z}$ and $\sigma_{a^*}$. 
    }
    \label{a_star_i_z}
\end{figure}

\begin{figure}
    \centering
    \includegraphics[width=7cm]{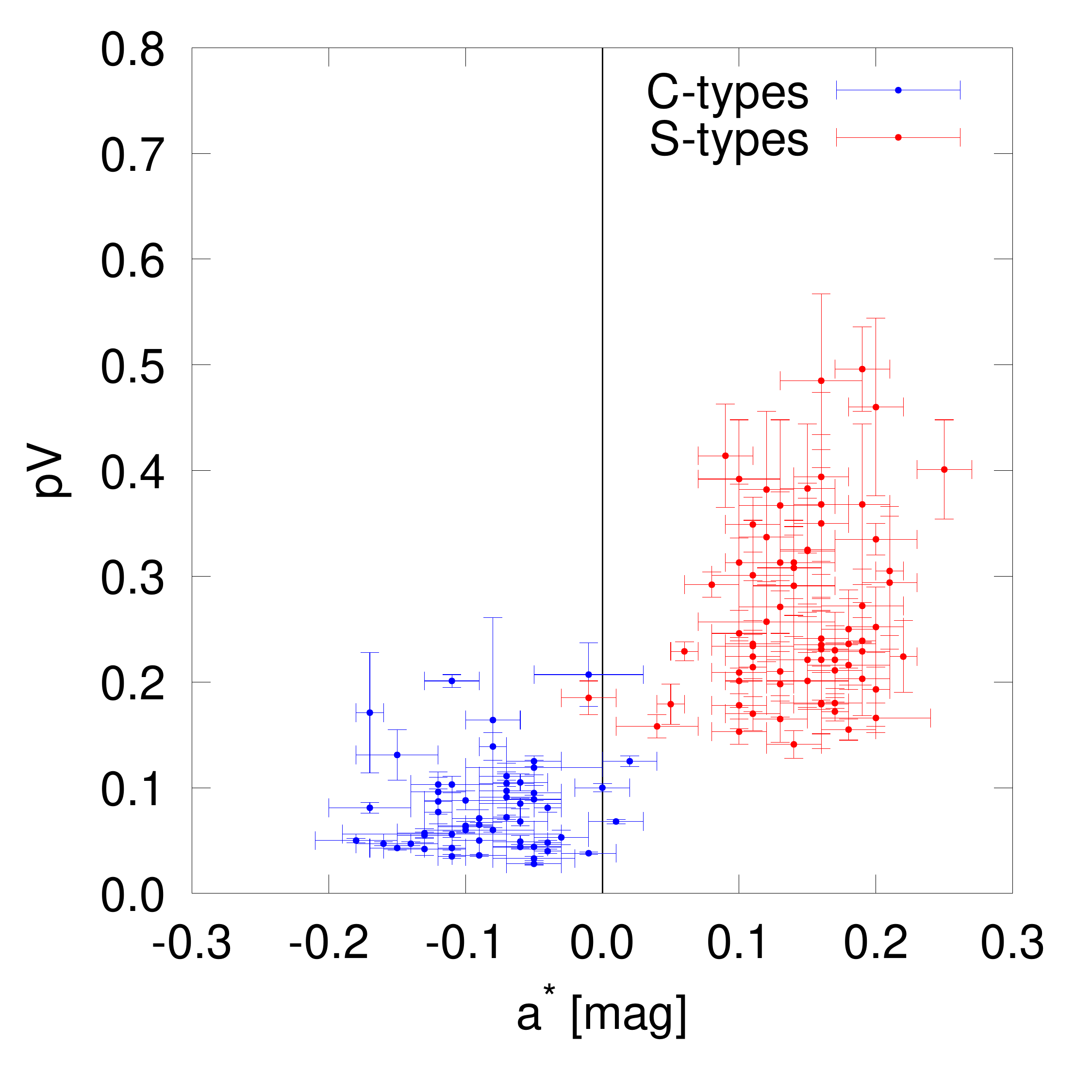}
    \caption{
    Same as Fig.~\ref{a_star_i_z}, but in the $pV$ vs. $a^*$ plane.
    }
    \label{a_star_pV}
\end{figure}

The C- or S-type was appended to the respective control sample if it satisfied three criteria: all values of $a^*$, $i-z$, and $pV$ known;
uncertainties $\sigma_{a^*}$, $\sigma_{i-z}$, and $\sigma_{pV}$ below 0.1;
and semimajor axis (proper or osculating) in the range 2.05–3.8\,au.
Applying these constraints yielded 56 C-types and 72 S-types.  
The scatter plots of both control samples are shown
in Figs.~\ref{a_star_i_z} and \ref{a_star_pV}.
Within the uncertainties,
C-types have negative $a^*$, and
S-types have positive $a^*$,
in agreement with \citet{Parker_2008Icar..198..138P}. 
A few C-types, however, have $pV > 0.1$, which is more typical of X-types, especially M-types in Tholen's taxonomy
\citep{Tholen_1984PhDT.........3T}
with a median of $0.17$
\citep{Bus_2002Icar..158..146B}. 
These cases may represent other types
but were retained in the C-type control sample.

\subsection{Algorithm for an assignment of C-types, S-types, and other types}\label{2.2}

We implemented a Monte Carlo algorithm to assign C-, S-,
or neither C- nor S-type (hereinafter, other type)
to every asteroid with unknown taxonomy, using our control samples.
The details are described in \cite{Vavra_2024}
and in Appendix~\ref{algorithm}.

\begin{figure}
    \centering
    \includegraphics[width=9cm]{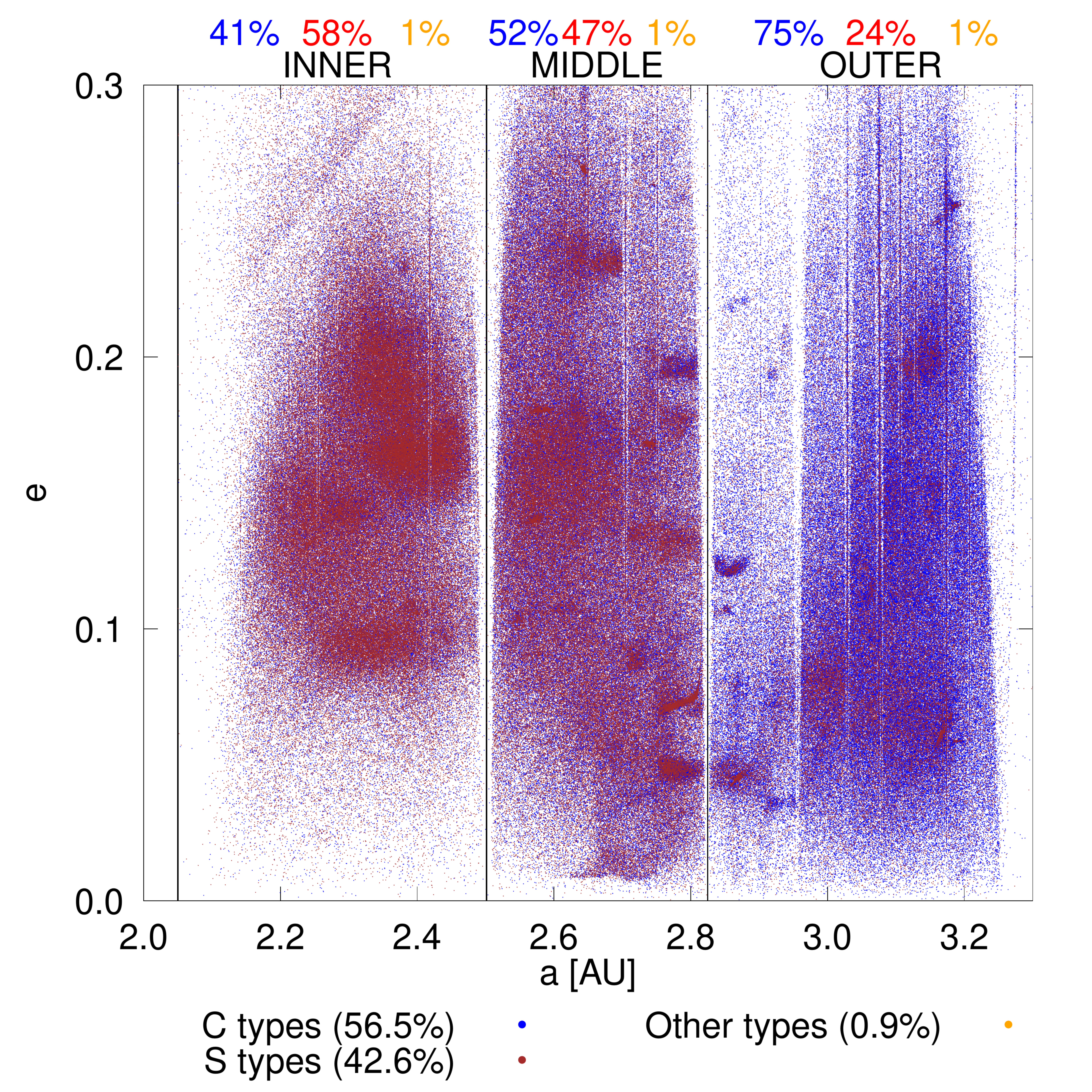}
    \caption{Scatter plot of eccentricity vs. semimajor axis (both proper or osculating) for all resulted $601\,213$ C-, $453\,998$ S-, and $9\,823$ other types across different parts of the MB (inner, middle, and outer) separated by vertical black lines.
    Above the plot are the relative numbers of each taxonomy (\textcolor{blue}{C-}, \textcolor{red}{S-}, and \textcolor{orange}{ other types}) within the specific parts of the MB.
    }
    \label{a_e_taxonomy}
\end{figure}

As a result, from $1\,065\,034$ asteroids,
we obtained $601\,213$ C- and $453\,998$ S-types.
Their spatial distribution is shown in Fig.~\ref{a_e_taxonomy}. 
S-types prevail in the inner part of the MB, but as the semimajor
axis increases, they are gradually overtaken by C-types,
in agreement with
\citet{Ivezic_2001AJ....122.2749I},
\citet{Mothe-Diniz_2003Icar..162...10M}, and
\citet{Marsset_2022AJ....163..165M}.
Specifically,
S-types comprise $58\%$ of the inner MB, while
C-types comprise $52\%$ and $75\%$ of the middle and outer parts,
respectively.
Other types represent only $1\%$ of all asteroids.

\subsection{Observational incompleteness} \label{obs_bias}

We estimated the observational limit size from the differential distribution
of absolute magnitudes~$H$, as in \citet{Hendler_2020PSJ.....1...75H}.
The method is based on the maximum absolute magnitude $H_{\text{max}}$ of the differential distribution.
Below this value ($H < H_{\text{max}}$), a population is considered complete, i.e., unbiased.
The differential magnitude distributions $N(H)$
for C-, S-, and other types are shown in Fig.~\ref{H_dist_tax}.
We set the binning of the differential distributions to $\dd H = \un{0.1}{mag}$.
The most populated bins for C-, S-, and other types
are located at $17.00$, $17.63$, and $\un{17.37}{mag}$, respectively.

We then computed corresponding sizes for all absolute magnitudes
within the most populated bin.
For C-types, for example, $H$ ranged from $16.95$ to $\un{17.05}{mag}$.
To compute $D$, we used the standard formula
\citep{Bowell_1989aste.conf..524B}
\begin{equation} \label{diam}
D = 10^{0.5[6.259-\log_{10}pV-0.4H]}\,.
\end{equation}
From these sizes, we constructed a "tiny" differential SFD $N(D)$,
with binning $\dd D=\un{0.1}{km}$,
as shown in Fig.~\ref{sizes_dist_tax}.
If the differential SFD was unimodal, 
we used the 16\% and 84\% percentiles (to suppress outliers).
If it was bimodal --
because it contained a "mix" of other taxonomies --
we used the 8\% and 92\% percentiles.
We assumed that the observational limit size,
or rather a transition from completeness to incompleteness,
is between these percentiles.
Hence, the observational limits are as follows:
For C-types, $1.7$ and $\un{2.6}{km}$;
S-types, $0.7$ and $\un{1.0}{km}$;
other types, $0.7$ and $\un{2.1}{km}$;
the whole MB, $1.0$~and $\un{2.5}{km}$;
the inner part, $0.5$~and $\un{1.3}{km}$;
the middle part, $0.7$~and $\un{2.0}{km}$; and
the outer part, $1.2$~and $\un{2.8}{km}$.
In the following, the upper bound of each size range is adopted as the respective observational limit.

\subsection{Distributions of sizes} \label{Distribution_of_sizes}

\begin{figure}
    \centering
    \includegraphics[width=8cm]{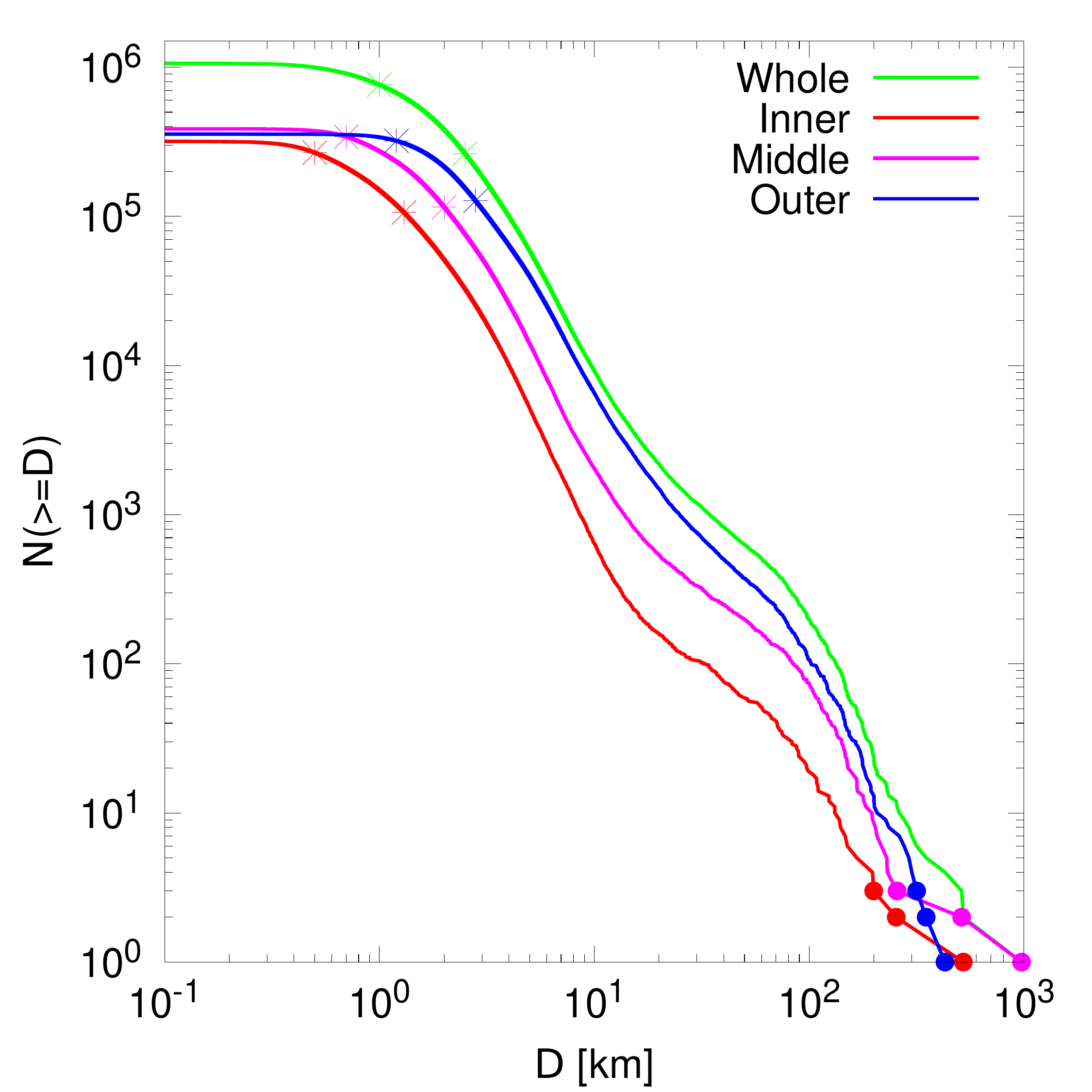}
    \caption{Observed SFD of the entire MB (green) and its individual parts:
inner (red), middle (magenta), and outer (blue).
    We highlight the three largest asteroids in each region: inner -- (19)~Fortuna, (7)~Iris, and (4)~Vesta (red points);
middle -- (15)~Eunomia, (2)~Pallas, and (1)~Ceres (magenta points);
    and outer -- (704) Interamnia, (52) Europa, and (10) Hygiea (blue points).
    The "star-shaped" points on each SFD mark the limit size ranges obtained in Sec.~\ref{obs_bias}, below which the data are observationally incomplete for the entire MB (green), inner (red), middle (magenta), and outer (blue) parts.
    The observed SFDs differ significantly,
    even above $100\,{\rm km}$,
    where collisional evolution is negligible,
   implying different initial conditions.
    }
    \label{SFD_whole_MB}
\end{figure}

\begin{figure}[hbt!]
    \centering
    \includegraphics[width=8cm]{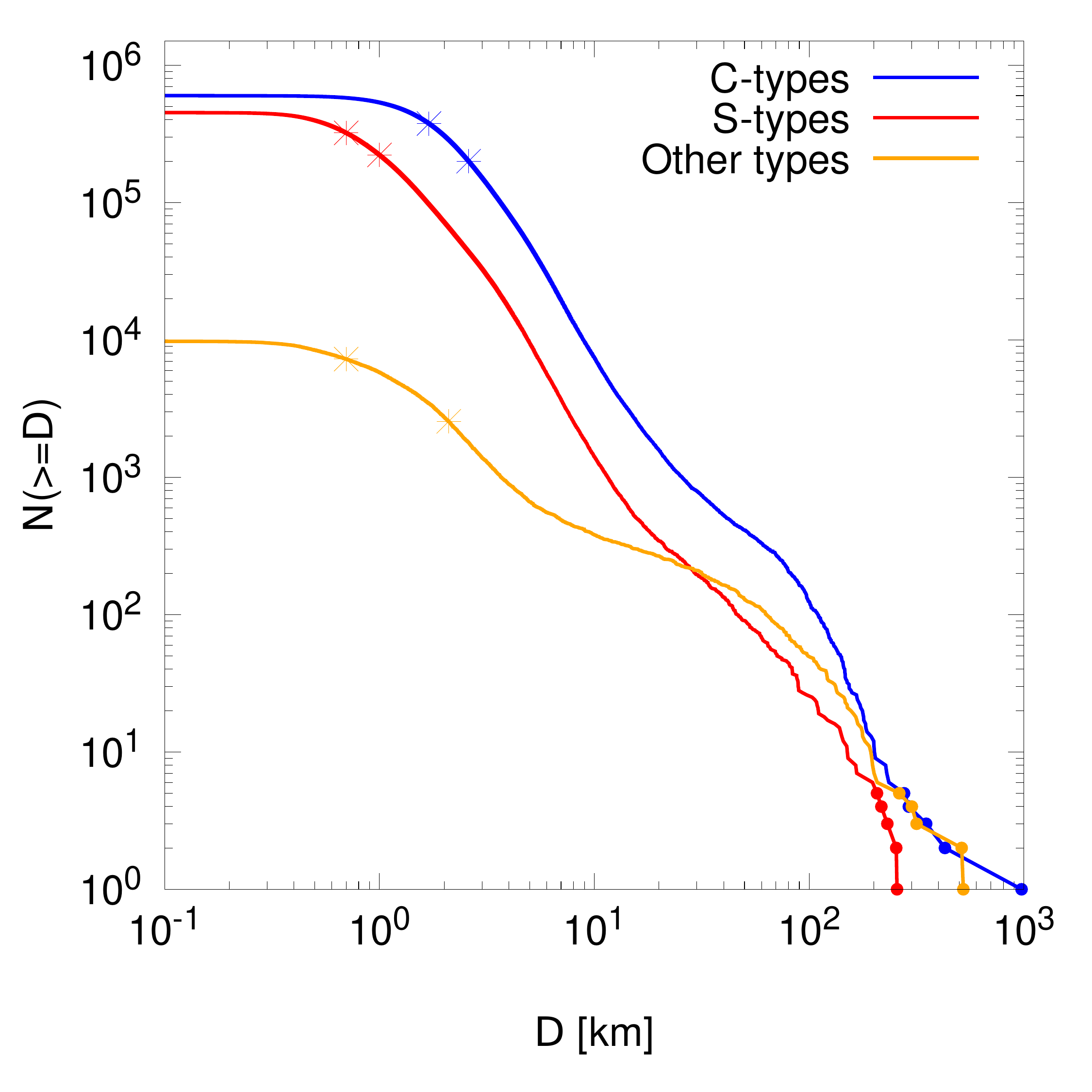}
    \caption{Same as Fig.~\ref{SFD_whole_MB}, but for C-types (blue), S-types (red), and other types (orange). 
     We include: the five largest C-types -- (1)~Ceres, (10)~Hygiea, (52)~Europa, (511)~Davida, and (31)~Euphrosyne  (blue points);
    S-types -- (15)~Eunomia, (7)~Iris, (3)~Juno, (532)~Herculina, and (29)~Amphitrite (red points); and
    other types (4)~Vesta, (2)~Pallas, (704)~Interamnia, (65)~Cybele, and (87)~Sylvia  (orange points).}
    \label{SFD_types}
\end{figure}

Finally, 
we computed sizes for all $1\,065\,034$ asteroids
and constructed the observed cumulative SFDs
for the whole MB, its parts (inner, middle, outer),
and C-type, S-type, and other taxonomies.
The borders of the parts are based on the positions of the Kirkwood gaps
\citep{Kirkwood_1860AJ......6..126K}.
In particular,
the inner borders lie at $2.05$ and $2.502\,\mathrm{au}$,
the middle at $2.502$ and $2.825\,\mathrm{au}$, and
the outer at $2.825$ and $3.25\,\mathrm{au}$.
Substantial differences appear among individual SFDs,
namely
the power-law slopes in different ranges,
the cumulative number of asteroids $N({\geq}D)$,
the presence of "knees" (where the SFDs change slopes),
and the observational limits.
The SFDs are shown in Figs.~\ref{SFD_whole_MB} and \ref{SFD_types}.

The differences are intrinsic properties of the populations.
For the inner, middle, and outer parts, this is obvious,
as they were split according to semimajor axis.
For the C-types, S-types, and other types,
the differences among SFDs are not "artifacts" of our method
(Sec.~\ref{2.2}),
because the surveys' completeness (0.5 to 2.5\,km) is already sufficient to observe them.
Using a different method (e.g., based only on $pV$) would lead to similar results.

\section{Collisional models with fixed rheology} \label{Collisional_models with_fixed_rheology}

Collisional models of the MB vary in complexity. 
For simplicity, single or "mean" rheology is assumed for all populations.
Here, we thoroughly test models with single rheology
but different numbers of populations (one to three).

For all main-belt populations, we prescribed the decay timescales $\tau_{\rm mb}(D)$
for various sizes, as in
\citet{Broz_2024A&A...689A.183B},
in agreement with the observed near-Earth object (NEO) population
(see their Fig. B.3).
The decay timescales $\tau_{\rm mb}$ directly determine the influx of NEOs, while the decay timescales $\tau_{\rm neo}$ of NEOs determine their outflux.
If the influx or outflux is incorrect, the synthetic versus observed
SFDs of NEOs do not match.
For simplicity, we assumed that $\tau_{\rm mb}$ is the same
for all populations. 
However, each population were required to have different
collisional probabilities and impact velocities.

For each model and each population involved,
we set the initial SFDs to follow a pattern similar to the observed SFDs
but with higher cumulative numbers $N({\geq}D)$,
i.e., their initial SFD lies above the observed one.
This excess was used because long-term collisional evolution
inevitably leads to a temporal decrease in $N({\geq}D)$.
The smaller the size $D$, the higher excess.
The observed SFD of each population is different
(cf. Figs.~\ref{SFD_whole_MB} and~\ref{SFD_types});
therefore, the initial SFD for each population must also be different.
In this context, the term "initial" means
after early bombardment by leftover planetesimals
\citep{Nesvorny_2023Icar..39915545N},
after giant-planet instabilities
\citep{Deienno_2024PSJ.....5..110D,Deienno_2025ApJ...986..146D}, and
after implantation of C-type bodies
\citep{Anderson_2025NatAs...9.1464A}.
We do not refer to post-accretion SFDs,
which could be very similar in different parts of the belt.

Using the Monte Carlo collisional code \texttt{Boulder}
\citep{Morbidelli_2009Icar..204..558M,Vernazza_2018A&A...618A.154V},
SFDs were collisionally evolved for up to 4.5~billion years.
Each was evolved 100 times (100 simulations) with different random seeds
due to the intrinsic stochasticity of large disruptions.
The set of synthetic SFDs was then compared to the observed SFD
discussed in Sec.~\ref{Distribution_of_sizes}.
We primarily focused on how many synthetic SFDs lay below or above the observed one.
If the majority (e.g., more than 60) lay significantly below,
we considered the simulation problematic.
This occurred especially at sizes $D < \un{4}{km}$,
i.e., close to the observational limit.
If more asteroids had been detected below this size,
the observed SFD would have increased and differed even more from
the synthetic SFDs.
We thus manually adjusted the initial SFDs
and ran the simulations again.

\subsection{Scaling~law} \label{Scaling~law}

To describe the tendency of a target to be disrupted,
we must directly measure or simulate collisions 
between targets and projectiles under different scenarios --
namely, with different impact velocities, impact angles, and different sizes.
In case of direct measurements, we are limited to very small target sizes; for example,
\cite{Gault_NASA} used centimeter-sized rocky targets, while
\citet{Fujiwara_1977Icar...31..277F}, \citet{Davis_1990Icar...83..156D}, and \citet{Nakamura_1992Icar..100..127N}  used up to 10$\,\mathrm{cm}$-sized, and \citet{Capaccioni_1986Icar...66..487C} 30$\,\mathrm{cm}$-sized ones.
On the other hand, as far as simulations are concerned, we can use results from hundreds of SPH simulations 
performed on much larger targets than those in laboratory experiments (e.g., \cite{Benz_1994Icar..107...98B,Benz_1995CoPhC..87..253B,Benz_1999Icar..142....5B,Love_1996Icar..124..141L,Durda_2007Icar..186..498D,Jutzi_2008Icar..198..242J,Jutzi_2010Icar..207...54J,Jutzi_2015P&SS..107....3J,Schwartz_2016AdSpR..57.1832S,Sevecek_2017Icar..296..239S,Sevecek_2019A&A...629A.122S,
Vernazza_2018A&A...618A.154V,Rozehnal_2022Icar..38315064R}).
As a result of such calibrated simulations, we can obtain the so-called scaling~law for the target.

We adopted the scaling~law from \cite{Benz_1999Icar..142....5B} in the form
\begin{equation} \label{scalling_law}
Q^*(D)\equiv {1\over q}\left[Q_0\left(\frac{D}{D_0}\right)^a+B\rho \left(\frac{D}{D_0}\right)^b\right],
\end{equation}
where $D\,[\mathrm{cm}]$ is the target size (diameter); 
$D_0\,[\mathrm{cm}]$ is the normalization size; 
$Q^*\,[\mathrm{erg\,g^{-1}}]$ is the energy of the impact per unit mass of the target required to disperse 50\% of the target \citep{Durda_1998Icar..135..431D,O'Brien_2003Icar..164..334O}; 
$a,\,b,\,B\,[\mathrm{erg \, g^{-2}}],\,q$, and $Q_0\,[\mathrm{erg \, g^{-1}}]$ are free parameters; and 
$\rho\,[\mathrm{g\,cm^{-3}}]$ is the density of the target.

As shown by \citet{Benz_1999Icar..142....5B}, the free parameters $a$, $b$, $B$, and $Q_0$ depend
on the impact velocity as well as on the material.
They performed several SPH simulations to derive the free parameters for basalt and ice.
We used their scaling~law for basalt,
which was slightly adjusted
(see the discussion in \citealt{Vernazza_2018A&A...618A.154V}).
The respective parameters were
$q = 1$,
$D_0 = \un{2}{cm}$,
$\rho = \un{3}{g\,cm^{-3}}$,
$Q_0 = \un{9\cdot10^7}{erg\,g^{-1}}$,
$B = \un{0.5}{erg\,g^{-2}\,cm^3}$,
$a = -0.53$, and
$b = 1.36$.

\subsection{Collisional probabilities and impact velocities} \label{Determine_collprob_and_impvel}

We determined the collisional probabilities and impact velocities between individual populations using
the Öpik formalism \citep{Opik_1951PRIA...54..165O,Greenberg_1982AJ.....87..184G,Bottke_1993GeoRL..20..879B}.
The quantities were computed for individual pairs of asteroid orbits,
based on their osculating semimajor axis, eccentricity, and inclination.
To reduce computational time, we selected 1000 non-repeating random orbits from each population.
The resulting mean values are summarized in Tab.~\ref{collprob_and_impvel_tab}.

\subsection{Initial differential SFDs} \label{generate_SFD}

We generated initial differential SFDs as a set of points:
$\{D_i,{\rm d}N(D_i)\}_{i=1}^n$. 
These points follow a pattern prescribed by a piece-wise power-law function,
with a negative slope(s) $\gamma$ for a specific size range(s),
${\rm d}N(D) = CD^\gamma$, where $C$ is a constant.
The number of slopes is three, four, or five; 
the slopes,
along with their respective size ranges, are considered free parameters.
The bins were logarithmically spaced,
and neighboring bin centers had a fixed ratio of 1.15.
The normalization number $n_{\text{norm}}$
for the normalization size $d_{\text{norm}}$
is given for the corresponding cumulative distribution,
$n_{\text{norm}} = N({\geq}d_{\text{norm}})$.
The respective relation is
$N(\geq\!\!D) = \int_D^\infty{\rm d}N(D)$,
and the power-law slope thus differs by one:
$N(\geq\!\!D) = C'D^{\kappa}$, where $C'$ is a constant and $\kappa=\gamma+1$.
Additionally, the initial SFDs did not include the largest asteroids;
therefore, we added them manually to respective populations.

\subsection{One main belt population} \label{One Main Belt population}

As a verification, we simulated the collisional evolution of the entire MB population.
We tested two different types of the initial SFDs
-- "no tail" and "with tail."
Both followed a pattern similar to the observed SFD, down to $D=\un{4}{km}$.
The three largest asteroids
(1) Ceres, (2) Pallas, and (4) Vesta were added manually.
Both types of initial SFDs, along with their respective synthetic SFDs, are shown in Figs.~\ref{SFD_whole_MB_2},~\ref{SFD_whole_MB_9}.
We see that both initial conditions
resulted in synthetic SFDs matching the observed SFD,
in agreement with \cite{Bottke_2005Icar..175..111B,Bottke_2020AJ....160...14B}.
Therefore, it does not matter which type is chosen; for the initial SFDs in subsequent models,
we used only the first type.

\subsection{Inner, middle, and outer parts} \label{parts_same_rheo}

\begin{figure*}
\centering
\begin{tabular}{c@{\kern-.3cm}c@{\kern-.3cm}c}
\kern.3cmInner &
\kern.3cmMiddle &
\kern.3cmOuter \\[-.2cm]
\includegraphics[width=6.5cm]{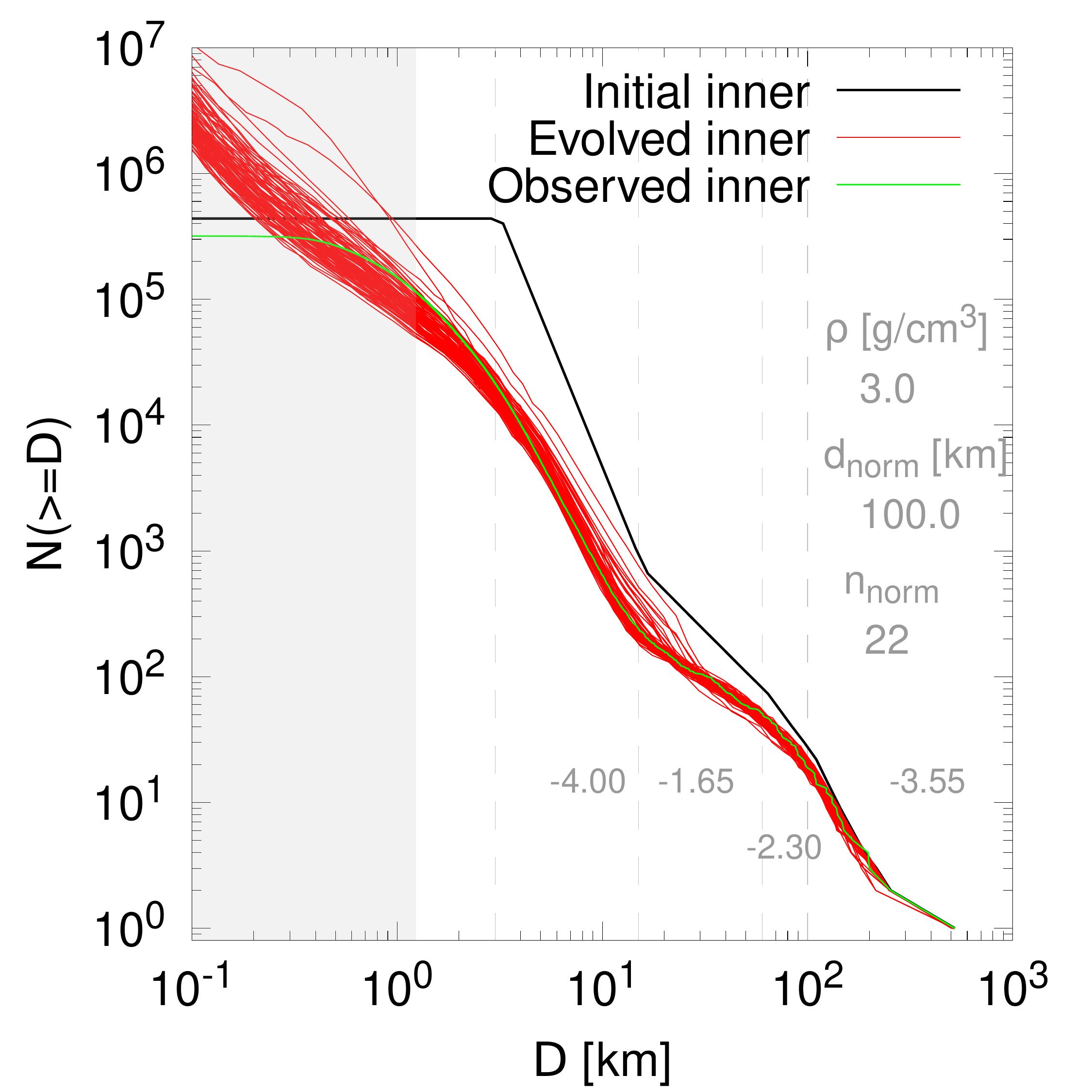} &
\includegraphics[width=6.5cm]{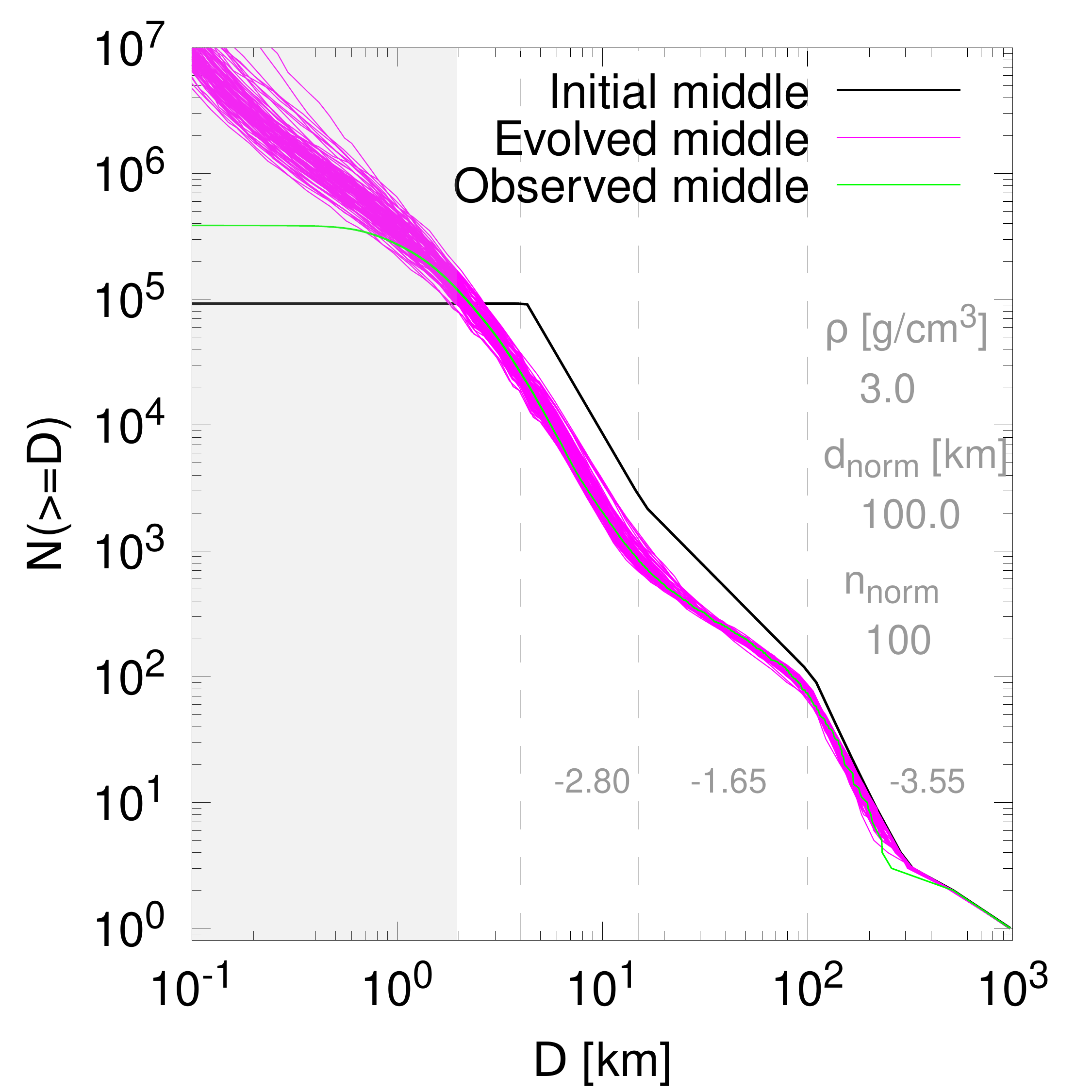} &
\includegraphics[width=6.5cm]{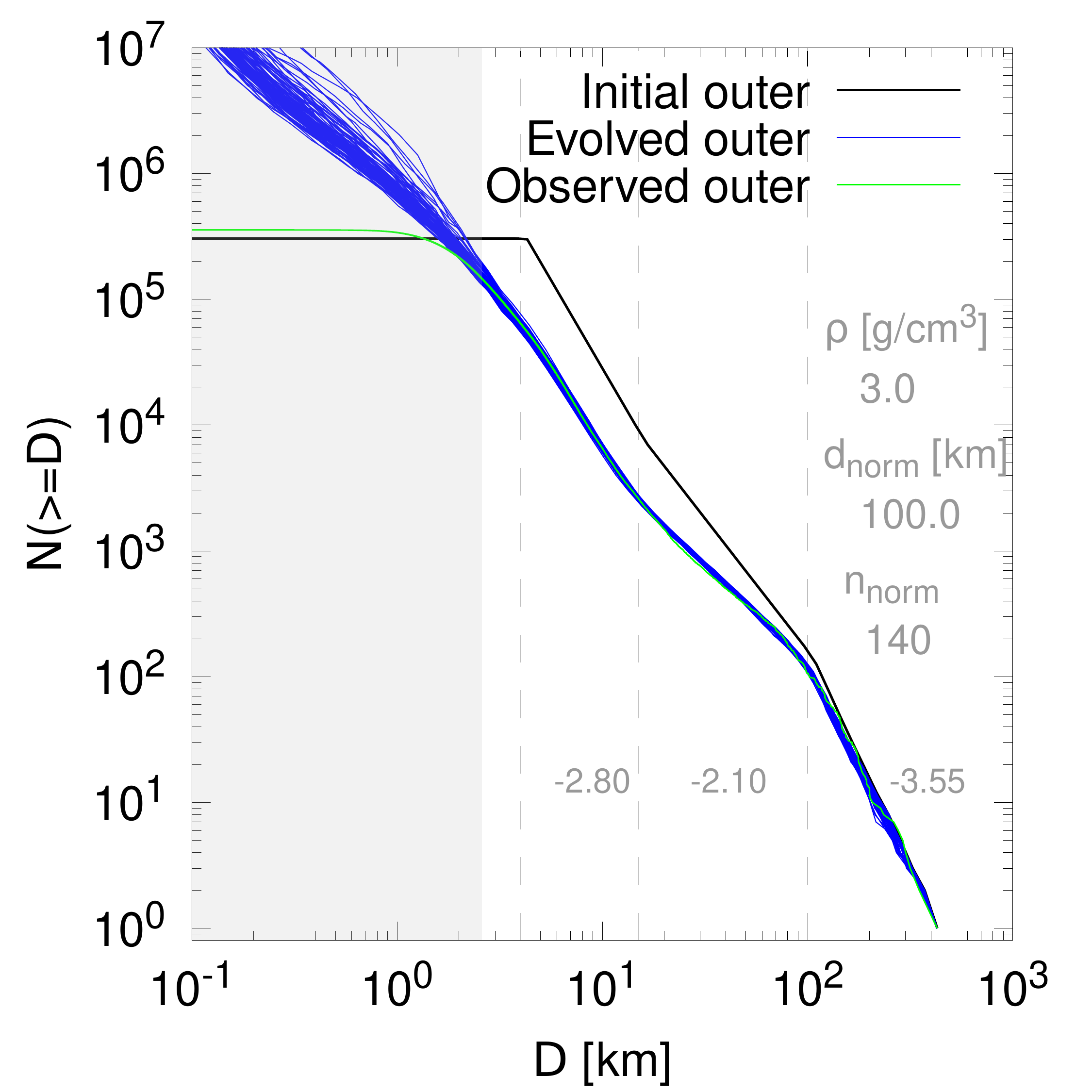} \\
\end{tabular}
\caption{Initial SFD (black), 100~evolved SFDs of the inner (red), middle (pink), and outer (blue) parts, and their respective observed SFD (green).
The vertical dashed gray lines represent size ranges
where the initial SFD was approximated by a different power law, i.e., a different slope.
For the inner part, for example, the slopes are
$-3.55$ above $\un{100}{km}$,
$-2.30$ between $100$ and $\un{60}{km}$,
$-1.65$ between $60$ and $\un{15}{km}$,
and $-4.00$ between $15$ and $\un{3}{km}$.
The last slope below $\un{3}{km}$ was set to zero,
i.e., a constant cumulative SFD ("no tail").
Individual slopes are shown in gray within the panels at the respective size ranges.
Density $\rho$, normalization size $d_{\text{norm}}$, and the normalization number $n_{\text{norm}}$ for each part are shown on the right side of the respective panel.
The gray areas correspond to the size ranges below the observational limits.
The problem in the inner part is that most evolved SFDs lie below the observed SFD below $D \sim 3\,{\rm km}$.
}
\label{sfd_4500_inner_43}
\end{figure*}

When we split the MB into three parts -- inner, middle, and outer -- the situation was different.
It was very difficult to set the correct initial SFDs for each part:
changing one parameter in one initial SFD affected the evolved~SFDs of all three parts
because the respective collisional probabilities are nonzero
(Tab.~\ref{collprob_and_impvel_tab}).
For the inner belt, in particular, we tested a number of different slopes:
$\kappa_4 \in \{-4.0, -4.5, -5.0\}$,
$\kappa_3 \in \{-1.8, -1.7, -1.65\}$, and
$\kappa_2 \in \{-2.2, -2.3, -2.4\}$.
Our best initial SFDs, along with each simulation, are shown in
Fig.~\ref{sfd_4500_inner_43}.

Overall, the sum of all three SFDs matches the observed SFD
(Fig.~\ref{sfd_4500_sum_43}).
However, there are imperfections in the individuals parts:
i)~every evolved~SFD of the outer part lies above the observed one
between $21$ and $\un{47}{km}$;
ii)~the majority of the evolved~SFDs of the middle part lies above the observed one between $5$ and $\un{10}{km}$; and 
iii)~the majority of the evolved~SFDs of the inner part lies below the observed one
below $D\approx\un{3}{km}$.
The last imperfection is the most crucial 
because more than $90\%$ of evolved~SFDs of the inner part lie below the observed SFD.
This imperfection persists to sizes below the observational limit at $\un{1.3}{km}$,
which is even worse, because the debiased observed SFD would lie even further above.
We found no "triad" of initial SFDs that  
resolves this deficiency.

\subsection{C-types, S-types, and other types} \label{C_S_other_def_rheo}

\begin{figure*}
\centering
\begin{tabular}{c@{\kern-.3cm}c@{\kern-.3cm}c}
\kern.3cm C-types &
\kern.3cm S-types &
\kern.3cm other types \\[-.2cm]
\includegraphics[width=6.5cm]{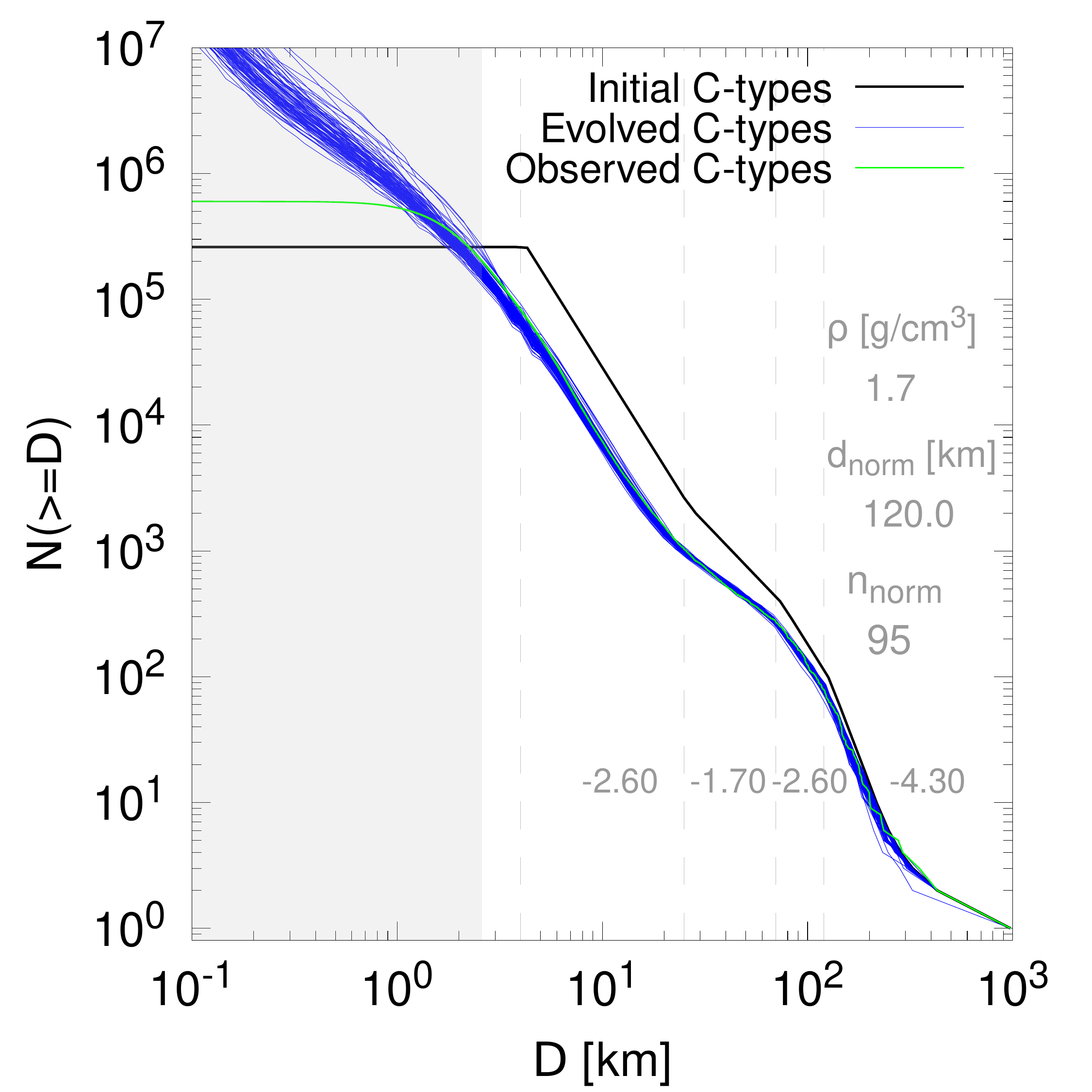} &
\includegraphics[width=6.5cm]{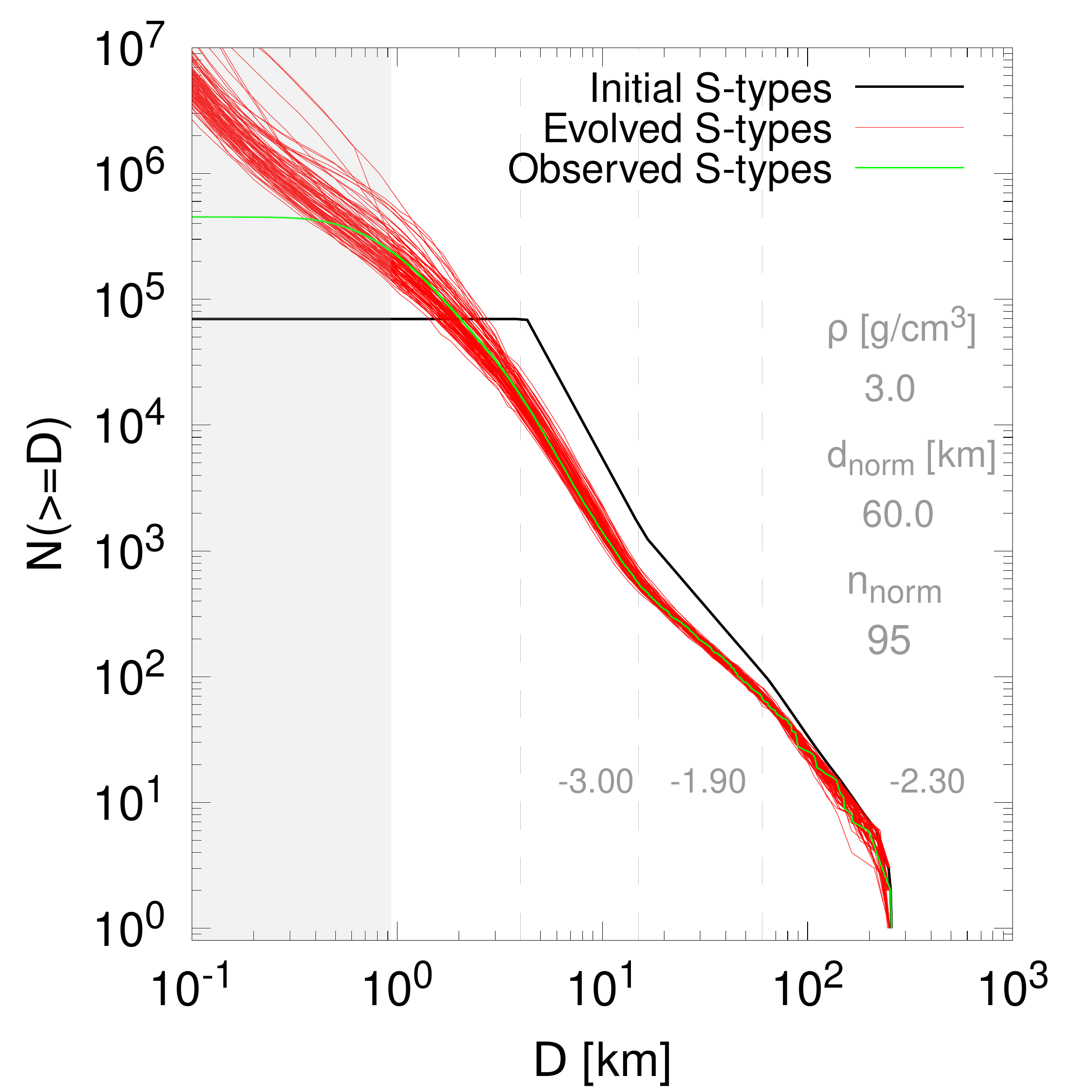} &
\includegraphics[width=6.5cm]{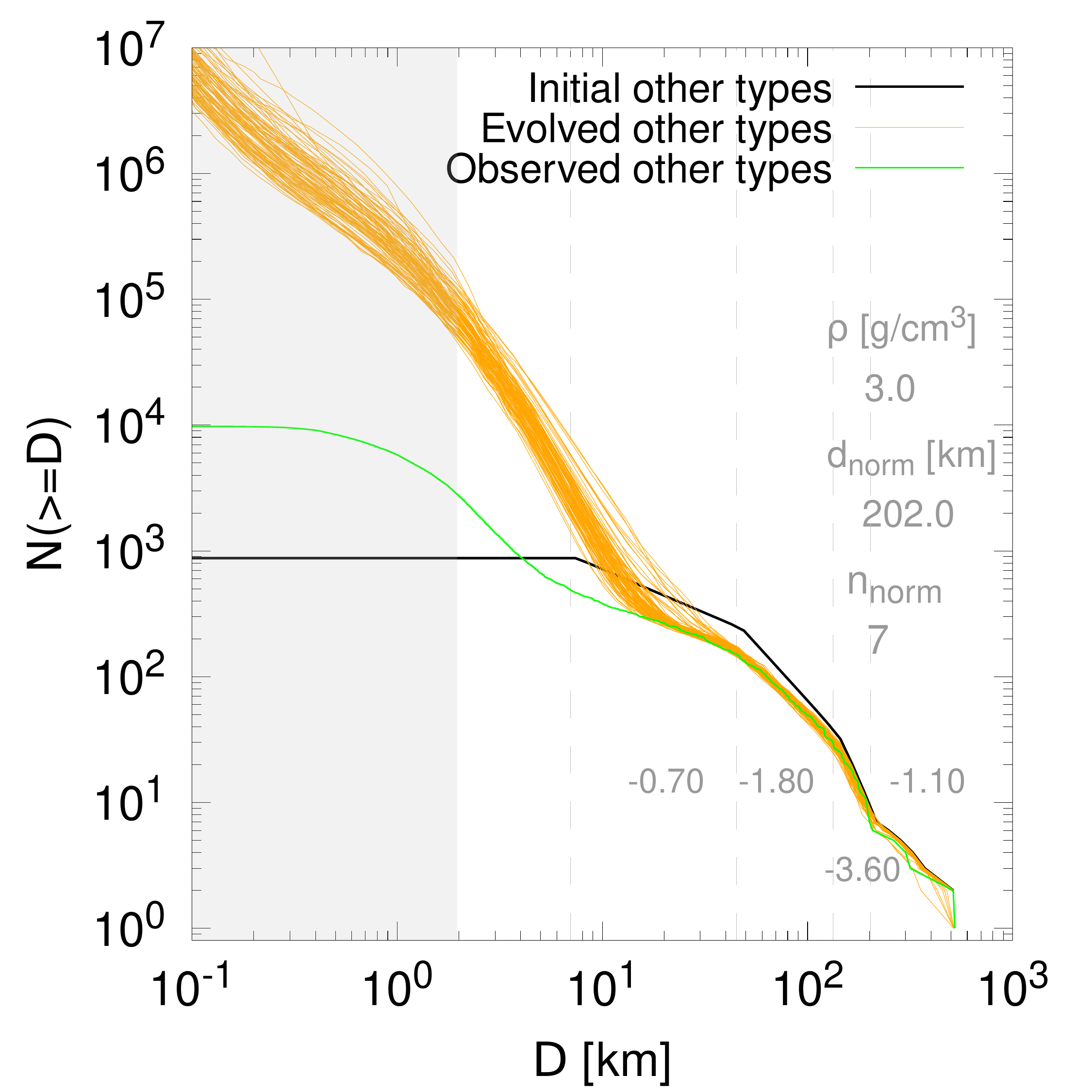} \\
\end{tabular}
\caption{
Same as Fig.~\ref{sfd_4500_inner_43}
for the C-types, S-types, and other types.
The problem with both C- and S-types is that most evolved SFDs
lie below the observed SFD
below $D \sim 7$ and 4\,km, respectively.
Conversely, evolved SFDs of other types always lie above the observed SFD
(see text for discussion).
}

\label{sfd_4500_C_final}
\end{figure*}

When we split the MB into three taxonomies -- C-types, S-types, and other types -- the situation was the same.
C-type density was set to $1.7\,\mathrm{g/cm^3}$, 
and S-type density to $3.0\,\mathrm{g/cm^3}$ \citep{Vernazza_2021A&A...654A..56V}.
Our best initial SFDs, along with individual simulations, are shown in Fig.~\ref{sfd_4500_C_final}.

Both C- and S-types exhibit the same problem as in the inner part.
In~particular, the most evolved~C- and S-types lie below the respective observed ones
below $7$ and $\un{4}{km}$, respectively.
The most crucial problem, though, is the fact that $94$ and $69$ evolved~SFDs of C- and S-types, respectively,
lie below their observed~SFDs at the observational limits
($\un{2.6}{km}$ for C-types; $\un{1}{km}$ for S-types).
We found no initial~SFDs
that resolved these deficiencies.

\paragraph{Problem with other types.}

Moreover, the observed~SFD of other types is clearly too shallow
(Fig.~\ref{sfd_4500_C_final}, right).
The "tail" should always be Dohnanyi-like, with a slope of approximately $-2.5$.
Indeed, evolved~SFDs of other types are always steep below $D\approx\un{10}{km}$.
The observed other types must therefore be incorrect,
i.e., some observed (assigned) C- and S-types must actually be other types.
Conversely,
including all other types in C- or S-types could potentially solve these deficiencies.

We thus tried to reassign all observed other types to C-types, creating a "mix" of C- and other types.
Of course, the initial SFDs of this "mix",
along with S-types, had to be modified
due to their mutual collisions.
This partially solved the problem,
but only for C-types.
Out of 100 simulations,
only 40\% were below the observed SFD (at or below 2.6\,km),
which is acceptable.
However, S-type deficiency remained:
out of 100 simulations,
70\% were below the observed SFD (at or below 1.0\,km).
Consequently, this is only a partial solution.

We also tried to reassign other types to S-types,
but this model was inconsistent. This "mix" exhibited
an excess at multi-kilometer sizes, which contradicts
our results for the inner belt (Sec.~\ref{parts_same_rheo}).
Moreover, SMASSII spectra of other types were mostly
B-types (with affinity to C-types) and X-types,
which appear distributed across the belt --
unlike S-types, which should prevail in the inner belt.

Another solution would be to modify our algorithm for assigning C- and S-types,
i.e., to tighten the condition for classifying asteroids with unknown taxonomy
as C- or S-type, making them less numerous. However, the "ultimate" solution is to modify the rheology (scaling~law)
for S-types
because SFD characteristics depend sensitively on the slope and, in particular,
the position of the scaling-law minimum \citep{Marschall_2022AJ....164..167M}.
On the other hand, the rheology of C-types
should remain similar to the entire MB,
because C-types comprise ${\sim}80\%$ of the multi-kilometer population
(Fig.~\ref{SFD_types}).

\section{Collisional models with variable rheology} \label{Multi-population models with variable rheology}

In more complex models,
each population can have distinct rheology.
In the case of C- and S-types,
differences arise due to their chemical and mineralogical composition:
C-types are carbonaceous, and S-types are silicate.
\citet{Vernazza_2021A&A...654A..56V} showed that C- and S-types differ further in
bulk density
($1.7$ vs. $3.0\,\mathrm{g\,cm^{-3}}$ for $D \gtrsim 100\,{\rm km}$);
porosity
($20$ vs. $8\%$); and
shape
(higher sphericity index for C-types).
Furthermore, the ratio between the critical periods (the spin barrier) of C- and S-types
depends on their cohesive strengths;
\citet{Carbognani_2017P&SS..147....1C} estimated that the cohesive strength
of C-types is higher than that of S-types by $40\%$.
All these properties were inferred from adaptive-optics imaging surveys
\citep{Vernazza_2021A&A...654A..56V},
spectroscopic surveys
\citep{Bus_2002Icar..158..106B,Binzel_2019Icar..324...41B},
and light curve inversion
\citep{Warner_2009Icar..202..134W,Durech_2010A&A...513A..46D,Durech_2019A&A...631A...2D}.

Regarding material properties, 
laboratory measurements were conducted on meteorites,
with compositions similar to C- and S-types --
namely, carbonaceous and ordinary chondrites.
They primarily contain different amounts of
carbon or silicates \citep{Hutchison2007} and
also differ in bulk and grain density and
hence porosity \citep{Consolmagno_2008ChEG...68....1C,Macke_2011M&PS...46.1842M}.
All these examples strongly suggest that C- and S-types have distinct rheology. 

\begin{figure}[hbt!]
    \centering
    \includegraphics[width=7.5cm]{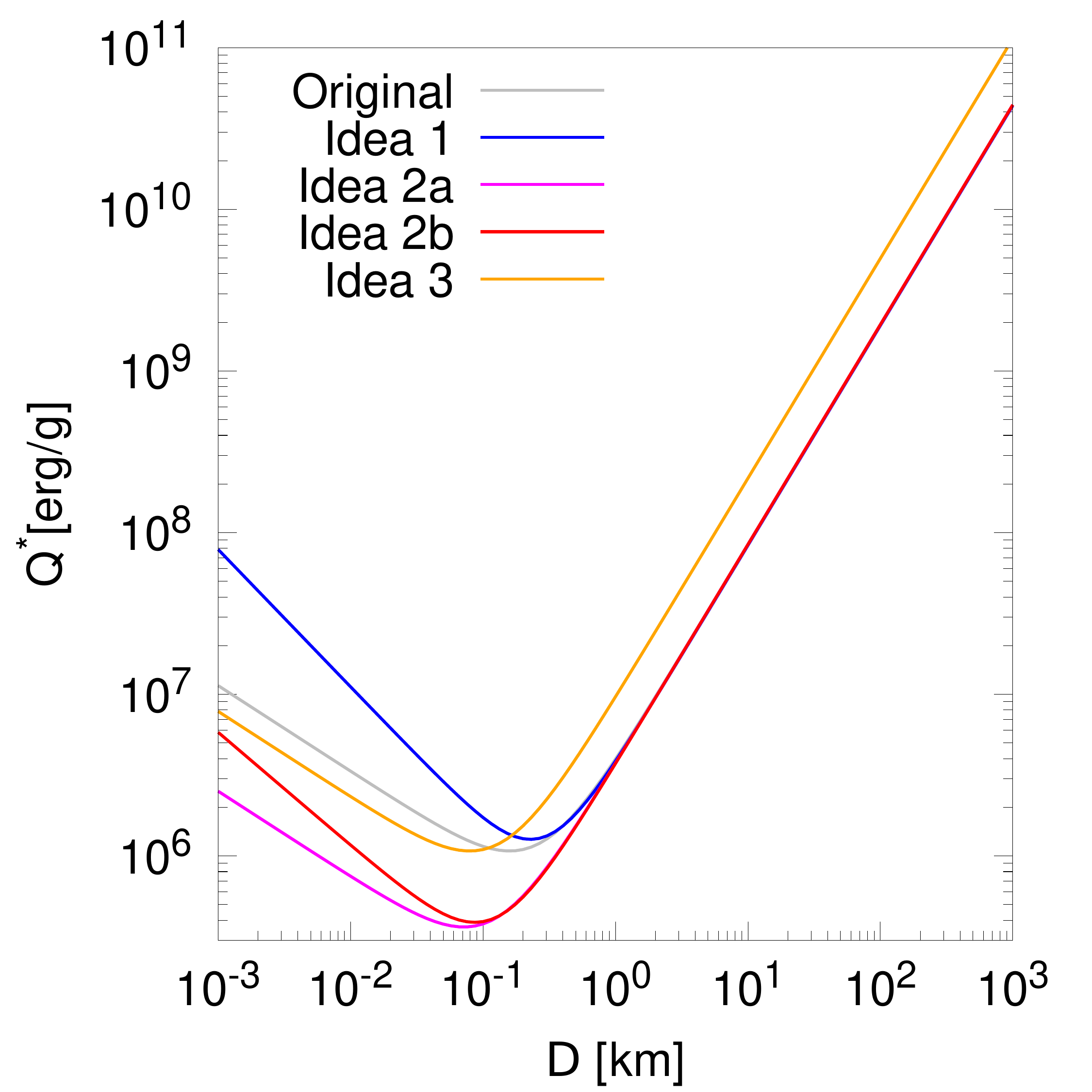}
    \caption{
Original scaling~law (gray) and its modifications (labeled here as "ideas"), plotted in blue, pink, red, and orange.
}
    \label{idea_scaling_laws}
\end{figure}

To solve the deficiencies of the inner part and S-types (Secs.~\ref{parts_same_rheo}, \ref{C_S_other_def_rheo}),
we tried modifying the scaling~law.
We proposed three ideas.
Specifically, we either made asteroids
\begin{enumerate}
    \item below the scaling-law minimum stronger,
    so that they more easily disrupted those at the minimum, 
    making them less numerous and thus unable to disrupt even bigger ones; or
    \item made asteroids at and below the minimum weaker, so that they did not disrupt those above the minimum; or
    \item made asteroids below the minimum weaker and those above stronger to preserve them.
\end{enumerate}
In the context of Eq.~(\ref{scalling_law}), these ideas correspond to
properly lowering $a$, increasing $D_0$, and lowering $q$ (i.e., idea~1);
lowering $Q_0$ (idea~2a) or $a$ (idea~2b); or
lowering $D_0$ (idea~3);
which simultaneously shifts the scaling-law minimum,
as shown in Fig.~\ref{idea_scaling_laws}.

We tested these ideas on the entire MB population to determine
which ones substantially increase the SFDs
at sizes near (and below) the observational limit.
We rejected idea~1, because it did not substantially increase the SFDs. 
We also rejected idea~2a because it resulted in roughly the same increase as idea 2b.
Hereinafter, we refer to idea~2b as idea~2.

\subsection{Fitting scaling laws -- inner, middle, and outer} \label{Fitting scaling laws --- inner, middle and outer}

\begin{figure}[hbt!]
    \centering
    \includegraphics[width=7.3cm]{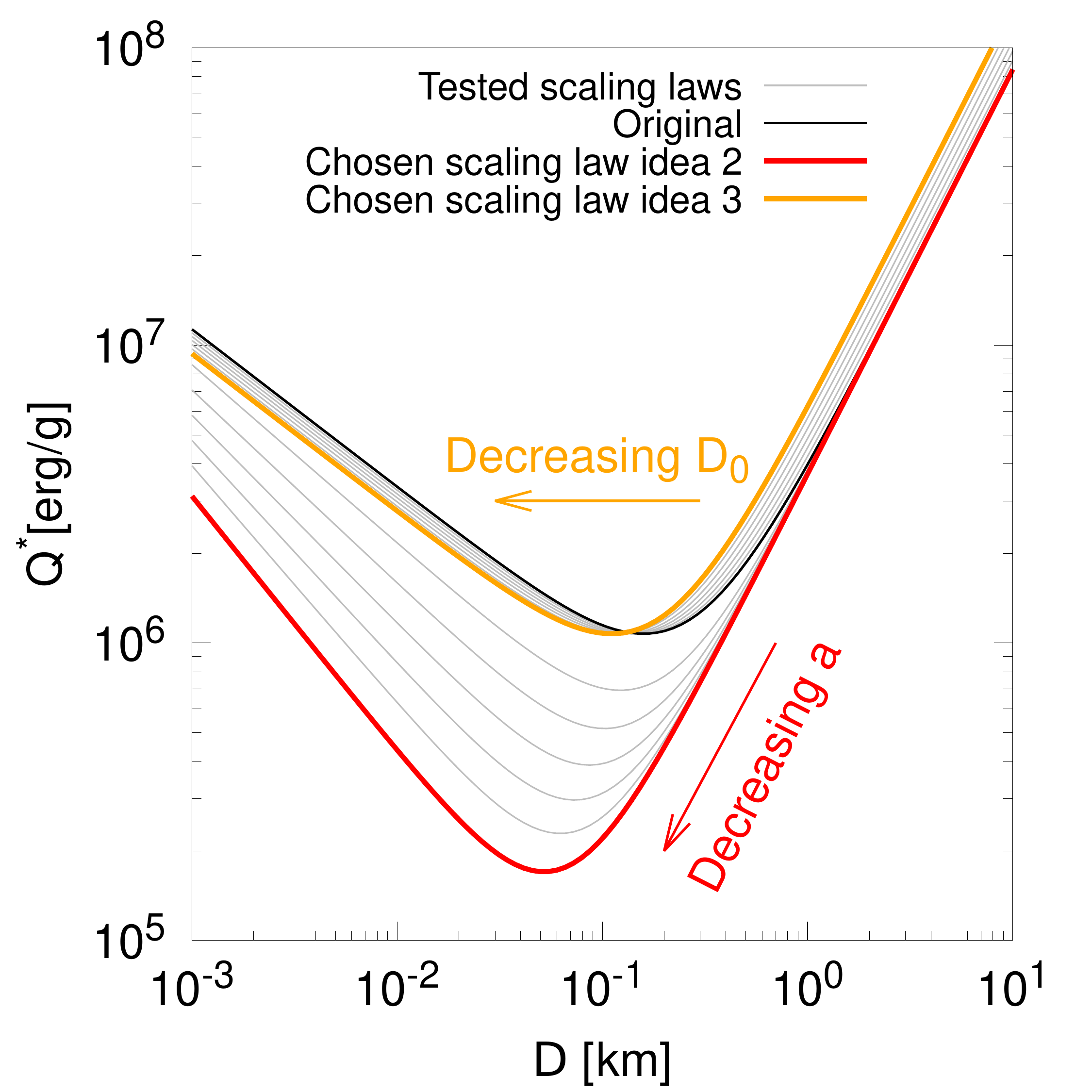}
    \caption{
    Twelve scaling laws (gray) tested for the inner belt,
    compared to the original (black) from Sec.~\ref{parts_same_rheo}. 
    Those, which substantially increased the evolved SFDs below the observational limit, are shown in red and orange.
   The directions of scaling~law minimum changes for decreasing $a$ and $D_0$ are shown by red and orange arrows, respectively.
    }    
    \label{scaling_law_inner}
\end{figure}

For the inner part of the MB,
we tested 12 different scaling~laws.
For idea 2, we used values
$a\in\{-0.60, -0.65, \dots, -0.86\}$,
while for the idea 3, we used 
$D\in\{1.9, 1.8, \dots, 1.4\}\,\mathrm{cm}$ (cf. $a=-0.53$ and $D_0=\un{2}{cm}$ in Sec.~\ref{Scaling~law}).
The corresponding set of scaling~laws is shown in Fig.~\ref{scaling_law_inner}.

\begin{figure}[hbt!]
\centering
\begin{tabular}{c}
\kern.5cm Inner \\[-0.2cm]
\includegraphics[width=8cm]{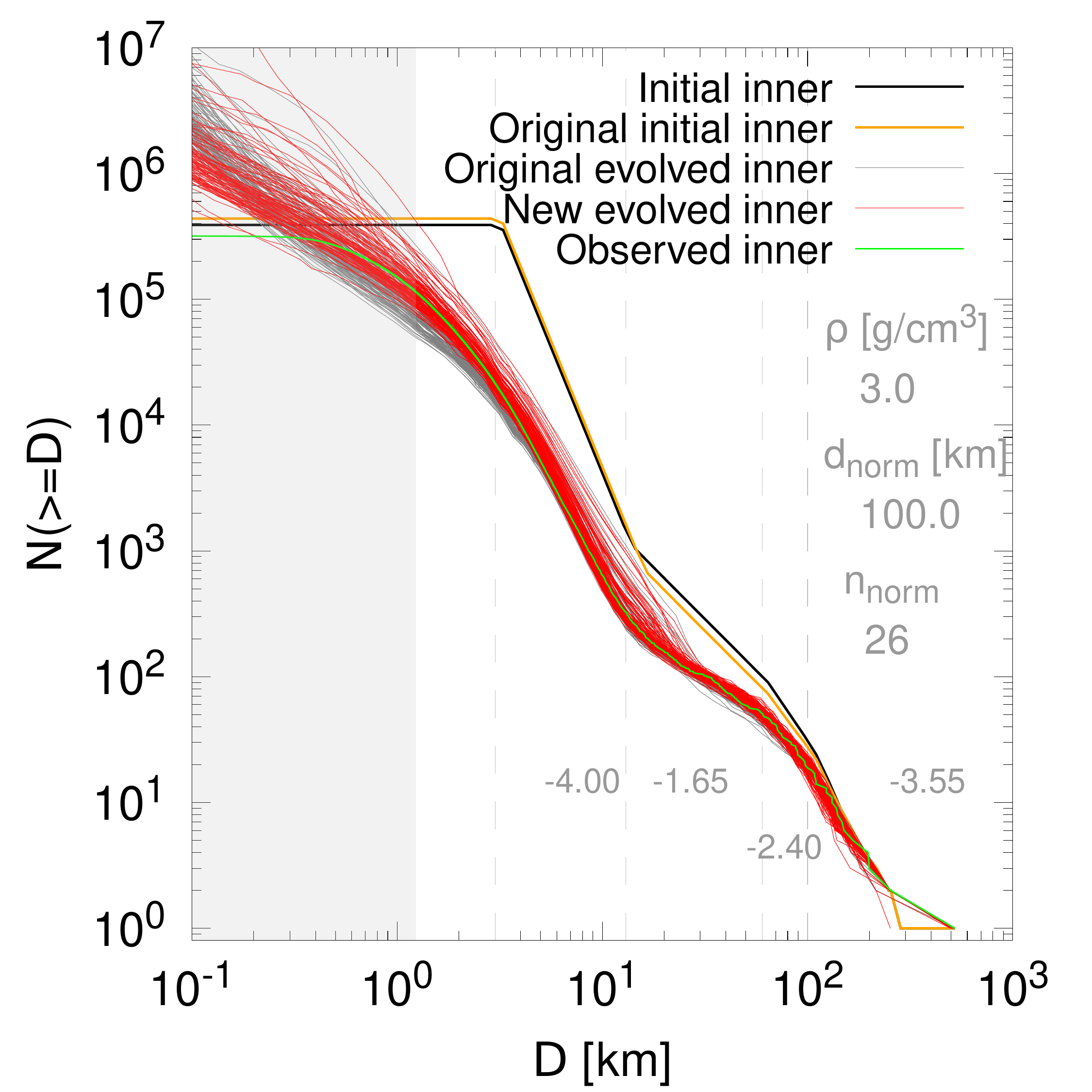} \\
\end{tabular}
\caption{
Same as Fig.~\ref{sfd_4500_inner_43},
for the inner part and a modified scaling~law
("idea 2").
For comparison, we also plot SFDs for the original scaling~law
(gray~lines, corresponding to the left panel in Fig.~\ref{sfd_4500_inner_43}).
The synthetic SFDs of the middle and outer belt did not change considerably.
}
\label{sfd_4500_inner_14_4_3_3_100_BEST}
\end{figure}

Compared to the SFDs for the original scaling~law
(left panel in Fig.~\ref{sfd_4500_inner_43}),
the SFD of the inner belt increased most for $a=-0.86$,
without considerable change in the SFDs of the other two parts.
However, the initial conditions were modified accordingly;
we tried 25 different initial SFDs.
The best final SFDs, in comparison to those observed, are shown in
Fig.~\ref{sfd_4500_inner_14_4_3_3_100_BEST}.
At the observational limit ($D=\un{1}{km}$), 
52\% of the final~SFDs lies below the observed SFD,
which is acceptable.
Unfortunately, a minor excess appears
between approximately $3$ and $\un{7}{km}$.
Some initial conditions resolved this excess; however,
they led to a substantial decrease
below the observational limit,
which would eventually bring us back to the original problem.
Still, we conclude that idea 2 is acceptable.
For the idea 3, we obtained comparable results if
$D_0=\un{1.4}{cm}$,
but at the expense of considerable excess of multi-kilometer bodies;
we thus consider idea 3 less acceptable.

\subsection{Fitting scaling~laws for C- and S-types}

Since the inner part is mostly composed of S-types, we used the same modified scaling law for them as for the inner part (idea~2)
to see if it solved the deficiency at small sizes.
For C-types, we kept the original
scaling~law, namely, the one used in Sec.~\ref{Collisional_models with_fixed_rheology}.
To select the best pair of initial SFDs for C- and S-types, 
we defined the goodness of the fit in terms of the "pseudo" $\chi^2$
\citep{Cibulkova_2014Icar..241..358C}.
It is called "pseudo" because we used only formal uncertainties
and did not perform a statistical $\chi^2$ test
to determine a statistical significance of the fit.
The definition of $\chi^2$ is discussed in Appendix~\ref{Definition of chi}.

We tested six different initial SFDs for C-types.
For S-types, we tried the original plus two modifications
because we expected an increase in C-types to create more projectiles that could potentially disrupt some S-types.
The total number of initial SFD pairs was 18
(Fig.~\ref{sfd_initial}).
For each pair, we computed the collisional evolution
and corresponding~$\chi^2$.
The $\chi^2$ was evaluated between the observational limit
and $\un{120}{km}$ and $\un{60}{km}$,
for C- and S-types, respectively.
We chose these limits 
because the observed SFDs are "jaggy" above these sizes,
which would bias $\chi^2$.

\begin{figure*}[hbt!]
\centering
\begin{tabular}{c@{\kern-.3cm}c@{\kern-.3cm}c}
\kern.5cm C-types &
\kern.5cm S-types &
\kern.5cm Best-fit scaling~laws\\[-.2cm]
\includegraphics[width=6.5cm]{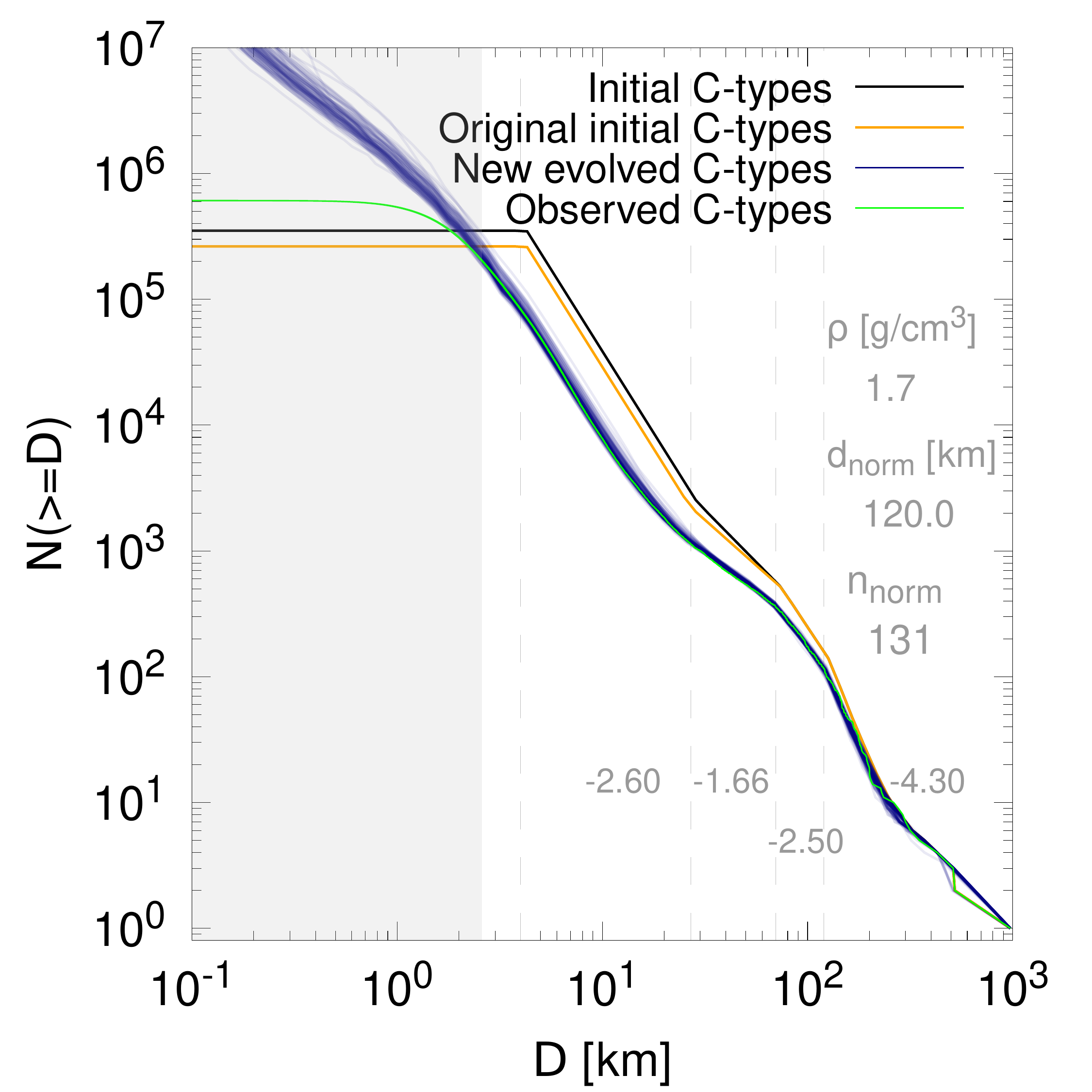} &
\includegraphics[width=6.5cm]{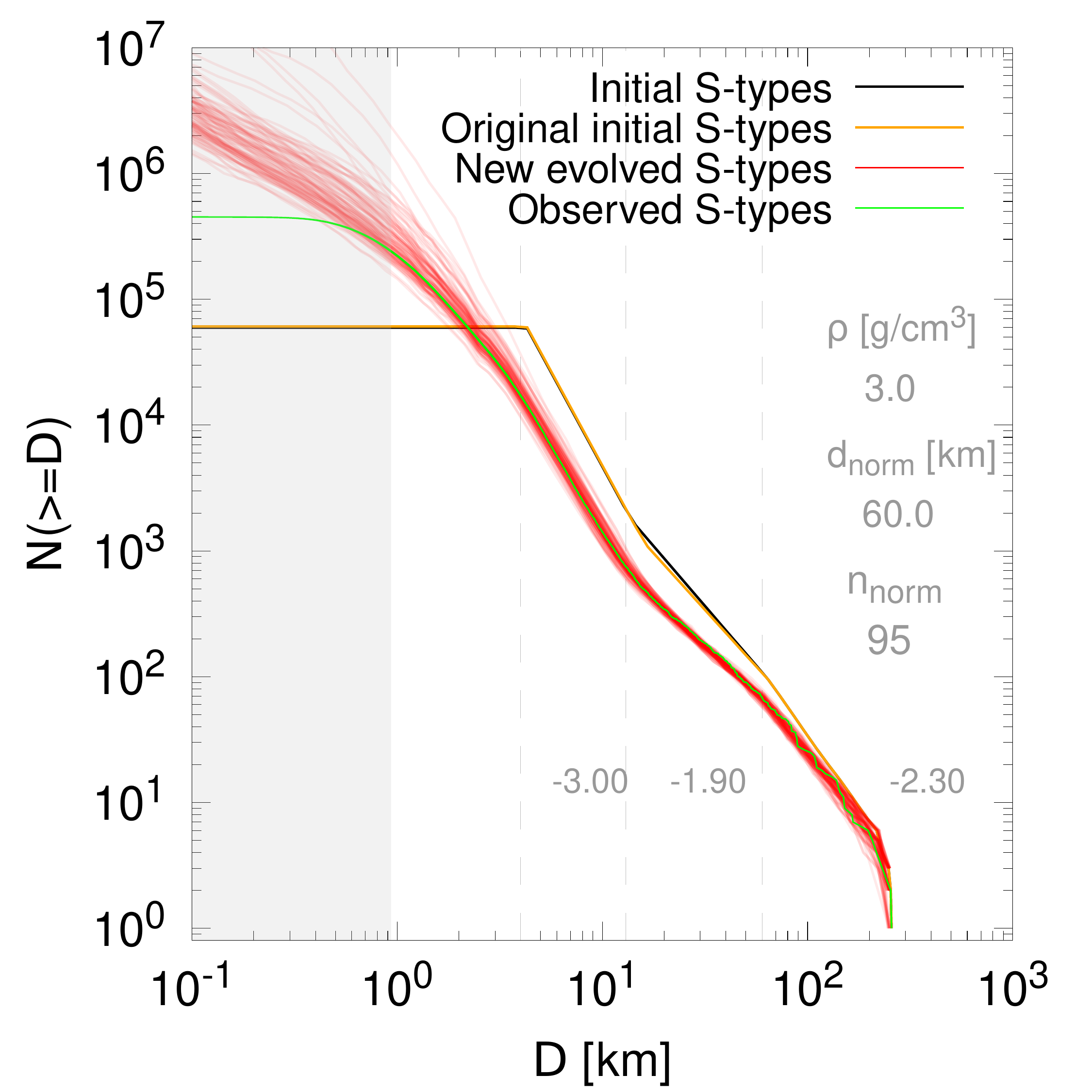} &
\includegraphics[width=5.8cm]{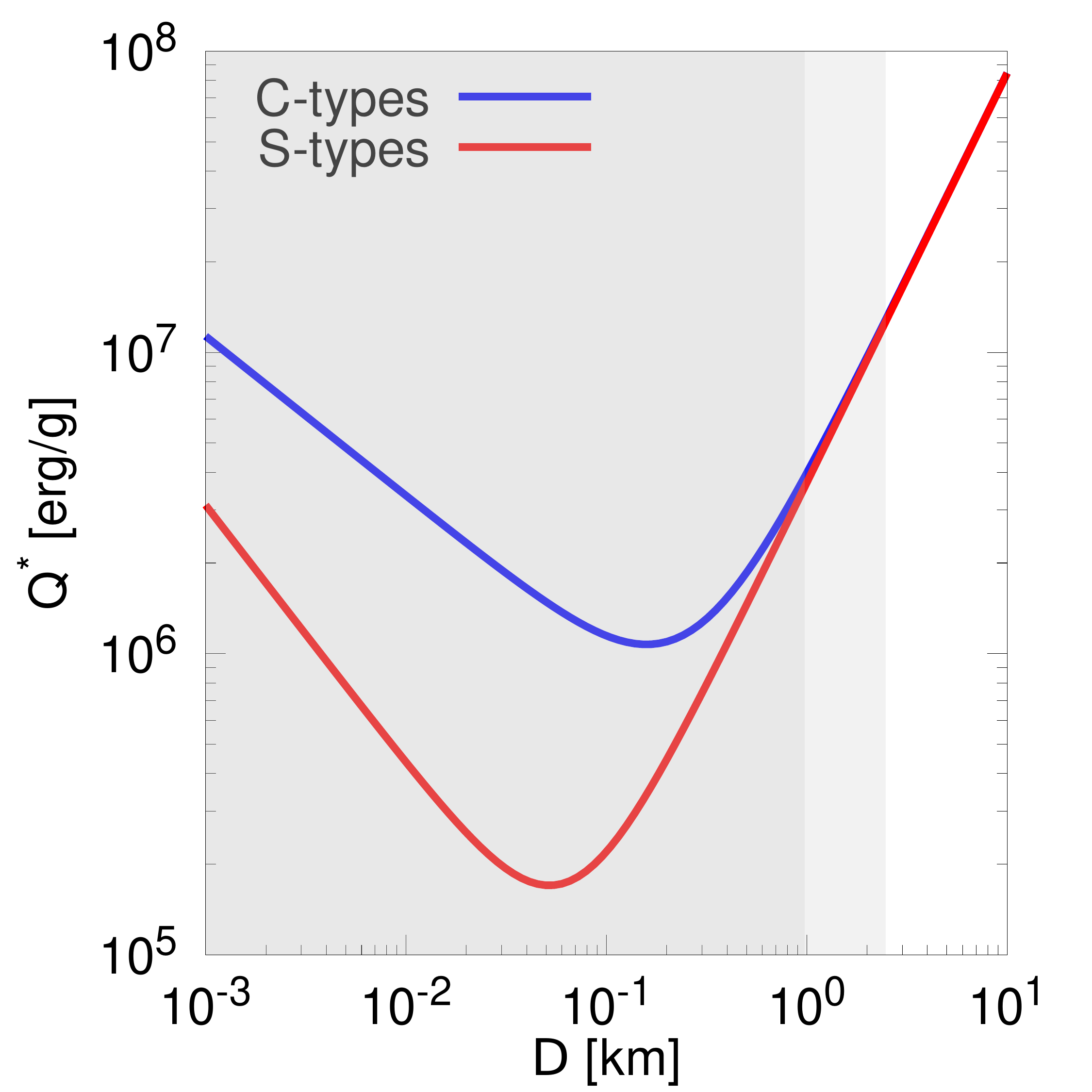} \\
\end{tabular}
\caption{
Same as Fig.~6 for C- and S-types with a modified scaling law
("idea 2").
After this modification, both synthetic SFDs agree very well with the observed SFDs
above the observational limits,
and almost none "undershoots" the observed SFDs
below the limits.
The respective pseudo $\chi^2$ was $50.0$.
For reference, we also plot the corresponding best-fit scaling laws
for the C- and S-types
(Eqs.~(\ref{eq3}) and (\ref{eq4})) with the gray areas corresponding to the range of sizes below the observational limits.
}
\label{sfd_4500_C_plus_other_idea2_final_9_100}
\end{figure*}

The best-fit collisional model,
with the lowest $\chi^2 = 50.0$,
is shown in Fig.~\ref{sfd_4500_C_plus_other_idea2_final_9_100}.
It is a significant improvement with respect to the original initial SFDs ($\chi^2 = 197.9$).
This corresponds to better than $10\%$ agreement with the observed SFDs.
The statistical distribution of 100 synthetic SFDs is even:
approximately 50 lie above, and 50 lie below the observed SFD
for sizes above the observational limit.
More importantly, almost no synthetic SFDs "undershoot"
the observed SFD 
for sizes below the observational limit.
In this sense, our collisional model is self-consistent
and explains the observed SFDs of both C- and S-types.

The explicit forms of our best-fit scaling~laws for C- and S-types
are as follows:
\begin{align}
Q^*_{\rm{C}}(D)&=9\cdot10^7 R^{-0.53} + 1.5 R^{1.36}\,,\label{eq3}\\
Q^*_{\rm{S}}(D)&=9\cdot10^7 R^{-0.86} + 1.5 R^{1.36}\,,\label{eq4}
\end{align}
where $Q^*_{\rm{C,S}}$ is in ergs per gram and $R=D/2$ is in centimeters.
These formul\ae\ demonstrate that S-types are significantly weaker
than C-types at sizes $D \lesssim 0.2\,{\rm km}$
below the current observational limits.
This weakness, however, is seen due to collisional cascade
at sizes $D \gtrsim 1\,{\rm km}$
above the limits.

\section{Conclusions}

In this work,
we studied the collisional evolution and equilibrium of the MB.
We distinguished either individual parts of the MB
(inner, middle, outer)
or taxonomic classes of asteroids
(C-, S-, other types),
which exhibit notable differences
in observed SFDs.

First, we studied models with fixed, single rheology.
For the inner MB and of S-types,
synthetic SFDs showed significant deficiencies
at sizes near or below the observational limit (1 to 2.5\,km).
It was impossible to solve this problem by modifying initial conditions
because the final state was largely determined by
collisional equilibrium between populations.

Therefore, we also studied models with variable, modified rheology.
For the inner MB or, equivalently, S-types, 
we found that these asteroids had to be 
substantially weaker than C-types,
for sizes below the current observational limit
($D\lesssim 0.2\,{\rm km}$).
As we explain in Appendices~\ref{Note_YORP} and~\ref{Note_Yarko},
this weakness is not related to the YORP effect \citep{Rubincam_2000Icar..148....2R}
or the Yarkovsky effect \citep{Vokrouhlicky_2015aste.book..509V},
but reflects the intrinsic rheology of materials.

\begin{figure}
\centering
\includegraphics[width=9cm]{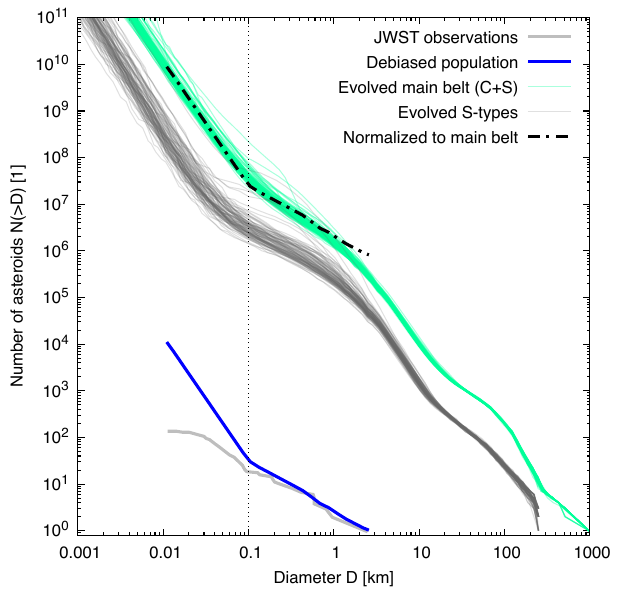}
\caption{
Comparison of our collisional model with JWST observations
\citep{Burdanov_2025Natur.638...74B}.
The SFD for the sum of C- and S-types (\textcolor{green}{green})
exhibits a change of slope from shallow to steep ($q = -2.66$)
at approximately 100\,m,
in agreement with the
observed (\textcolor{gray}{gray}),
debiased (\textcolor{blue}{blue}), and
scaled (dash-dotted)
SFD from JWST.
For reference, we also plot the synthetic SFD of S-types (\textcolor{gray}{gray});
these S-types are about five times less numerous than C-types
and exhibit a change of slope at a smaller size (${\sim}$50\,m).
}
\label{Burdanov_2025_Fig4}
\end{figure}

Our collisional model is in agreement with the most recent
James Webb Space Telescope (JWST) observations
of decameter MB asteroids
\citep{Burdanov_2025Natur.638...74B},
which show a break from shallow to steep slope ($-2.66$; debiased)
at approximately 100\,m
(see Fig.~\ref{Burdanov_2025_Fig4}).
While above this break, the SFDs of C- and S-types
exhibit similar shapes due to mutual collisions
(as observed by
\citealt{Maeda_2021AJ....162..280M,Gallegos_2023PSJ.....4..128G}),
we predict a difference between C- and S-types
for decameter bodies.
If these bodies are rubble-piles,
a transition to the strength regime
at smaller sizes is natural
\citep{Jutzi_2023LPICo2851.2439J,Walsh_2024NatCo..15.5653W}.
It will be soon possible to obtain the respective SFDs for these types separately,
as more and more JWST observations become available.

The fact that C-types are stronger and S-types are weaker seems counter-intuitive.
First, C-types are more porous than S-types
($20\%$ vs. $8\%$; \citet{Vernazza_2021A&A...654A..56V}).
However, \citet{Jutzi_2010Icar..207...54J,Jutzi_2015P&SS..107....3J} showed
that greater porosity allows pore compaction, 
which actually increases strength (see their fig.~5).
Second, most meteorites reaching Earth are ordinary chondrites,
meaning carbonaceous chondrite parent bodies
either disintegrate in Earth's atmosphere due to pre-existing fractures,
\citep{Borovicka_2019M&PS...54.1024B,Broz_2024A&A...689A.183B}
or disrupt in interplanetary space due to thermal cracking
\citep{Granvik_2016Natur.530..303G,Shober_2025NatAs...9..799S}.
However, this reflects a different type of weakness,
occurring at small, meter scales;
here, we addressed hundred-meter scales.

\begin{acknowledgements}
M.B. was supported by GACR grants no. 25-16507S, 25-16789S of the Czech Science Foundation.
We thank the referee Rogerio Deienno for a constructive review,
which helped us to clearly explain our results.
\end{acknowledgements}

\bibliographystyle{aa}
\bibliography{bibliography}

@ARTICLE{Vernazza_2021A&A...654A..56V,
       author = {{Vernazza}, P. and {Ferrais}, M. and {Jorda}, L. and {Hanu{\v{s}}}, J. and {Carry}, B. and {Marsset}, M. and {Bro{\v{z}}}, M. and {Fetick}, R. and {Viikinkoski}, M. and {Marchis}, F. and {Vachier}, F. and {Drouard}, A. and {Fusco}, T. and {Birlan}, M. and {Podlewska-Gaca}, E. and {Rambaux}, N. and {Neveu}, M. and {Bartczak}, P. and {Dudzi{\'n}ski}, G. and {Jehin}, E. and {Beck}, P. and {Berthier}, J. and {Castillo-Rogez}, J. and {Cipriani}, F. and {Colas}, F. and {Dumas}, C. and {{\v{D}}urech}, J. and {Grice}, J. and {Kaasalainen}, M. and {Kryszczynska}, A. and {Lamy}, P. and {Le Coroller}, H. and {Marciniak}, A. and {Michalowski}, T. and {Michel}, P. and {Santana-Ros}, T. and {Tanga}, P. and {Vigan}, A. and {Witasse}, O. and {Yang}, B. and {Antonini}, P. and {Audejean}, M. and {Aurard}, P. and {Behrend}, R. and {Benkhaldoun}, Z. and {Bosch}, J.~M. and {Chapman}, A. and {Dalmon}, L. and {Fauvaud}, S. and {Hamanowa}, Hiroko and {Hamanowa}, Hiromi and {His}, J. and {Jones}, A. and {Kim}, D. -H. and {Kim}, M. -J. and {Krajewski}, J. and {Labrevoir}, O. and {Leroy}, A. and {Livet}, F. and {Molina}, D. and {Montaigut}, R. and {Oey}, J. and {Payre}, N. and {Reddy}, V. and {Sabin}, P. and {Sanchez}, A.~G. and {Socha}, L.},
        title = "{VLT/SPHERE imaging survey of the largest main-belt asteroids: Final results and synthesis}",
      journal = {Astronomy and Astrophysics},
         year = 2021,
        month = oct,
       volume = {654},
          eid = {A56},
        pages = {A56},
          doi = {10.1051/0004-6361/202141781},
       adsurl = {https://ui.adsabs.harvard.edu/abs/2021A&A...654A..56V}
}

@ARTICLE{Benz_1999Icar..142....5B,
       author = {{Benz}, Willy and {Asphaug}, Erik},
        title = "{Catastrophic Disruptions Revisited}",
      journal = {Icarus},
         year = 1999,
        month = nov,
       volume = {142},
       number = {1},
        pages = {5-20},
          doi = {10.1006/icar.1999.6204},
archivePrefix = {arXiv},
       eprint = {astro-ph/9907117},
 primaryClass = {astro-ph},
       adsurl = {https://ui.adsabs.harvard.edu/abs/1999Icar..142....5B}
}

@ARTICLE{Dohnanyi_1969JGR....74.2531D,
       author = {{Dohnanyi}, J.~S.},
        title = {Collisional Model of Asteroids and Their Debris},
      journal = {Journal of Geophysics Research},
         year = 1969,
        month = may,
       volume = {74},
        pages = {2531-2554},
          doi = {10.1029/JB074i010p02531},
       adsurl = {https://ui.adsabs.harvard.edu/abs/1969JGR....74.2531D}
}

@ARTICLE{Benz_1995CoPhC..87..253B,
       author = {{Benz}, W. and {Asphaug}, E.},
        title = "{Simulations of brittle solids using smooth particle hydrodynamics}",
      journal = {Computer Physics Communications},
         year = 1995,
        month = may,
       volume = {87},
       number = {1-2},
        pages = {253-265},
          doi = {10.1016/0010-4655(94)00176-3},
       adsurl = {https://ui.adsabs.harvard.edu/abs/1995CoPhC..87..253B}
}

@BOOK{Rubinstein2016-ju,
  title     = "Simulation and the Monte Carlo method",
  author    = "Rubinstein, Reuven Y and Kroese, Dirk P",
  publisher = "John Wiley \& Sons",
  series    = "Wiley Series in Probability and Statistics",
  edition   =  3,
  month     =  oct,
  year      =  2016,
  address   = "Nashville, TN",
  language  = "en"
}

@ARTICLE{Jutzi_2015P&SS..107....3J,
       author = {{Jutzi}, Martin},
        title = "{SPH calculations of asteroid disruptions: The role of pressure dependent failure models}",
      journal = {\planss},
         year = 2015,
        month = mar,
       volume = {107},
        pages = {3-9},
          doi = {10.1016/j.pss.2014.09.012},
archivePrefix = {arXiv},
       eprint = {1502.01860},
 primaryClass = {astro-ph.EP},
       adsurl = {https://ui.adsabs.harvard.edu/abs/2015P&SS..107....3J}
}

@ARTICLE{DeMeo_2009Icar..202..160D,
       author = {{DeMeo}, Francesca E. and {Binzel}, Richard P. and {Slivan}, Stephen M. and {Bus}, Schelte J.},
        title = "{An extension of the Bus asteroid taxonomy into the near-infrared}",
      journal = {Icarus},
         year = 2009,
        month = jul,
       volume = {202},
       number = {1},
        pages = {160-180},
          doi = {10.1016/j.icarus.2009.02.005},
       adsurl = {https://ui.adsabs.harvard.edu/abs/2009Icar..202..160D}
}

@ARTICLE{Bus_2002Icar..158..146B,
       author = {{Bus}, Schelte J. and {Binzel}, Richard P.},
        title = "{Phase II of the Small Main-Belt Asteroid Spectroscopic Survey. A Feature-Based Taxonomy}",
      journal = {\icarus},
         year = 2002,
        month = jul,
       volume = {158},
       number = {1},
        pages = {146-177},
          doi = {10.1006/icar.2002.6856},
       adsurl = {https://ui.adsabs.harvard.edu/abs/2002Icar..158..146B}
}

@ARTICLE{Bus_2002Icar..158..106B,
       author = {{Bus}, Schelte J. and {Binzel}, Richard P.},
        title = "{Phase II of the Small Main-Belt Asteroid Spectroscopic Survey. The Observations}",
      journal = {Icarus},
         year = 2002,
        month = jul,
       volume = {158},
       number = {1},
        pages = {106-145},
          doi = {10.1006/icar.2002.6857},
       adsurl = {https://ui.adsabs.harvard.edu/abs/2002Icar..158..106B}
}

@ARTICLE{Parker_2008Icar..198..138P,
       author = {{Parker}, A. and {Ivezi{\'c}}, {\v{Z}}. and {Juri{\'c}}, M. and {Lupton}, R. and {Sekora}, M.~D. and {Kowalski}, A.},
        title = "{The size distributions of asteroid families in the SDSS Moving Object Catalog 4}",
      journal = {\icarus},
         year = 2008,
        month = nov,
       volume = {198},
       number = {1},
        pages = {138-155},
          doi = {10.1016/j.icarus.2008.07.002},
archivePrefix = {arXiv},
       eprint = {0807.3762},
 primaryClass = {astro-ph},
       adsurl = {https://ui.adsabs.harvard.edu/abs/2008Icar..198..138P}
}

@ARTICLE{Mothe-Diniz_2003Icar..162...10M,
       author = {{Moth{\'e}-Diniz}, Thais and {Carvano}, Jorge M. {\'a}rcio and {Lazzaro}, Daniela},
        title = "{Distribution of taxonomic classes in the main belt of asteroids}",
      journal = {Icarus},
         year = 2003,
        month = mar,
       volume = {162},
       number = {1},
        pages = {10-21},
          doi = {10.1016/S0019-1035(02)00066-0},
       adsurl = {https://ui.adsabs.harvard.edu/abs/2003Icar..162...10M}
}

@ARTICLE{Gradie_1982Sci...216.1405G,
       author = {{Gradie}, J. and {Tedesco}, E.},
        title = "{Compositional Structure of the Asteroid Belt}",
      journal = {Science},
         year = 1982,
        month = jun,
       volume = {216},
       number = {4553},
        pages = {1405-1407},
          doi = {10.1126/science.216.4553.1405},
       adsurl = {https://ui.adsabs.harvard.edu/abs/1982Sci...216.1405G}
}

@ARTICLE{Cibulkova_2014Icar..241..358C,
       author = {{Cibulkov{\'a}}, H. and {Bro{\v{z}}}, M. and {Benavidez}, P.~G.},
        title = "{A six-part collisional model of the main asteroid belt}",
      journal = {Icarus},
         year = 2014,
        month = oct,
       volume = {241},
        pages = {358-372},
          doi = {10.1016/j.icarus.2014.07.016},
archivePrefix = {arXiv},
       eprint = {1407.6143},
 primaryClass = {astro-ph.EP},
       adsurl = {https://ui.adsabs.harvard.edu/abs/2014Icar..241..358C}
}

@ARTICLE{Sevecek_2019A&A...629A.122S,
       author = {{\v{S}eve{\v{c}}ek}, P. and {Bro{\v{z}}}, M. and {Jutzi}, M.},
        title = "{Impacts into rotating targets: angular momentum draining and efficient formation of synthetic families}",
      journal = {\aap},
         year = 2019,
        month = sep,
       volume = {629},
          eid = {A122},
        pages = {A122},
          doi = {10.1051/0004-6361/201935690},
archivePrefix = {arXiv},
       eprint = {1908.03248},
 primaryClass = {astro-ph.EP},
       adsurl = {https://ui.adsabs.harvard.edu/abs/2019A&A...629A.122S}
}

@misc{Cossins_2010PhDT.......301C,
  doi = {10.48550/ARXIV.1007.1245},
  
  url = {https://arxiv.org/abs/1007.1245},
  
  author = {Cossins, Peter J.},
  
  year={2010},
  
  title = {Smoothed Particle Hydrodynamics},
}

@ARTICLE{Kirkwood_1860AJ......6..126K,
       author = {{Kirkwood}, Daniel},
        title = "{Letter to the editor [inequalities of long period in the motion of asteroids and satellites of Saturn]}",
      journal = {The Astronomical Journal},
         year = 1860,
        month = jul,
       volume = {6},
        pages = {126-126},
          doi = {10.1086/100779},
       adsurl = {https://ui.adsabs.harvard.edu/abs/1860AJ......6..126K}
}

@ARTICLE{Michel_2001Sci...294.1696M,
       author = {{Michel}, Patrick and {Benz}, Willy and {Tanga}, Paolo and {Richardson}, Derek C.},
        title = "{Collisions and Gravitational Reaccumulation: Forming Asteroid Families and Satellites}",
      journal = {Science},
         year = 2001,
        month = nov,
       volume = {294},
       number = {5547},
        pages = {1696-1700},
          doi = {10.1126/science.1065189},
       adsurl = {https://ui.adsabs.harvard.edu/abs/2001Sci...294.1696M}
}

@misc{Gault_NASA,
  author       = {{Gault}, D.~E. and {Moore}, H.~J. and {Shoemaker}, E.~M.},
  title        = {Spray Ejected from the Lunar Surface by Meteoroid Impact},
  month        = {jan},
  year         = {1963},
}

@ARTICLE{Fujiwara_1977Icar...31..277F,
       author = {{Fujiwara}, A. and {Kamimoto}, G. and {Tsukamoto}, A.},
        title = "{Destruction of basaltic bodies by high-velocity impact}",
      journal = {\icarus},
         year = 1977,
        month = jun,
       volume = {31},
       number = {2},
        pages = {277-288},
          doi = {10.1016/0019-1035(77)90038-0},
       adsurl = {https://ui.adsabs.harvard.edu/abs/1977Icar...31..277F}
}

@ARTICLE{Davis_1990Icar...83..156D,
       author = {{Davis}, D.~R. and {Ryan}, E.~V.},
        title = "{On collisional disruption: Experimental results and scaling laws}",
      journal = {\icarus},
         year = 1990,
        month = jan,
       volume = {83},
       number = {1},
        pages = {156-182},
          doi = {10.1016/0019-1035(90)90012-X},
       adsurl = {https://ui.adsabs.harvard.edu/abs/1990Icar...83..156D}
}

@ARTICLE{Jutzi_2010Icar..207...54J,
       author = {{Jutzi}, Martin and {Michel}, Patrick and {Benz}, Willy and {Richardson}, Derek C.},
        title = "{Fragment properties at the catastrophic disruption threshold: The effect of the parent body{\textquoteright}s internal structure}",
      journal = {Icarus},
         year = 2010,
        month = may,
       volume = {207},
       number = {1},
        pages = {54-65},
          doi = {10.1016/j.icarus.2009.11.016},
archivePrefix = {arXiv},
       eprint = {0911.3937},
 primaryClass = {astro-ph.EP},
       adsurl = {https://ui.adsabs.harvard.edu/abs/2010Icar..207...54J}
}

@ARTICLE{Jutzi_2008Icar..198..242J,
       author = {{Jutzi}, Martin and {Benz}, Willy and {Michel}, Patrick},
        title = "{Numerical simulations of impacts involving porous bodies. I. Implementing sub-resolution porosity in a 3D SPH hydrocode}",
      journal = {Icarus},
         year = 2008,
        month = nov,
       volume = {198},
       number = {1},
        pages = {242-255},
          doi = {10.1016/j.icarus.2008.06.013},
archivePrefix = {arXiv},
       eprint = {0807.1264},
 primaryClass = {astro-ph},
       adsurl = {https://ui.adsabs.harvard.edu/abs/2008Icar..198..242J}
}

@ARTICLE{Schwartz_2016AdSpR..57.1832S,
       author = {{Schwartz}, Stephen R. and {Yu}, Yang and {Michel}, Patrick and {Jutzi}, Martin},
        title = "{Small-body deflection techniques using spacecraft: Techniques in simulating the fate of ejecta}",
      journal = {Advances in Space Research},
         year = 2016,
        month = apr,
       volume = {57},
       number = {8},
        pages = {1832-1846},
          doi = {10.1016/j.asr.2015.12.042},
archivePrefix = {arXiv},
       eprint = {1601.05844},
 primaryClass = {astro-ph.EP},
       adsurl = {https://ui.adsabs.harvard.edu/abs/2016AdSpR..57.1832S}
}

@ARTICLE{Durda_2007Icar..186..498D,
       author = {{Durda}, Daniel D. and {Bottke}, William F. and {Nesvorný}, David and {Enke}, Brian L. and {Merline}, William J. and {Asphaug}, Erik and {Richardson}, Derek C.},
        title = "{Size-frequency distributions of fragments from SPH/ N-body simulations of asteroid impacts: Comparison with observed asteroid families}",
      journal = {Icarus},
         year = 2007,
        month = feb,
       volume = {186},
       number = {2},
        pages = {498-516},
          doi = {10.1016/j.icarus.2006.09.013},
       adsurl = {https://ui.adsabs.harvard.edu/abs/2007Icar..186..498D}
}

@ARTICLE{Capaccioni_1986Icar...66..487C,
       author = {{Capaccioni}, F. and {Cerroni}, P. and {Coradini}, M. and {Di Martino}, M. and {Farinella}, P. and {Flamini}, E. and {Martelli}, G. and {Paolicchi}, P. and {Smith}, P.~N. and {Woodward}, A. and {Zappala}, V.},
        title = "{Asteroidal catastrophic collisions simulated by hypervelocity impact experiments}",
      journal = {\icarus},
         year = 1986,
        month = jun,
       volume = {66},
       number = {3},
        pages = {487-514},
          doi = {10.1016/0019-1035(86)90087-4},
       adsurl = {https://ui.adsabs.harvard.edu/abs/1986Icar...66..487C}
}

@ARTICLE{Nakamura_1992Icar..100..127N,
       author = {{Nakamura}, Akiko and {Suguiyama}, Kohji and {Fujiwara}, Akira},
        title = "{Velocity and spin of fragments from impact disruptions  I. An experimental approach to a general law between mass and velocity}",
      journal = {\icarus},
         year = 1992,
        month = nov,
       volume = {100},
       number = {1},
        pages = {127-135},
          doi = {10.1016/0019-1035(92)90023-Z},
       adsurl = {https://ui.adsabs.harvard.edu/abs/1992Icar..100..127N}
}

@ARTICLE{Benz_1994Icar..107...98B,
       author = {{Benz}, W. and {Asphaug}, E.},
        title = "{Impact Simulations with Fracture. I. Method and Tests}",
      journal = {Icarus},
         year = 1994,
        month = jan,
       volume = {107},
       number = {1},
        pages = {98-116},
          doi = {10.1006/icar.1994.1009},
       adsurl = {https://ui.adsabs.harvard.edu/abs/1994Icar..107...98B}
}

@ARTICLE{Moskovitz_2022A&C....4100661M,
       author = {{Moskovitz}, N.~A. and {Wasserman}, L. and {Burt}, B. and {Schottland}, R. and {Bowell}, E. and {Bailen}, M. and {Granvik}, M.},
        title = "{The astorb database at Lowell Observatory}",
      journal = {Astronomy and Computing},
         year = 2022,
        month = oct,
       volume = {41},
          eid = {100661},
        pages = {100661},
          doi = {10.1016/j.ascom.2022.100661},
archivePrefix = {arXiv},
       eprint = {2210.10217},
 primaryClass = {astro-ph.EP},
       adsurl = {https://ui.adsabs.harvard.edu/abs/2022A&C....4100661M}
}

@ARTICLE{Masiero_2011ApJ...741...68M,
       author = {{Masiero}, Joseph R. and {Mainzer}, A.~K. and {Grav}, T. and {Bauer}, J.~M. and {Cutri}, R.~M. and {Dailey}, J. and {Eisenhardt}, P.~R.~M. and {McMillan}, R.~S. and {Spahr}, T.~B. and {Skrutskie}, M.~F. and {Tholen}, D. and {Walker}, R.~G. and {Wright}, E.~L. and {DeBaun}, E. and {Elsbury}, D. and {Gautier}, T., IV and {Gomillion}, S. and {Wilkins}, A.},
        title = "{Main Belt Asteroids with WISE/NEOWISE. I. Preliminary Albedos and Diameters}",
      journal = {The Astrophysical Journal},
         year = 2011,
        month = nov,
       volume = {741},
       number = {2},
          eid = {68},
        pages = {68},
          doi = {10.1088/0004-637X/741/2/68},
archivePrefix = {arXiv},
       eprint = {1109.4096},
 primaryClass = {astro-ph.EP},
       adsurl = {https://ui.adsabs.harvard.edu/abs/2011ApJ...741...68M}
}

@ARTICLE{Blanton_2017AJ....154...28B,
       author = {{Blanton}, Michael R. and {Bershady}, Matthew A. and {Abolfathi}, Bela and {Albareti}, Franco D. and {Allende Prieto}, Carlos and {Almeida}, Andres and {Alonso-Garc{\'\i}a}, Javier and {Anders}, Friedrich and {Anderson}, Scott F. and {Andrews}, Brett and et al.},
        title = "{Sloan Digital Sky Survey IV: Mapping the Milky Way, Nearby Galaxies, and the Distant Universe}",
      journal = {The Astronomical Journal},
         year = 2017,
        month = jul,
       volume = {154},
       number = {1},
          eid = {28},
        pages = {28},
          doi = {10.3847/1538-3881/aa7567},
archivePrefix = {arXiv},
       eprint = {1703.00052},
 primaryClass = {astro-ph.GA},
       adsurl = {https://ui.adsabs.harvard.edu/abs/2017AJ....154...28B}
}

@ARTICLE{Yamauchi_2011PASP..123..852Y,
       author = {{Yamauchi}, C. and {Fujishima}, S. and {Ikeda}, N. and {Inada}, K. and {Katano}, M. and {Kataza}, H. and {Makiuti}, S. and {Matsuzaki}, K. and {Takita}, S. and {Yamamoto}, Y. and {Yamamura}, I. and {Ishihara}, D. and {Oyabu}, S.},
        title = "{AKARI-CAS{\textemdash}Online Service for AKARI All-Sky Catalogues}",
      journal = {Publications of the Astronomical Society of the Pacific},
         year = 2011,
        month = jul,
       volume = {123},
       number = {905},
        pages = {852},
          doi = {10.1086/660926},
archivePrefix = {arXiv},
       eprint = {1107.5385},
 primaryClass = {astro-ph.IM},
       adsurl = {https://ui.adsabs.harvard.edu/abs/2011PASP..123..852Y}
}

@INPROCEEDINGS{Knezevic_2012IAUJD...7P..18K,
       author = {{Knezevic}, Zoran and {Milani}, Andrea},
        title = "{Asteroids Dynamic Site-AstDyS}",
    booktitle = {IAU Joint Discussion},
         year = 2012,
       series = {IAU Joint Discussion},
        month = aug,
          eid = {P18},
        pages = {P18},
       adsurl = {https://ui.adsabs.harvard.edu/abs/2012IAUJD...7P..18K}
}

@INCOLLECTION{Nesvorny_2015aste.book..297N,
       author = {{Nesvorný}, D. and {Bro{\v{z}}}, M. and {Carruba}, V.},
        title = "{Identification and Dynamical Properties of Asteroid Families}",
    booktitle = {Asteroids IV},
         year = 2015,
        pages = {297-321},
          doi = {10.2458/azu_uapress_9780816532131-ch016},
       adsurl = {https://ui.adsabs.harvard.edu/abs/2015aste.book..297N}
}

@INCOLLECTION{Zellner_1979aste.book..783Z,
       author = {{Zellner}, B.},
        title = "{Asteroid taxonomy and the distribution of the compositional types.}",
    booktitle = {Asteroids},
         year = 1979,
       editor = {{Gehrels}, Tom and {Matthews}, Mildred Shapley},
        pages = {783-806},
       adsurl = {https://ui.adsabs.harvard.edu/abs/1979aste.book..783Z}
}

@ARTICLE{Durda_1998Icar..135..431D,
       author = {{Durda}, Daniel D. and {Greenberg}, Richard and {Jedicke}, Robert},
        title = "{Collisional Models and Scaling Laws: A New Interpretation of the Shape of the Main-Belt Asteroid Size Distribution}",
      journal = {Icarus},
         year = 1998,
        month = oct,
       volume = {135},
       number = {2},
        pages = {431-440},
          doi = {10.1006/icar.1998.5960},
       adsurl = {https://ui.adsabs.harvard.edu/abs/1998Icar..135..431D}
}

@ARTICLE{Rozehnal_2022Icar..38315064R,
       author = {{Rozehnal}, J. and {Bro{\v{z}}}, M. and {Nesvorný}, D. and {Walsh}, K.~J. and {Durda}, D.~D. and {Richardson}, D.~C. and {Asphaug}, E.},
        title = "{SPH simulations of high-speed collisions between asteroids and comets}",
      journal = {Icarus},
         year = 2022,
        month = sep,
       volume = {383},
          eid = {115064},
        pages = {115064},
          doi = {10.1016/j.icarus.2022.115064},
archivePrefix = {arXiv},
       eprint = {2205.14370},
 primaryClass = {astro-ph.EP},
       adsurl = {https://ui.adsabs.harvard.edu/abs/2022Icar..38315064R}
}

@PHDTHESIS{Delbo_2004PhDT.......371D,
       author = {{Delbo}, Marco},
        title = "{The nature of near-earth asteroids from the study of their thermal infrared emission}",
       school = {Free University of Berlin, Germany},
         year = 2004,
        month = jan,
       adsurl = {https://ui.adsabs.harvard.edu/abs/2004PhDT.......371D}
}

@ARTICLE{Tedesco_2002AJ....123.2070T,
       author = {{Tedesco}, Edward F. and {Desert}, Fran{\c{c}}ois-Xavier},
        title = "{The Infrared Space Observatory Deep Asteroid Search}",
      journal = {\aj},
         year = 2002,
        month = apr,
       volume = {123},
       number = {4},
        pages = {2070-2082},
          doi = {10.1086/339482},
       adsurl = {https://ui.adsabs.harvard.edu/abs/2002AJ....123.2070T}
}

@ARTICLE{Nesvorny_2002Icar..157..155N,
       author = {{Nesvorný}, D. and {Morbidelli}, A. and {Vokrouhlický}, D. and {Bottke}, W.~F. and {Bro{\v{z}}}, M.},
        title = "{The Flora Family: A Case of the Dynamically Dispersed Collisional Swarm?}",
      journal = {\icarus},
         year = 2002,
        month = may,
       volume = {157},
       number = {1},
        pages = {155-172},
          doi = {10.1006/icar.2002.6830},
       adsurl = {https://ui.adsabs.harvard.edu/abs/2002Icar..157..155N}
}

@ARTICLE{Vokrouhlicky_2006Icar..182...92V,
       author = {{Vokrouhlický}, D. and {Bro{\v{z}}}, M. and {Morbidelli}, A. and {Bottke}, W.~F. and {Nesvorný}, D. and {Lazzaro}, D. and {Rivkin}, A.~S.},
        title = "{Yarkovsky footprints in the Eos family}",
      journal = {\icarus},
         year = 2006,
        month = may,
       volume = {182},
       number = {1},
        pages = {92-117},
          doi = {10.1016/j.icarus.2005.12.011},
       adsurl = {https://ui.adsabs.harvard.edu/abs/2006Icar..182...92V}
}

@ARTICLE{Nesvorny_2006Icar..183..296N,
       author = {{Nesvorný}, David and {Enke}, Brian L. and {Bottke}, William F. and {Durda}, Daniel D. and {Asphaug}, Erik and {Richardson}, Derek C.},
        title = "{Karin cluster formation by asteroid impact}",
      journal = {Icarus},
         year = 2006,
        month = aug,
       volume = {183},
       number = {2},
        pages = {296-311},
          doi = {10.1016/j.icarus.2006.03.008},
       adsurl = {https://ui.adsabs.harvard.edu/abs/2006Icar..183..296N}
}

@ARTICLE{Bottke_2005Icar..175..111B,
       author = {{Bottke}, William F. and {Durda}, Daniel D. and {Nesvorný}, David and {Jedicke}, Robert and {Morbidelli}, Alessandro and {Vokrouhlický}, David and {Levison}, Hal},
        title = "{The fossilized size distribution of the main asteroid belt}",
      journal = {Icarus},
         year = 2005,
        month = may,
       volume = {175},
       number = {1},
        pages = {111-140},
          doi = {10.1016/j.icarus.2004.10.026},
       adsurl = {https://ui.adsabs.harvard.edu/abs/2005Icar..175..111B}
}

@INCOLLECTION{Davis_1979aste.book..528D,
       author = {{Davis}, D.~R. and {Chapman}, C.~R. and {Greenberg}, R. and {Weidenschilling}, S.~J. and {Harris}, A.~W.},
        title = "{Collisional evolution of asteroids: populations, rotations, and velocities.}",
    booktitle = {Asteroids},
         year = 1979,
       editor = {{Gehrels}, Tom and {Matthews}, Mildred Shapley},
        pages = {528-557},
       adsurl = {https://ui.adsabs.harvard.edu/abs/1979aste.book..528D}
}

@ARTICLE{Davis_1985Icar...62...30D,
       author = {{Davis}, D.~R. and {Chapman}, C.~R. and {Weidenschilling}, S.~J. and {Greenberg}, R.},
        title = "{Collisional history of asteroids: Evidence from Vesta and the Hirayama families}",
      journal = {\icarus},
         year = 1985,
        month = apr,
       volume = {62},
       number = {1},
        pages = {30-53},
          doi = {10.1016/0019-1035(85)90170-8},
       adsurl = {https://ui.adsabs.harvard.edu/abs/1985Icar...62...30D}
}

@ARTICLE{Love_1996Icar..124..141L,
       author = {{Love}, Stanley G. and {Ahrens}, Thomas J.},
        title = "{Catastrophic Impacts on Gravity Dominated Asteroids}",
      journal = {Icarus},
         year = 1996,
        month = nov,
       volume = {124},
       number = {1},
        pages = {141-155},
          doi = {10.1006/icar.1996.0195},
       adsurl = {https://ui.adsabs.harvard.edu/abs/1996Icar..124..141L}
}

@ARTICLE{Anders_1965Icar....4..399A,
       author = {{Anders}, Edward},
        title = "{Fragmentation history of asteroids}",
      journal = {Icarus},
         year = 1965,
        month = sep,
       volume = {4},
       number = {4},
        pages = {399-408},
          doi = {10.1016/0019-1035(65)90044-8},
       adsurl = {https://ui.adsabs.harvard.edu/abs/1965Icar....4..399A}
}

@ARTICLE{Campo_1994P&SS...42.1079C,
       author = {{Campo Bagatin}, A. and {Cellino}, A. and {Davis}, D.~R. and {Farinella}, P. and {Paolicchi}, P.},
        title = "{Wavy size distributions for collisional systems with a small-size cutoff}",
      journal = {\planss},
         year = 1994,
        month = dec,
       volume = {42},
       number = {12},
        pages = {1079-1092},
          doi = {10.1016/0032-0633(94)90008-6},
       adsurl = {https://ui.adsabs.harvard.edu/abs/1994P&SS...42.1079C}
}

@ARTICLE{Zappala_1990AJ....100.2030Z,
       author = {{Zappalà}, Vincenzo and {Cellino}, Alberto and {Farinella}, Paolo and {Knezevic}, Zoran},
        title = "{Asteroid Families. I. Identification by Hierarchical Clustering and Reliability Assessment}",
      journal = {\aj},
         year = 1990,
        month = dec,
       volume = {100},
        pages = {2030},
          doi = {10.1086/115658},
       adsurl = {https://ui.adsabs.harvard.edu/abs/1990AJ....100.2030Z}
}

@ARTICLE{Opik_1951PRIA...54..165O,
       author = {{\"Opik}, E.~J.},
        title = "{Collision probability with the planets and the distribution of planetary matter}",
      journal = {Proc. R. Irish Acad. Sect. A},
         year = 1951,
        month = jan,
       volume = {54},
        pages = {165-199},
       adsurl = {https://ui.adsabs.harvard.edu/abs/1951PRIA...54..165O}
}

@ARTICLE{Greenberg_1982AJ.....87..184G,
       author = {{Greenberg}, R.},
        title = "{Orbital interactions - A new geometrical formalism}",
      journal = {\aj},
         year = 1982,
        month = jan,
       volume = {87},
        pages = {184-195},
          doi = {10.1086/113095},
       adsurl = {https://ui.adsabs.harvard.edu/abs/1982AJ.....87..184G}
}

@INCOLLECTION{Bottke_2015aste.book..701B,
       author = {{Bottke}, W.~F. and {Bro{\v{z}}}, M. and {O'Brien}, D.~P. and {Campo Bagatin}, A. and {Morbidelli}, A. and {Marchi}, S.},
        title = "{The Collisional Evolution of the Main Asteroid Belt}",
    booktitle = {Asteroids IV},
         year = 2015,
        pages = {701-724},
          doi = {10.2458/azu_uapress_9780816532131-ch036},
       adsurl = {https://ui.adsabs.harvard.edu/abs/2015aste.book..701B}
}

@ARTICLE{Morbidelli_2009Icar..204..558M,
       author = {{Morbidelli}, Alessandro and {Bottke}, William F. and {Nesvorný}, David and {Levison}, Harold F.},
        title = "{Asteroids were born big}",
      journal = {Icarus},
         year = 2009,
        month = dec,
       volume = {204},
       number = {2},
        pages = {558-573},
          doi = {10.1016/j.icarus.2009.07.011},
archivePrefix = {arXiv},
       eprint = {0907.2512},
 primaryClass = {astro-ph.EP},
       adsurl = {https://ui.adsabs.harvard.edu/abs/2009Icar..204..558M}
}

@ARTICLE{Vokrouhlicky_2010AJ....139.2148V,
       author = {{Vokrouhlický}, David and {Nesvorný}, David and {Bottke}, William F. and {Morbidelli}, Alessandro},
        title = "{Collisionally Born Family About 87 Sylvia}",
      journal = {\aj},
         year = 2010,
        month = jun,
       volume = {139},
       number = {6},
        pages = {2148-2158},
          doi = {10.1088/0004-6256/139/6/2148},
       adsurl = {https://ui.adsabs.harvard.edu/abs/2010AJ....139.2148V}
}

@ARTICLE{Marschall_2022AJ....164..167M,
       author = {{Marschall}, Raphael and {Nesvorný}, David and {Deienno}, Rogerio and {Wong}, Ian and {Levison}, Harold F. and {Bottke}, William F.},
        title = "{Implications for the Collisional Strength of Jupiter Trojans from the Eurybates Family}",
      journal = {\aj},
         year = 2022,
        month = oct,
       volume = {164},
       number = {4},
          eid = {167},
        pages = {167},
          doi = {10.3847/1538-3881/ac8d6b},
archivePrefix = {arXiv},
       eprint = {2208.10505},
 primaryClass = {astro-ph.EP},
       adsurl = {https://ui.adsabs.harvard.edu/abs/2022AJ....164..167M}
}

@ARTICLE{Vernazza_2018A&A...618A.154V,
       author = {{Vernazza}, P. and {Bro{\v{z}}}, M. and {Drouard}, A. and {Hanu{\v{s}}}, J. and {Viikinkoski}, M. and {Marsset}, M. and {Jorda}, L. and {Fetick}, R. and {Carry}, B. and {Marchis}, F. and {Birlan}, M. and {Fusco}, T. and {Santana-Ros}, T. and {Podlewska-Gaca}, E. and {Jehin}, E. and {Ferrais}, M. and {Bartczak}, P. and {Dudzi{\'n}ski}, G. and {Berthier}, J. and {Castillo-Rogez}, J. and {Cipriani}, F. and {Colas}, F. and {Dumas}, C. and {{\v{D}}urech}, J. and {Kaasalainen}, M. and {Kryszczynska}, A. and {Lamy}, P. and {Le Coroller}, H. and {Marciniak}, A. and {Michalowski}, T. and {Michel}, P. and {Pajuelo}, M. and {Tanga}, P. and {Vachier}, F. and {Vigan}, A. and {Warner}, B. and {Witasse}, O. and {Yang}, B. and {Asphaug}, E. and {Richardson}, D.~C. and {{\v{S}}eve{\v{c}}ek}, P. and {Gillon}, M. and {Benkhaldoun}, Z.},
        title = "{The impact crater at the origin of the Julia family detected with VLT/SPHERE?}",
      journal = {\aap},
         year = 2018,
        month = oct,
       volume = {618},
          eid = {A154},
        pages = {A154},
          doi = {10.1051/0004-6361/201833477},
       adsurl = {https://ui.adsabs.harvard.edu/abs/2018A&A...618A.154V}
}

@ARTICLE{Marchis_2021A&A...653A..57M,
       author = {{Marchis}, F. and {Jorda}, L. and {Vernazza}, P. and {Bro{\v{z}}}, M. and {Hanu{\v{s}}}, J. and {Ferrais}, M. and {Vachier}, F. and {Rambaux}, N. and {Marsset}, M. and {Viikinkoski}, M. and {Jehin}, E. and {Benseguane}, S. and {Podlewska-Gaca}, E. and {Carry}, B. and {Drouard}, A. and {Fauvaud}, S. and {Birlan}, M. and {Berthier}, J. and {Bartczak}, P. and {Dumas}, C. and {Dudzi{\'n}ski}, G. and {{\v{D}}urech}, J. and {Castillo-Rogez}, J. and {Cipriani}, F. and {Colas}, F. and {Fetick}, R. and {Fusco}, T. and {Grice}, J. and {Kryszczynska}, A. and {Lamy}, P. and {Marciniak}, A. and {Michalowski}, T. and {Michel}, P. and {Pajuelo}, M. and {Santana-Ros}, T. and {Tanga}, P. and {Vigan}, A. and {Witasse}, O. and {Yang}, B.},
        title = "{(216) Kleopatra, a low density critically rotating M-type asteroid}",
      journal = {\aap},
         year = 2021,
        month = sep,
       volume = {653},
          eid = {A57},
        pages = {A57},
          doi = {10.1051/0004-6361/202140874},
archivePrefix = {arXiv},
       eprint = {2108.07207},
 primaryClass = {astro-ph.EP},
       adsurl = {https://ui.adsabs.harvard.edu/abs/2021A&A...653A..57M}
}

@ARTICLE{Sevecek_2017Icar..296..239S,
       author = {{\v{S}eve{\v{c}}ek}, P. and {Bro{\v{z}}}, M. and {Nesvorný}, D. and {Enke}, B. and {Durda}, D. and {Walsh}, K. and {Richardson}, D.~C.},
        title = "{SPH/N-Body simulations of small (D = 10km) asteroidal breakups and improved parametric relations for Monte-Carlo collisional models}",
      journal = {Icarus},
         year = 2017,
        month = nov,
       volume = {296},
        pages = {239-256},
          doi = {10.1016/j.icarus.2017.06.021},
archivePrefix = {arXiv},
       eprint = {1803.10666},
 primaryClass = {astro-ph.EP},
       adsurl = {https://ui.adsabs.harvard.edu/abs/2017Icar..296..239S}
}

@ARTICLE{Broz_2022A&A...657A..76B,
       author = {{Bro{\v{z}}}, M. and {{\v{D}}urech}, J. and {Carry}, B. and {Vachier}, F. and {Marchis}, F. and {Hanu{\v{s}}}, J. and {Jorda}, L. and {Vernazza}, P. and {Vokrouhlický}, D. and {Walterov{\'a}}, M. and {Behrend}, R.},
        title = "{Observed tidal evolution of Kleopatra's outer satellite}",
      journal = {\aap},
         year = 2022,
        month = jan,
       volume = {657},
          eid = {A76},
        pages = {A76},
          doi = {10.1051/0004-6361/202142055},
archivePrefix = {arXiv},
       eprint = {2110.12702},
 primaryClass = {astro-ph.EP},
       adsurl = {https://ui.adsabs.harvard.edu/abs/2022A&A...657A..76B}
}

@INPROCEEDINGS{Bottke_2020DPS....5240202B,
       author = {{Bottke}, W. and {Walsh}, K. and {Vokrouhlický}, D. and {Nesvorný}, D.},
        title = "{The Mix is In! Exploring the Origin of Exogenous Material in Asteroids and Meteorites}",
    booktitle = {AAS/Division for Planetary Sciences Meeting Abstracts},
         year = 2020,
       series = {AAS/Division for Planetary Sciences Meeting Abstracts},
       volume = {52},
        month = oct,
          eid = {402.02},
        pages = {402.02},
       adsurl = {https://ui.adsabs.harvard.edu/abs/2020DPS....5240202B}
}

@ARTICLE{Landgraf_2002AJ....123.2857L,
       author = {{Landgraf}, M. and {Liou}, J. -C. and {Zook}, H.~A. and {Gr{\"u}n}, E.},
        title = "{Origins of Solar System Dust beyond Jupiter}",
      journal = {Astronomical Journal},
         year = 2002,
        month = may,
       volume = {123},
       number = {5},
        pages = {2857-2861},
          doi = {10.1086/339704},
archivePrefix = {arXiv},
       eprint = {astro-ph/0201291},
 primaryClass = {astro-ph},
       adsurl = {https://ui.adsabs.harvard.edu/abs/2002AJ....123.2857L}
}

@INPROCEEDINGS{Bottke_2022DPS....5430403B,
       author = {{Bottke}, William and {Vokrouhlický}, David and {Marschall}, Raphael and {Nesvorný}, David and {Morbidelli}, Alessandro and {Marchi}, Simone and {Deienno}, Rogerio and {Dones}, Henry and {Levison}, Harold},
        title = "{The Size Distribution and Impact Flux of Comets in the Outer Solar System}",
    booktitle = {AAS/Division for Planetary Sciences Meeting Abstracts},
         year = 2022,
       series = {AAS/Division for Planetary Sciences Meeting Abstracts},
       volume = {54},
        month = dec,
          eid = {304.03},
        pages = {304.03},
       adsurl = {https://ui.adsabs.harvard.edu/abs/2022DPS....5430403B}
}

@ARTICLE{Ivezic_2001AJ....122.2749I,
       author = {{Ivezi{\'c}}, {\v{Z}}eljko and {Tabachnik}, Serge and {Rafikov}, Roman and {Lupton}, Robert H. and {Quinn}, Tom and {Hammergren}, Mark and {Eyer}, Laurent and {Chu}, Jennifer and {Armstrong}, John C. and {Fan}, Xiaohui and {Finlator}, Kristian and {Geballe}, Tom R. and {Gunn}, James E. and {Hennessy}, Gregory S. and {Knapp}, Gillian R. and {Leggett}, Sandy K. and {Munn}, Jeffrey A. and {Pier}, Jeffrey R. and {Rockosi}, Constance M. and {Schneider}, Donald P. and {Strauss}, Michael A. and {Yanny}, Brian and {Brinkmann}, Jonathan and {Csabai}, Istv{\'a}n and {Hindsley}, Robert B. and {Kent}, Stephen and {Lamb}, Don Q. and {Margon}, Bruce and {McKay}, Timothy A. and {Smith}, J. Allyn and {Waddel}, Patrick and {York}, Donald G. and {SDSS Collaboration}},
        title = "{Solar System Objects Observed in the Sloan Digital Sky Survey Commissioning Data}",
      journal = {\aj},
         year = 2001,
        month = nov,
       volume = {122},
       number = {5},
        pages = {2749-2784},
          doi = {10.1086/323452},
archivePrefix = {arXiv},
       eprint = {astro-ph/0105511},
 primaryClass = {astro-ph},
       adsurl = {https://ui.adsabs.harvard.edu/abs/2001AJ....122.2749I}
}

@ARTICLE{Bottke_2020AJ....160...14B,
       author = {{Bottke}, W.~F. and {Vokrouhlický}, D. and {Ballouz}, R. -L. and {Barnouin}, O.~S. and {Connolly}, H.~C., Jr. and {Elder}, C. and {Marchi}, S. and {McCoy}, T.~J. and {Michel}, P. and {Nolan}, M.~C. and {Rizk}, B. and {Scheeres}, D.~J. and {Schwartz}, S.~R. and {Walsh}, K.~J. and {Lauretta}, D.~S.},
        title = "{Interpreting the Cratering Histories of Bennu, Ryugu, and Other Spacecraft-explored Asteroids}",
      journal = {\aj},
         year = 2020,
        month = jul,
       volume = {160},
       number = {1},
          eid = {14},
        pages = {14},
          doi = {10.3847/1538-3881/ab88d3},
       adsurl = {https://ui.adsabs.harvard.edu/abs/2020AJ....160...14B}
}

@ARTICLE{Marsset_2022AJ....163..165M,
       author = {{Marsset}, Micha{\"e}l and {DeMeo}, Francesca E. and {Burt}, Brian and {Polishook}, David and {Binzel}, Richard P. and {Granvik}, Mikael and {Vernazza}, Pierre and {Carry}, Benoit and {Bus}, Schelte J. and {Slivan}, Stephen M. and {Thomas}, Cristina A. and {Moskovitz}, Nicholas A. and {Rivkin}, Andrew S.},
        title = "{The Debiased Compositional Distribution of MITHNEOS: Global Match between the Near-Earth and Main-belt Asteroid Populations, and Excess of D-type Near-Earth Objects}",
      journal = {The Astronomical Journal},
         year = 2022,
        month = apr,
       volume = {163},
       number = {4},
          eid = {165},
        pages = {165},
          doi = {10.3847/1538-3881/ac532f},
archivePrefix = {arXiv},
       eprint = {2202.13796},
 primaryClass = {astro-ph.EP},
       adsurl = {https://ui.adsabs.harvard.edu/abs/2022AJ....163..165M}
}

@ARTICLE{Granvik_2016Natur.530..303G,
       author = {{Granvik}, Mikael and {Morbidelli}, Alessandro and {Jedicke}, Robert and {Bolin}, Bryce and {Bottke}, William F. and {Beshore}, Edward and {Vokrouhlický}, David and {Delb{\`o}}, Marco and {Michel}, Patrick},
        title = "{Super-catastrophic disruption of asteroids at small perihelion distances}",
      journal = {Nature},
         year = 2016,
        month = feb,
       volume = {530},
       number = {7590},
        pages = {303-306},
          doi = {10.1038/nature16934},
       adsurl = {https://ui.adsabs.harvard.edu/abs/2016Natur.530..303G}
}

@INPROCEEDINGS{Bowell_1989aste.conf..524B,
       author = {{Bowell}, Edward and {Hapke}, Bruce and {Domingue}, Deborah and {Lumme}, Kari and {Peltoniemi}, Jouni and {Harris}, Alan W.},
        title = "{Application of photometric models to asteroids.}",
    booktitle = {Asteroids II},
         year = 1989,
       editor = {{Binzel}, Richard P. and {Gehrels}, Tom and {Matthews}, Mildred Shapley},
        month = jan,
        pages = {524-556},
       adsurl = {https://ui.adsabs.harvard.edu/abs/1989aste.conf..524B}
}

@PHDTHESIS{Tholen_1984PhDT.........3T,
       author = {{Tholen}, David James},
        title = "{Asteroid Taxonomy from Cluster Analysis of Photometry.}",
       school = {University of Arizona},
         year = 1984,
        month = sep,
       adsurl = {https://ui.adsabs.harvard.edu/abs/1984PhDT.........3T}
}

@ARTICLE{Bottke_2006AREPS..34..157B,
       author = {{Bottke}, William F., Jr. and {Vokrouhlický}, David and {Rubincam}, David P. and {Nesvorný}, David},
        title = "{The Yarkovsky and Yorp Effects: Implications for Asteroid Dynamics}",
      journal = {Annual Review of Earth and Planetary Sciences},
         year = 2006,
        month = may,
       volume = {34},
        pages = {157-191},
          doi = {10.1146/annurev.earth.34.031405.125154},
       adsurl = {https://ui.adsabs.harvard.edu/abs/2006AREPS..34..157B}
}

@ARTICLE{Hendler_2020PSJ.....1...75H,
       author = {{Hendler}, Nathanial P. and {Malhotra}, Renu},
        title = "{Observational Completion Limit of Minor Planets from the Asteroid Belt to Jupiter Trojans}",
      journal = {The Planetary Science Journal},
         year = 2020,
        month = dec,
       volume = {1},
       number = {3},
          eid = {75},
        pages = {75},
          doi = {10.3847/PSJ/abbe25},
archivePrefix = {arXiv},
       eprint = {2010.07822},
 primaryClass = {astro-ph.EP},
       adsurl = {https://ui.adsabs.harvard.edu/abs/2020PSJ.....1...75H}
}

@ARTICLE{Carbognani_2017P&SS..147....1C,
       author = {{Carbognani}, Albino},
        title = "{The spin-barrier ratio for S and C-type main asteroids belt}",
      journal = {Planetary and Space Science},
         year = 2017,
        month = nov,
       volume = {147},
        pages = {1-5},
          doi = {10.1016/j.pss.2017.07.019},
archivePrefix = {arXiv},
       eprint = {1903.11876},
 primaryClass = {astro-ph.EP},
       adsurl = {https://ui.adsabs.harvard.edu/abs/2017P&SS..147....1C}
}

@ARTICLE{Warner_2009Icar..202..134W,
       author = {{Warner}, Brian D. and {Harris}, Alan W. and {Pravec}, Petr},
        title = "{The asteroid lightcurve database}",
      journal = {Icarus},
         year = 2009,
        month = jul,
       volume = {202},
       number = {1},
        pages = {134-146},
          doi = {10.1016/j.icarus.2009.02.003},
       adsurl = {https://ui.adsabs.harvard.edu/abs/2009Icar..202..134W}
}

@ARTICLE{Durech_2010A&A...513A..46D,
       author = {{\v{D}urech}, J. and {Sidorin}, V. and {Kaasalainen}, M.},
        title = "{DAMIT: a database of asteroid models}",
      journal = {Astronomy and Astrophysics},
         year = 2010,
        month = apr,
       volume = {513},
          eid = {A46},
        pages = {A46},
          doi = {10.1051/0004-6361/200912693},
       adsurl = {https://ui.adsabs.harvard.edu/abs/2010A&A...513A..46D}
}

@ARTICLE{Durech_2019A&A...631A...2D,
       author = {{\v{D}urech}, J. and {Hanu{\v{s}}}, J. and {Van{\v{c}}o}, R.},
        title = "{Inversion of asteroid photometry from Gaia DR2 and the Lowell Observatory photometric database}",
      journal = {Astronomy and Astrophysics},
         year = 2019,
        month = nov,
       volume = {631},
          eid = {A2},
        pages = {A2},
          doi = {10.1051/0004-6361/201936341},
archivePrefix = {arXiv},
       eprint = {1909.09395},
 primaryClass = {astro-ph.EP},
       adsurl = {https://ui.adsabs.harvard.edu/abs/2019A&A...631A...2D}
}

@ARTICLE{Binzel_2019Icar..324...41B,
       author = {{Binzel}, R.~P. and {DeMeo}, F.~E. and {Turtelboom}, E.~V. and {Bus}, S.~J. and {Tokunaga}, A. and {Burbine}, T.~H. and {Lantz}, C. and {Polishook}, D. and {Carry}, B. and {Morbidelli}, A. and {Birlan}, M. and {Vernazza}, P. and {Burt}, B.~J. and {Moskovitz}, N. and {Slivan}, S.~M. and {Thomas}, C.~A. and {Rivkin}, A.~S. and {Hicks}, M.~D. and {Dunn}, T. and {Reddy}, V. and {Sanchez}, J.~A. and {Granvik}, M. and {Kohout}, T.},
        title = "{Compositional distributions and evolutionary processes for the near-Earth object population: Results from the MIT-Hawaii Near-Earth Object Spectroscopic Survey (MITHNEOS)}",
      journal = {Icarus},
         year = 2019,
        month = may,
       volume = {324},
        pages = {41-76},
          doi = {10.1016/j.icarus.2018.12.035},
archivePrefix = {arXiv},
       eprint = {2004.05090},
 primaryClass = {astro-ph.EP},
       adsurl = {https://ui.adsabs.harvard.edu/abs/2019Icar..324...41B}
}

@BOOK{Hutchison2007,
  title     = "Cambridge planetary science: Meteorites: A petrologic, chemical
               and isotopic synthesis series number 2",
  author    = "Hutchison, Robert",
  publisher = "Cambridge University Press",
  month     =  jan,
  year      =  2007,
  address   = "Cambridge, England"
}

@ARTICLE{Macke_2011M&PS...46.1842M,
       author = {{Macke}, Robert J. and {Consolmagno}, Guy J. and {Britt}, Daniel T.},
        title = "{Density, porosity, and magnetic susceptibility of carbonaceous chondrites}",
      journal = {Meteoritics and Planetary Science},
         year = 2011,
        month = dec,
       volume = {46},
       number = {12},
        pages = {1842-1862},
          doi = {10.1111/j.1945-5100.2011.01298.x},
       adsurl = {https://ui.adsabs.harvard.edu/abs/2011M&PS...46.1842M}
}

@ARTICLE{Consolmagno_2008ChEG...68....1C,
       author = {{Consolmagno}, G. and {Britt}, D. and {Macke}, R.},
        title = "{The significance of meteorite density and porosity}",
      journal = {Chemie der Erde / Geochemistry},
         year = 2008,
        month = apr,
       volume = {68},
       number = {1},
        pages = {1-29},
          doi = {10.1016/j.chemer.2008.01.003},
       adsurl = {https://ui.adsabs.harvard.edu/abs/2008ChEG...68....1C}
}

@ARTICLE{Borovicka_2019M&PS...54.1024B,
       author = {{Borovi{\v{c}}ka}, Ji{\v{r}}{\'\i} and {Popova}, Olga and {Spurný}, Pavel},
        title = "{The Maribo CM2 meteorite fall{\textemdash}Survival of weak material at high entry speed}",
      journal = {Meteoritics and Planetary Science},
         year = 2019,
        month = may,
       volume = {54},
       number = {5},
        pages = {1024-1041},
          doi = {10.1111/maps.13259},
archivePrefix = {arXiv},
       eprint = {1902.01112},
 primaryClass = {astro-ph.EP},
       adsurl = {https://ui.adsabs.harvard.edu/abs/2019M&PS...54.1024B}
}

@ARTICLE{Gladman_2009Icar..202..104G,
       author = {{Gladman}, Brett J. and {Davis}, Donald R. and {Neese}, Carol and {Jedicke}, Robert and {Williams}, Gareth and {Kavelaars}, J.~J. and {Petit}, Jean-Marc and {Scholl}, Hans and {Holman}, Matthew and {Warrington}, Ben and {Esquerdo}, Gil and {Tricarico}, Pasquale},
        title = "{On the asteroid belt's orbital and size distribution}",
      journal = {\icarus},
         year = 2009,
        month = jul,
       volume = {202},
       number = {1},
        pages = {104-118},
          doi = {10.1016/j.icarus.2009.02.012},
       adsurl = {https://ui.adsabs.harvard.edu/abs/2009Icar..202..104G}
}

@INCOLLECTION{Vokrouhlicky_2015aste.book..509V,
       author = {{Vokrouhlick{\'y}}, D. and {Bottke}, W.~F. and {Chesley}, S.~R. and {Scheeres}, D.~J. and {Statler}, T.~S.},
        title = "{The Yarkovsky and YORP Effects}",
    booktitle = {Asteroids IV},
         year = 2015,
       editor = {{Michel}, Patrick and {DeMeo}, Francesca E. and {Bottke}, William F.},
        pages = {509-531},
          doi = {10.2458/azu_uapress_9780816532131-ch027},
       adsurl = {https://ui.adsabs.harvard.edu/abs/2015aste.book..509V}
}

@MASTERSTHESIS{Vavra_2024,
       author = {{Vávra}, Michael},
        title = "{Extensions of the main belt collisional model}",
       school = {Charles University, Prague, Czech Republic},
         year = 2024,
          url = {https://dspace.cuni.cz/handle/20.500.11956/188494}
}

@ARTICLE{Broz_2024A&A...689A.183B,
       author = {{Bro{\v{z}}}, M. and {Vernazza}, P. and {Marsset}, M. and {Binzel}, R.~P. and {DeMeo}, F. and {Birlan}, M. and {Colas}, F. and {Anghel}, S. and {Bouley}, S. and {Blanpain}, C. and {Gattacceca}, J. and {Jeanne}, S. and {Jorda}, L. and {Lecubin}, J. and {Malgoyre}, A. and {Steinhausser}, A. and {Vaubaillon}, J. and {Zanda}, B.},
        title = "{Source regions of carbonaceous meteorites and near-Earth objects}",
      journal = {\aap},
         year = 2024,
        month = sep,
       volume = {689},
          eid = {A183},
        pages = {A183},
          doi = {10.1051/0004-6361/202450532},
archivePrefix = {arXiv},
       eprint = {2406.19727},
 primaryClass = {astro-ph.EP},
       adsurl = {https://ui.adsabs.harvard.edu/abs/2024A&A...689A.183B}
}

@ARTICLE{Shober_2025NatAs...9..799S,
       author = {{Shober}, Patrick M. and {Devillepoix}, Hadrien A.~R. and {Vaubaillon}, Jeremie and {Anghel}, Simon and {Deam}, Sophie E. and {Sansom}, Eleanor K. and {Colas}, Francois and {Zanda}, Brigitte and {Vernazza}, Pierre and {Bland}, Phil},
        title = "{Perihelion history and atmospheric survival as primary drivers of the Earth's meteorite record}",
      journal = {Nature Astronomy},
         year = 2025,
        month = jun,
       volume = {9},
        pages = {799-812},
          doi = {10.1038/s41550-025-02526-6},
archivePrefix = {arXiv},
       eprint = {2504.10690},
 primaryClass = {astro-ph.EP},
       adsurl = {https://ui.adsabs.harvard.edu/abs/2025NatAs...9..799S}
}

@ARTICLE{Maeda_2021AJ....162..280M,
       author = {{Maeda}, Natsuho and {Terai}, Tsuyoshi and {Ohtsuki}, Keiji and {Yoshida}, Fumi and {Ishihara}, Kosuke and {Deyama}, Takuto},
        title = "{Size Distributions of Bluish and Reddish Small Main-belt Asteroids Obtained by Subaru/Hyper Suprime-Cam}",
      journal = {\aj},
         year = 2021,
        month = dec,
       volume = {162},
       number = {6},
          eid = {280},
        pages = {280},
          doi = {10.3847/1538-3881/ac2c6e},
archivePrefix = {arXiv},
       eprint = {2110.00178},
 primaryClass = {astro-ph.EP},
       adsurl = {https://ui.adsabs.harvard.edu/abs/2021AJ....162..280M}
}

@ARTICLE{Gallegos_2023PSJ.....4..128G,
       author = {{Gallegos}, Cesar and {Fuentes}, Cesar and {Pe{\~n}a}, Jos{\'e}},
        title = "{Physical Properties of the Asteroid Belts from Brightness-limited Surveys}",
      journal = {\psj},
         year = 2023,
        month = jul,
       volume = {4},
       number = {7},
          eid = {128},
        pages = {128},
          doi = {10.3847/PSJ/ace116},
       adsurl = {https://ui.adsabs.harvard.edu/abs/2023PSJ.....4..128G}
}

@ARTICLE{Burdanov_2025Natur.638...74B,
       author = {{Burdanov}, Artem Y. and {de Wit}, Julien and {Bro{\v{z}}}, Miroslav and {M{\"u}ller}, Thomas G. and {Hoffmann}, Tobias and {Ferrais}, Marin and {Micheli}, Marco and {Jehin}, Emmanuel and {Parrott}, Daniel and {Hasler}, Samantha N. and {Binzel}, Richard P. and {Ducrot}, Elsa and {Kreidberg}, Laura and {Gillon}, Micha{\"e}l and {Greene}, Thomas P. and {Grundy}, Will M. and {Kareta}, Theodore and {Lagage}, Pierre-Olivier and {Moskovitz}, Nicholas and {Thirouin}, Audrey and {Thomas}, Cristina A. and {Zieba}, Sebastian},
        title = "{JWST sighting of decametre main-belt asteroids and view on meteorite sources}",
      journal = {\nat},
         year = 2025,
        month = feb,
       volume = {638},
       number = {8049},
        pages = {74-78},
          doi = {10.1038/s41586-024-08480-z},
archivePrefix = {arXiv},
       eprint = {2502.01744},
 primaryClass = {astro-ph.EP},
       adsurl = {https://ui.adsabs.harvard.edu/abs/2025Natur.638...74B}
}

@ARTICLE{Rubincam_2000Icar..148....2R,
       author = {{Rubincam}, David Parry},
        title = "{Radiative Spin-up and Spin-down of Small Asteroids}",
      journal = {\icarus},
         year = 2000,
        month = nov,
       volume = {148},
       number = {1},
        pages = {2-11},
          doi = {10.1006/icar.2000.6485},
       adsurl = {https://ui.adsabs.harvard.edu/abs/2000Icar..148....2R}
}

@ARTICLE{Walsh_2024NatCo..15.5653W,
       author = {{Walsh}, Kevin J. and {Ballouz}, R. -L. and {Bottke}, W.~F. and {Avdellidou}, C. and {Connolly}, Jr., H.~C. and {Delbo}, M. and {DellaGiustina}, D.~N. and {Jawin}, E.~R. and {McCoy}, T. and {Michel}, P. and {Morota}, T. and {Nolan}, M.~C. and {Schwartz}, S.~R. and {Sugita}, S. and {Lauretta}, D.~S. .},
        title = "{Numerical simulations suggest asteroids (101955) Bennu and (162173) Ryugu are likely second or later generation rubble piles.}",
      journal = {Nature Communications},
         year = 2024,
        month = jul,
       volume = {15},
          eid = {5653},
        pages = {5653},
          doi = {10.1038/s41467-024-49310-0},
       adsurl = {https://ui.adsabs.harvard.edu/abs/2024NatCo..15.5653W}
}

@INPROCEEDINGS{Jutzi_2023LPICo2851.2439J,
       author = {{Jutzi}, M. and {Raducan}, S.~D. and {Landeck}, A. and {Blum}, J. and {Michel}, P.},
        title = "{Collisional Disruptions of Small Bodies: At What Size Are They the Weakest Against Impacts?}",
    booktitle = {Asteroids, Comets, Meteors Conference},
         year = 2023,
       series = {LPI Contributions},
       volume = {2851},
        month = aug,
          eid = {2439},
        pages = {2439},
       adsurl = {https://ui.adsabs.harvard.edu/abs/2023LPICo2851.2439J}
}

@ARTICLE{Walsh_2008Natur.454..188W,
       author = {{Walsh}, Kevin J. and {Richardson}, Derek C. and {Michel}, Patrick},
        title = "{Rotational breakup as the origin of small binary asteroids}",
      journal = {\nat},
         year = 2008,
        month = jul,
       volume = {454},
       number = {7201},
        pages = {188-191},
          doi = {10.1038/nature07078},
       adsurl = {https://ui.adsabs.harvard.edu/abs/2008Natur.454..188W}
}

@ARTICLE{Marzari_2011Icar..214..622M,
       author = {{Marzari}, F. and {Rossi}, A. and {Scheeres}, D.~J.},
        title = "{Combined effect of YORP and collisions on the rotation rate of small Main Belt asteroids}",
      journal = {\icarus},
         year = 2011,
        month = aug,
       volume = {214},
       number = {2},
        pages = {622-631},
          doi = {10.1016/j.icarus.2011.05.033},
       adsurl = {https://ui.adsabs.harvard.edu/abs/2011Icar..214..622M}
}

@ARTICLE{Capek_2004Icar..172..526C,
       author = {{{\v{C}}apek}, D. and {Vokrouhlick{\'y}}, D.},
        title = "{The YORP effect with finite thermal conductivity}",
      journal = {\icarus},
         year = 2004,
        month = dec,
       volume = {172},
       number = {2},
        pages = {526-536},
          doi = {10.1016/j.icarus.2004.07.003},
       adsurl = {https://ui.adsabs.harvard.edu/abs/2004Icar..172..526C}
}

@ARTICLE{Deienno_2025ApJ...986..146D,
       author = {{Deienno}, Rogerio and {Izidoro}, Andr{\'e} and {Nesvorn{\'y}}, David and {Bottke}, William F. and {Roig}, Fernando and {Marchi}, Simone},
        title = "{Size{\textendash}Frequency Distribution of Terrestrial Leftover Planetesimals and S-complex Implanted Asteroids}",
      journal = {\apj},
         year = 2025,
        month = jun,
       volume = {986},
       number = {2},
          eid = {146},
        pages = {146},
          doi = {10.3847/1538-4357/adccbb},
archivePrefix = {arXiv},
       eprint = {2504.11405},
 primaryClass = {astro-ph.EP},
       adsurl = {https://ui.adsabs.harvard.edu/abs/2025ApJ...986..146D}
}

@ARTICLE{Nesvorny_2023Icar..39915545N,
       author = {{Nesvorn{\'y}}, David and {Roig}, Fernando V. and {Vokrouhlick{\'y}}, David and {Bottke}, William F. and {Marchi}, Simone and {Morbidelli}, Alessandro and {Deienno}, Rogerio},
        title = "{Early bombardment of the moon: Connecting the lunar crater record to the terrestrial planet formation}",
      journal = {\icarus},
         year = 2023,
        month = jul,
       volume = {399},
          eid = {115545},
        pages = {115545},
          doi = {10.1016/j.icarus.2023.115545},
archivePrefix = {arXiv},
       eprint = {2303.17736},
 primaryClass = {astro-ph.EP},
       adsurl = {https://ui.adsabs.harvard.edu/abs/2023Icar..39915545N}
}

@ARTICLE{Deienno_2024PSJ.....5..110D,
       author = {{Deienno}, Rogerio and {Nesvorn{\'y}}, David and {Clement}, Matthew S. and {Bottke}, William F. and {Izidoro}, Andr{\'e} and {Walsh}, Kevin J.},
        title = "{Accretion and Uneven Depletion of the Main Asteroid Belt}",
      journal = {The Planetary Science Journal},
         year = 2024,
        month = may,
       volume = {5},
       number = {5},
          eid = {110},
        pages = {110},
          doi = {10.3847/PSJ/ad3a68},
archivePrefix = {arXiv},
       eprint = {2404.03791},
 primaryClass = {astro-ph.EP},
       adsurl = {https://ui.adsabs.harvard.edu/abs/2024PSJ.....5..110D}
}

@ARTICLE{Bottke_1993GeoRL..20..879B,
       author = {{Bottke}, W.~F. and {Greenberg}, R.},
        title = "{Asteroidal collision probabilities}",
      journal = {Geophysical Research Letters},
         year = 1993,
        month = may,
       volume = {20},
       number = {10},
        pages = {879-881},
          doi = {10.1029/92GL02713},
       adsurl = {https://ui.adsabs.harvard.edu/abs/1993GeoRL..20..879B}
}

@ARTICLE{Nesvorny_2024Icar..41716110N,
       author = {{Nesvorn{\'y}}, David and {Vokrouhlick{\'y}}, David and {Shelly}, Frank and {Deienno}, Rogerio and {Bottke}, William F. and {Fuls}, Carson and {Jedicke}, Robert and {Naidu}, Shantanu and {Chesley}, Steven R. and {Chodas}, Paul W. and {Farnocchia}, Davide and {Delbo}, Marco},
        title = "{NEOMOD 3: The debiased size distribution of Near Earth Objects}",
      journal = {\icarus},
         year = 2024,
        month = jul,
       volume = {417},
          eid = {116110},
        pages = {116110},
          doi = {10.1016/j.icarus.2024.116110},
archivePrefix = {arXiv},
       eprint = {2404.18805},
 primaryClass = {astro-ph.EP},
       adsurl = {https://ui.adsabs.harvard.edu/abs/2024Icar..41716110N}
}

@ARTICLE{Broz_2026,
author = {
{Bro\v{z}}, Miroslav and
{Binzel}, R.P. and
{Vernazza}, P. and
{Marsset}, M. and
{Chrenko}, O. and
{\v{D}urech}, J. and
{Herald}, D.
},
title = "{Apophis source population and Earth encounter frequency of Apophis-like bodies}",
journal = {\aap, submit.},
year = 2026,
}

@ARTICLE{Anderson_2025NatAs...9.1464A,
       author = {{Anderson}, Sarah E. and {Vernazza}, Pierre and {Bro{\v{z}}}, Miroslav},
        title = "{Different arrival times of CM- and CI-like bodies from the outer Solar System in the asteroid belt}",
      journal = {Nature Astronomy},
         year = 2025,
        month = sep,
       volume = {9},
        pages = {1464-1475},
          doi = {10.1038/s41550-025-02635-2},
archivePrefix = {arXiv},
       eprint = {2507.22649},
 primaryClass = {astro-ph.EP},
       adsurl = {https://ui.adsabs.harvard.edu/abs/2025NatAs...9.1464A}
}

\appendix

\section{Note on the YORP effect} \label{Note_YORP}

Alternatively,
weakness at sub-kilometer sizes might be related to the YORP effect
\citep{Rubincam_2000Icar..148....2R}.
If bodies spin up over the spin barrier, they disrupt
\citep{Walsh_2008Natur.454..188W,Marzari_2011Icar..214..622M}
and such disruptions are not accounted for in our model.
However, the scaling of the YORP effect
\citep{Capek_2004Icar..172..526C}
with the semimajor axis ($a^{-2}$),
the bulk density ($\rho^{-1}$),
suggests that C- and S-types should be comparable
in this regard.
Consequently, the differences between SFDs
of C- and S-types
should be attributed to something else.

\section{Note on the Yarkovsky effect} \label{Note_Yarko}

Alternatively, the deficiency of S-types at sub-km sizes
might be related to the Yarkovsky drift
(e.g., \citealt{Vokrouhlicky_2015aste.book..509V}),
or equivalently the decay timescales~$\tau_{\rm mb}$.
In principle, they might be different for C- and S-types.
We constrained them by the debiased SFD of C- and S-type NEOs,
as determined by \citet{Nesvorny_2024Icar..41716110N}.
According to our collisional model,
$\tau_{\rm mb}$ for C-types should be substantially increased
(by a factor of ${\approx}10$),
but for S-types, which are more numerous among NEOs
\citep{Marsset_2022AJ....163..165M},
it was impossible to reach an equilibrium
at the observed level.

\begin{figure}[h!] 
\centering
\includegraphics[width=7.5cm]{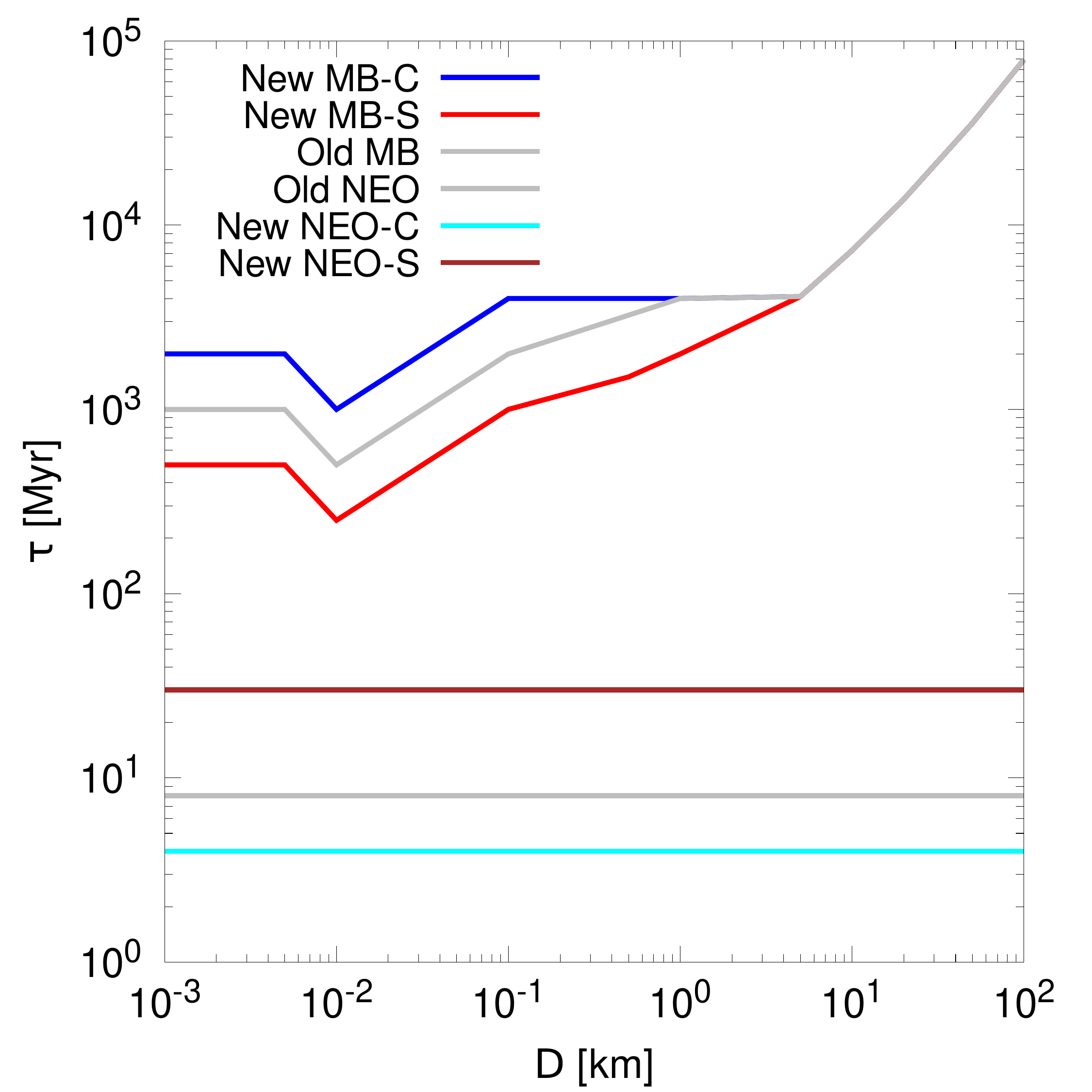}
\caption{
Decay timescales $\tau_{\rm mb}$ and $\tau_{\rm neo}$ of C- and S-types
for our modified collisional model,
aiming at fitting the SFDs of C- and S-type NEOs
\citep{Nesvorny_2024Icar..41716110N}.
The new timescales are plotted in color, the old ones in gray.
}
\label{yarko_2_1}
\end{figure}

In order to avoid this discrepancy,
we adjusted the decay timescales~$\tau_{\rm neo}$.
For C-types, we decreased the value to $3\,{\rm My}$,
in agreement with dynamical simulations of NEOs
originating from C-type families
\citep{Broz_2024A&A...689A.183B}.
For S-types, we increased the value up to $30\,{\rm My}$,
which is possible for long-lived objects among NEOs,
originating preferentially from S-type families;
the well-known example is (99924) Apophis
\citep{Broz_2026}.
This solves the discrepancy;
$\tau_{\rm mb}$ for C-types is increased
only by a factor of~2,
and for S-types it is decreased
by a factor of~2
(Fig.~\ref{yarko_2_1}),
which is sufficient to reach the equilibrium of S-type NEOs.
However, our collisional model also indicates
that these adjustments further decreased
the main-belt population of S-types,
so their deficiency
should be attributed to something else.

\section{Definition of $\chi^2$} \label{Definition of chi}

For individual populations (C- or S-types),
we define the contributions to $\chi^2$ as
\begin{equation}
    \chi_{\rm{C,S}}^2\equiv\sum_{i=1}^n \left(\frac{N(\geq\!\!D_i)_{\text{med}}-N(\geq\!\!D_i)_{\text{obs}}}{\sigma_i}\right)^2,
\end{equation}
where $N(\geq\!\!D_i)_{\text{med}}$ is the median of the set
$\{N(\geq\!\!D_i)_k\}_{k=1}^{100}$,
whose members (labeled by $k$) are 100 individual evolved SFDs at size $D_i$;
$N(\geq\!\!D_i)_{\text{obs}}$ is the observed SFD, and
$n$ is the number of individual bins (we chose $n=10^2$).
We assumed formal uncertainties
$\sigma_i\equiv 0.1N(\geq\!\!D_i)_{\text{obs}}$,
similarly as in \citet{Cibulkova_2014Icar..241..358C}.
The smallest size $D_1$ corresponds to our estimate of the observational limit
and the biggest size $D_n$ to the maximum size for evaluating $\chi^2$. 
The bin size $\log{[\dd D]}$ in the log space is constant,
therefore $\log{[\dd D]}=(\log{[D_n]}-\log{[D_1]})/(n-1)$.
We finally evaluate the total $\chi^2$ as a sum for C- and S-types
\begin{equation}
    \chi^2=\chi_{\rm{C}}^2+\chi_{\rm{S}}^2.
\end{equation}
Since individual points in the evolved SFDs,
as well as in the observed cumulative SFD,
do not have exactly the same resolution in sizes,
all $\{N(\geq\!\!D_i)_k\}_{k=1}^{100}$ and $N(\geq\!\!D_i)_{\text{obs}}$
were interpolated linearly in the log-log plane.

\section{Algorithm for an assignment of C– and S–types}\label{algorithm}

The algorithm, 
we implemented to assign C-, S-, or other types
to each asteroid with unknown taxonomies
is described as follows.
The asteroid were identified in five catalogs, namely,
Astorb (osculating orbital elements $a$, $e$, $i$ and absolute magnitudes $H$, \citealt{Moskovitz_2022A&C....4100661M});
Astdys (proper orbital elements $a_{\rm p}$, $e_{\rm p}$, $i_{\rm p}$, \citealt{Knezevic_2012IAUJD...7P..18K});
AKARI (albedo $pV$, \citealt{Yamauchi_2011PASP..123..852Y});
WISE (albedo $pV$, \citealt{Masiero_2011ApJ...741...68M}) and 
SDSS (colours $a^*$ and $i-z$, \citealt{Blanton_2017AJ....154...28B}).
If $pV$ was found in both AKARI and WISE,
we preferred the one from AKARI.

We split asteroids into five groups, 
which differ from each other according to the knowledge of
$a^*$, $i-z$ and $pV$.
The most reliable asteroids are in the first group (I),
as they have all three quantities known and all their uncertainties are below~$0.1$.
The second (II) has at least one uncertainty above~$0.1$.
The third (III) has only $a^*$ and $i-z$ known,
the fourth (IV) has only $pV$ known and
the fifth (V) has all three quantities unknown.

We then computed the individual distributions $\dd N$
for our control sample of C- or S-types and
for individual quantities
($a^*$, $i-z$ and $pV$).
We assumed each quantity is normally distributed, thus the formula is
\begin{equation} \label{dist}
    \dd N(x)=\sum_{i=1}^n \frac{1}{\sqrt{2\pi}\sigma_i}\exp[-\,\frac{1}{2}\left(\frac{x-q_i}{\sigma_i}\right)^2],
\end{equation}
where the summation is performed over all asteroids in either C- ($n=56$) or S-type ($n=72$) control sample;
$q_i$ is the specific value of quantity of $i-$th asteroid and $\sigma_i$ is its uncertainty;
$x$ is the independent variable where the distribution is evaluated.
We denote these control-sample distributions with sub- and superscripts
to distinguish C- and S-types as well as individual quantities,
for example, $\dd N^{\text{C}}_{pV}$.

For an asteroid with unknown taxonomy, we assumed that individual quantities
are also normally distributed,
for example, for the albedo
\begin{equation}
G_{pV}(x)\equiv\frac{1}{\sqrt{2\pi}\sigma_{\!pV}}\exp[-\frac{1}{2}\left(\frac{x-pV}{\sigma_{\!pV}}\right)^2]\,.\label{G}
\end{equation}
The five groups are inspected asteroid by asteroid, 
separately in the inner, middle, and outer parts of the MB.
If an asteroid was present in the SMASSII database
\citep{Bus_2002Icar..158..146B},
its taxonomy was directly assigned (C-, S- or other types).
If not present, the observed distributions of our control samples
(Figs.~\ref{a_star_i_z}, \ref{a_star_pV})
and the distributions defined in Eq.~(\ref{G})
were used to assign taxonomy.
The way of assignment differs for different groups:

\begin{enumerate}[I.]
\item The quantity $C_{\text{tot}}$ is calculated
as the product of three integrals
\begin{equation}  \label{3INT}
\int\displaylimits_{-\infty}^{+\infty}\!\! G_{a^*}(x)\dd N^{\rm C}_{a^*}(x)\dd x\!
\int\displaylimits_{-\infty}^{+\infty}\!\! G_{i-z}(x)\dd N^{\rm C}_{i-z}(x)\dd x\!  
\int\displaylimits_{-\infty}^{+\infty}\!\! G_{pV} (x)\dd N^{\rm C}_{pV}(x)\dd x\!
\end{equation}
and similarly $S_{\text{tot}}$.
If $C_{\text{tot}}<0.1$ and $S_{\text{tot}}<0.1$, then other taxonomy is assigned.
Else a pseudo-random number $r$ between 0 and 1 is generated and if 
\begin{equation} \label{condition}
\frac{C_{\text{tot}}}{C_{\text{tot}}+S_{\text{tot}}}\leq r,
\end{equation}
then C-type is assigned, 
else S-type is assigned.
If the assigned taxonomy is C/S/other,
then the tuple of asteroid's $a^*$, $\sigma_{a^*}$, $i-z$, $\sigma_{i-z}$, $pV$, $\sigma_{pV}$
is appended to a set denoted as $\mathcal{C}$/$\mathcal{S}$/$\mathcal{O}$.

\item Same as in the previous group.
However, now and in the following, $a^*$, $i-z$, $pV$, $\sigma_{a^*}$, $\sigma_{i-z}$, $\sigma_{pV}$
is not appended to the sets $\mathcal{C}$, $\mathcal{S}$ or $\mathcal{O}$.

\item $C_{\text{tot}}$ and $S_{\text{tot}}$ are calculated as the product of the first two integrals in Eq.~(\ref{3INT}). 
If $C_{\text{tot}}<0.1^{2/3}$ and $S_{\text{tot}}<0.1^{2/3}$, then other taxonomy is assigned.
Else, the pseudo-random number generator is called and the inequality Eq.~(\ref{condition}) is tested again. 
If assigned taxonomy is C/S/other, then $pV$ and it's corresponding uncertainty $\sigma_{pV}$ is randomly chosen from the set $\mathcal{C}$/$\mathcal{S}$/$\mathcal{O}$. 
The size $D$ is calculated according to Eq.~(\ref{diam}).

\item $C_{\text{tot}}$ and $S_{\text{tot}}$ are calculated with the last integral in Eq.~(\ref{3INT}).
If $C_{\text{tot}}<0.1^{1/3}$ and $S_{\text{tot}}<0.1^{1/3}$, then other taxonomy is assigned.
Else, again, a pseudo-random number generator is called.
If assigned taxonomy is C/S/other, then the tuple of $a^*$, $i-z$, $\sigma_{a^*}$ and $\sigma_{i-z}$ is randomly chosen from the set $\mathcal{C}$/$\mathcal{S}$/$\mathcal{O}$.

\item The tuple of $a^*$, $\sigma_{a^*}$, $i-z$, $\sigma_{i-z}$, $pV$ and $\sigma_{pV}$ is randomly chosen from the union of the sets $\mathcal{C}$, $\mathcal{S}$ and $\mathcal{O}$ and the size $D$ is calculated according to Eq.~(\ref{diam}).
The assigned taxonomy is C/S/other if the tuple is chosen from the set $\mathcal{C}$/$\mathcal{S}$/$\mathcal{O}$.

\end{enumerate}

\section{Supplementary tables}

\begingroup
\footnotesize
\centering
\captionof{table}{Collisional probabilities $p$ and impact velocities $v_{\text{imp}}$ between different populations. 
The value of the uncertainty corresponds to the uncertainty of the mean.
}
\label{collprob_and_impvel_tab}
\begin{tabular}{ccc}
\toprule
Populations & $p\ [10^{-18}\, \mathrm{km^{-2}\, yr^{-1}}]$ & $v_{\text{imp}} [\mathrm{km\, s^{-1}}]$\\
\midrule
MB-MB         & $2.93817 \pm 0.00235$ & $5.51327 \pm 0.00172$ \\
inner-inner   & $8.79184 \pm 0.00723$ & $5.70502 \pm 0.00317$ \\
inner-outer   & $1.47454 \pm 0.00242$ & $5.91617 \pm 0.00325$ \\ 
middle-inner  & $4.82357 \pm 0.00299$ & $5.99825 \pm 0.00246$ \\
middle-middle & $4.53761 \pm 0.00190$ & $6.07131 \pm 0.00188$ \\
outer-middle  & $2.58882 \pm 0.00166$ & $5.71286 \pm 0.00186$ \\
outer-outer   & $3.03996 \pm 0.00226$ & $5.52466 \pm 0.00166$ \\
C-C           & $2.85482 \pm 0.00201$ & $5.67254 \pm 0.00177$ \\
S-S           & $3.45479 \pm 0.00245$ & $5.25217 \pm 0.00171$ \\
S-C           & $2.94338 \pm 0.00209$ & $5.48628 \pm 0.00180$ \\
other-other   & $3.60880 \pm 0.00655$ & $5.54823 \pm 0.00579$ \\
other-S       & $3.45240 \pm 0.00372$ & $5.41329 \pm 0.00321$ \\
other-C       & $2.95665 \pm 0.00337$ & $5.62517 \pm 0.00328$ \\
\bottomrule
\end{tabular}
\endgroup

\section{Supplementary figures}

\begin{figure}[!h]
    \centering
    \vskip-.3cm
    \includegraphics[width=7.1cm]{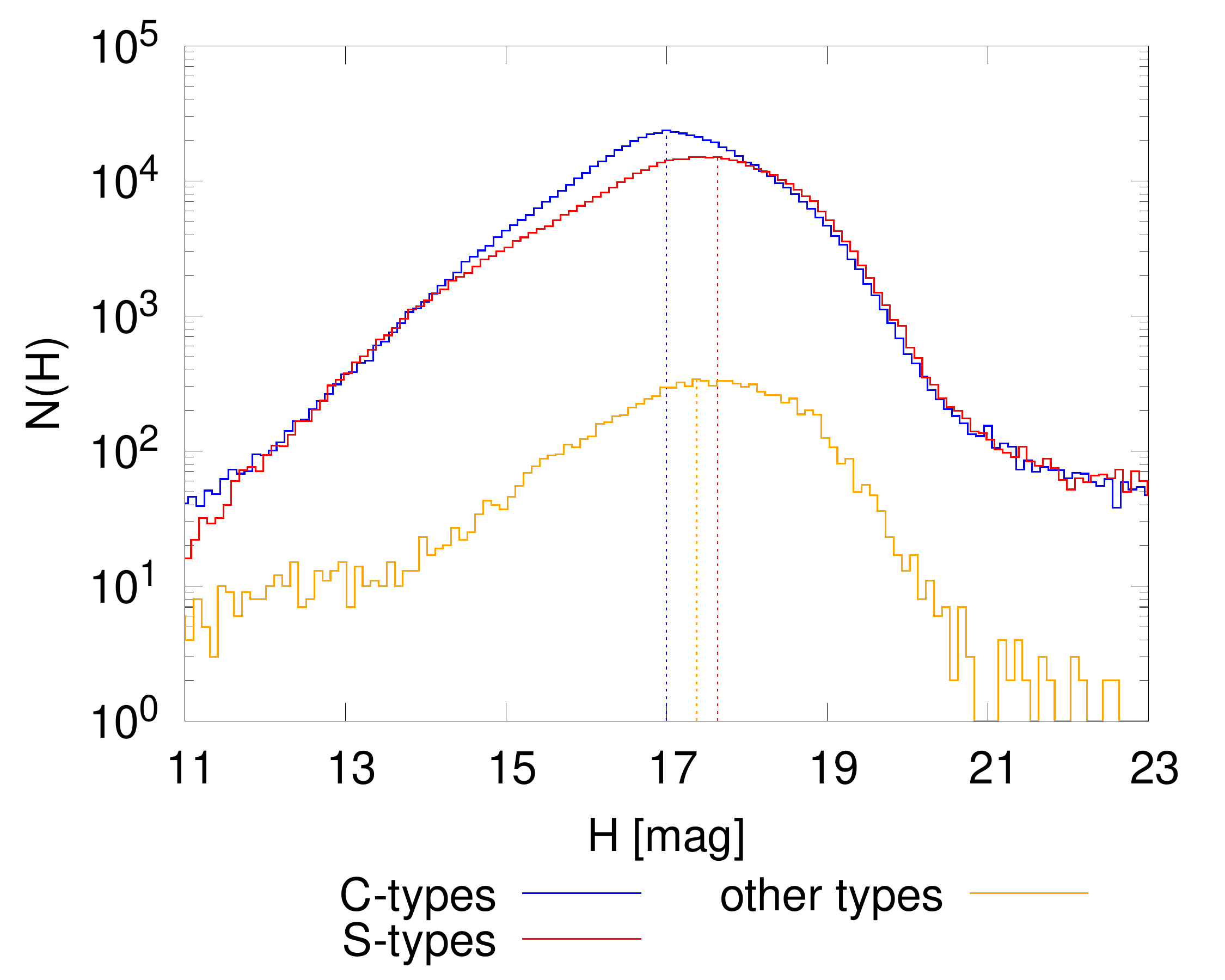}
    \caption{The differential distribution of absolute magnitude $H$ with binning $\dd H=\un{0.1}{mag}$ for C- (blue), S- (red) and other types (orange).
    The positions of $H_{\text{max}}$ of individual distributions are highlighted by vertical dotted lines -- blue for C-, red for S-, and orange for other types.
    }
    \label{H_dist_tax}
\end{figure}

\begin{figure}[!h]
    \centering
    \vskip-.3cm
    \includegraphics[width=7.2cm]{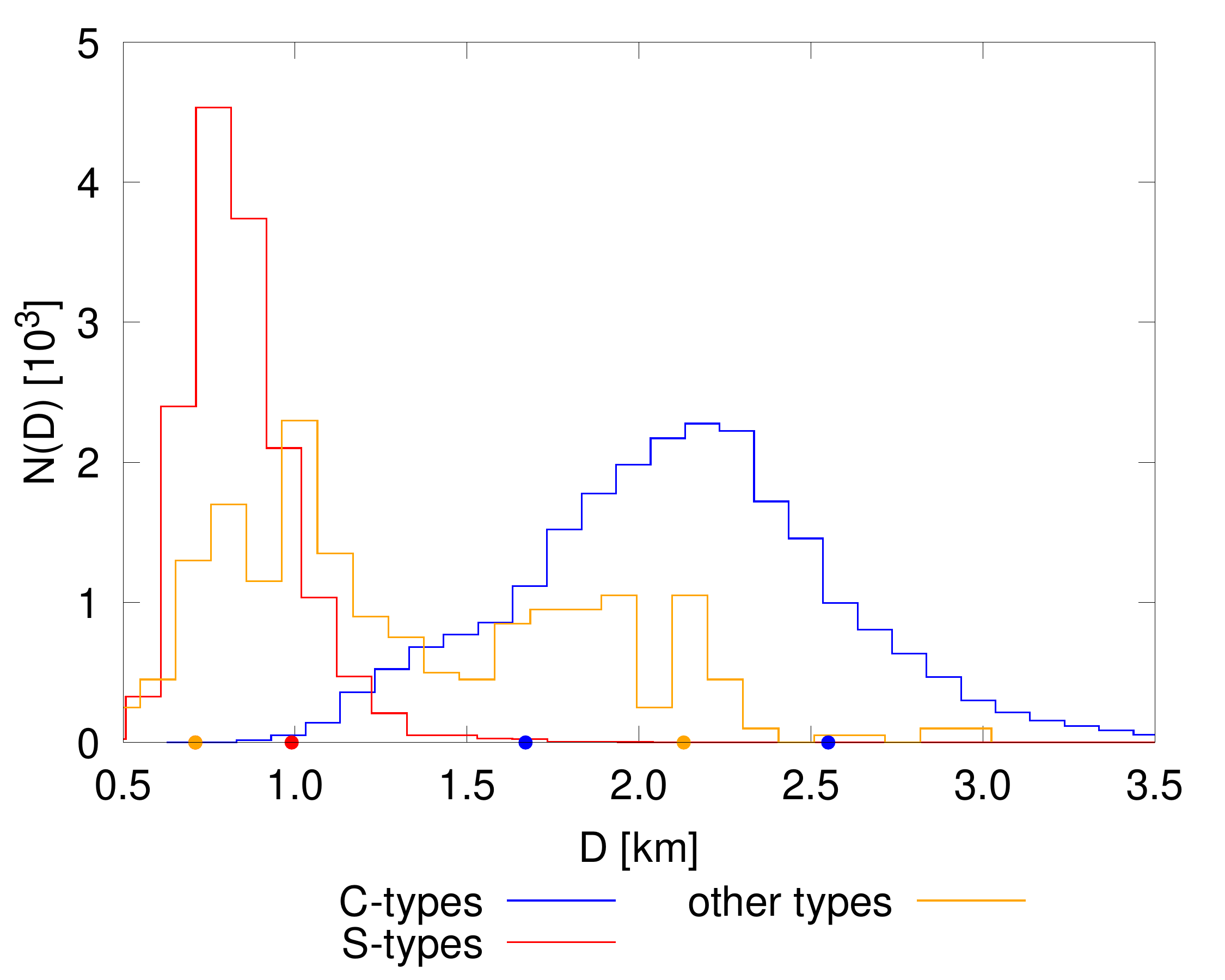}
    \caption{The differential SFDs with binning $\dd D=\un{0.1}{km}$ for C- (blue), S- (red), and other types (orange) within the most populated absolute magnitude bin.
    The blue (for C-), red (for S-) and orange (for other types) points correspond to the borders of size ranges within the observational limit sizes lie. 
    The lower bound of S- and other types is the same, therefore the lower bound of S-types is not visible.
    }
    \label{sizes_dist_tax}
\end{figure}

\begin{figure}[!b]
\centering
\includegraphics[width=9.5cm]{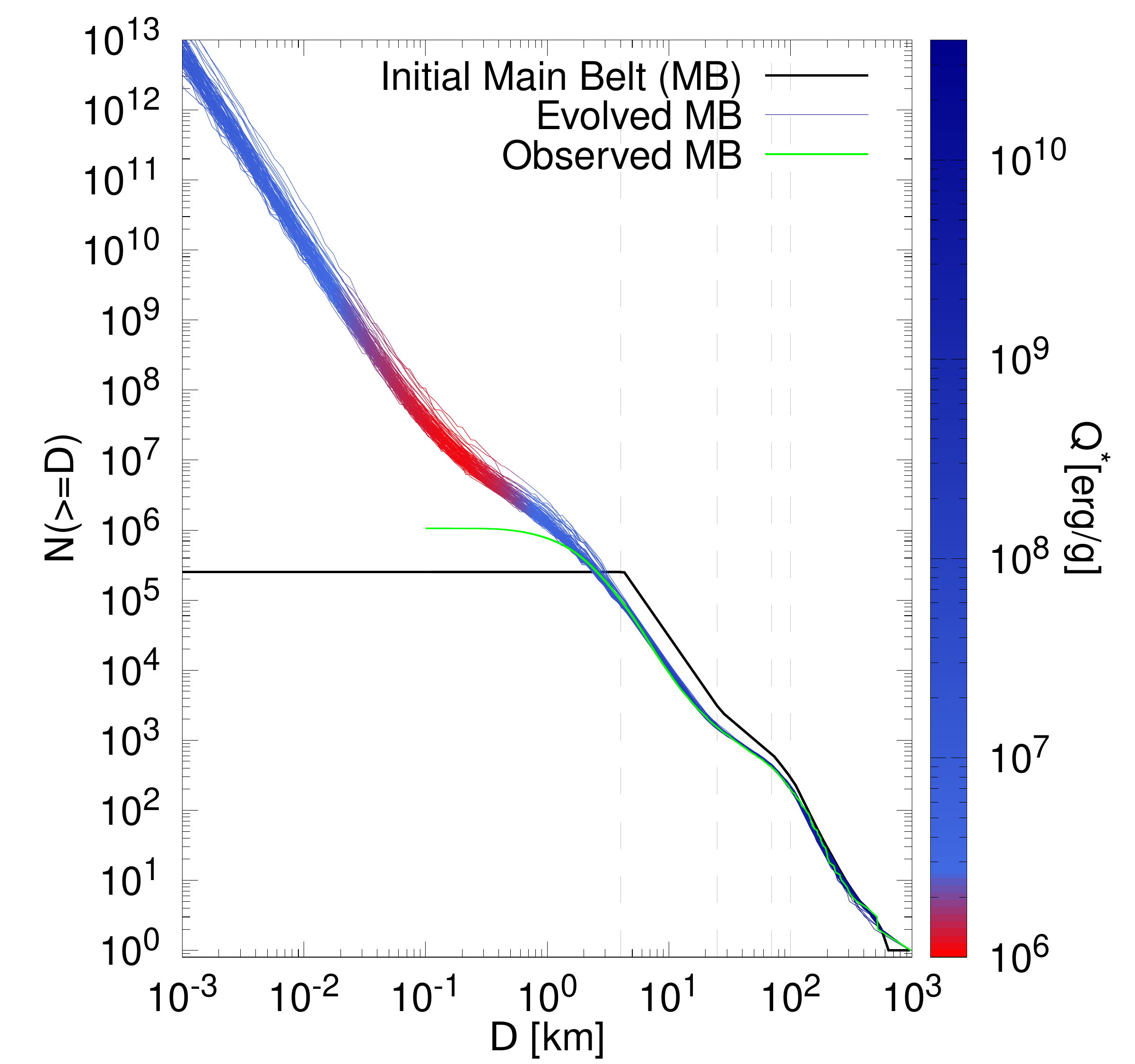}
\caption{
The first type of the initial size-frequency distribution (SFD) (black line),
100~evolved~(synthetic)~SFDs (color lines) and 
the observed SFD (green line) of the whole MB~population. 
The strength $Q^*(D)$ (scaling~law) is highlighted by the color bar,
emphasized in the vicinity of the minimum (red color).
}
\label{SFD_whole_MB_2}
\end{figure}

\begin{figure}[hbt!]
\centering
\includegraphics[width=9.5cm]{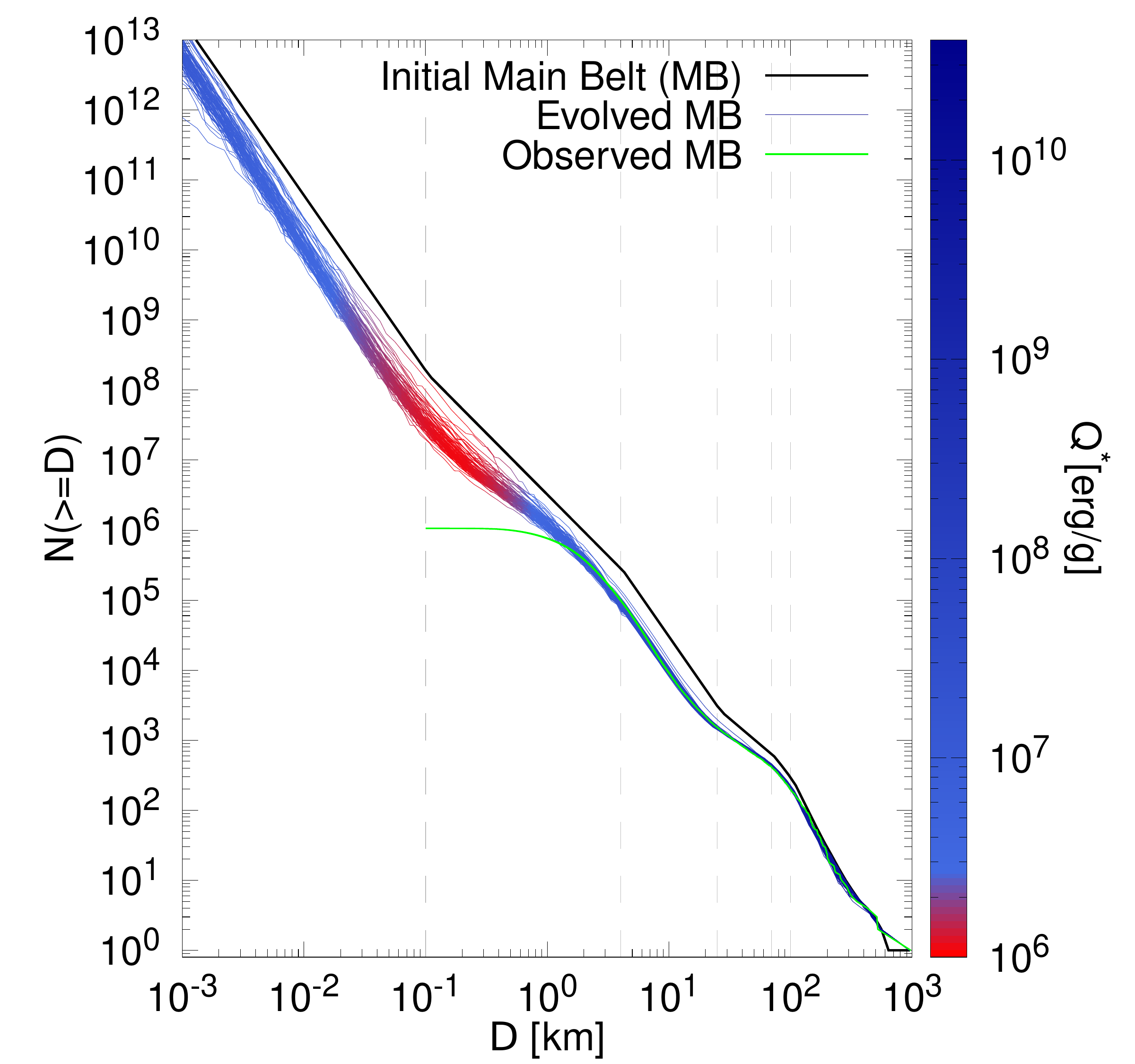}
\caption{
Same as Fig.~\ref{SFD_whole_MB_2},
computed for the second kind of initial SFD.
The slopes between $4$ and $\un{15}{km}$ as well as below $\un{0.1}{km}$,
i.e., $-2.5$, correspond to the
\citet{Dohnanyi_1969JGR....74.2531D} equilibrium slope.
}
\label{SFD_whole_MB_9}
\end{figure}

\begin{figure}[hbt!]
\centering
\begin{tabular}{c}
\kern.5cm Inner + Middle + Outer\\[-.2cm]
\includegraphics[width=9.5cm]{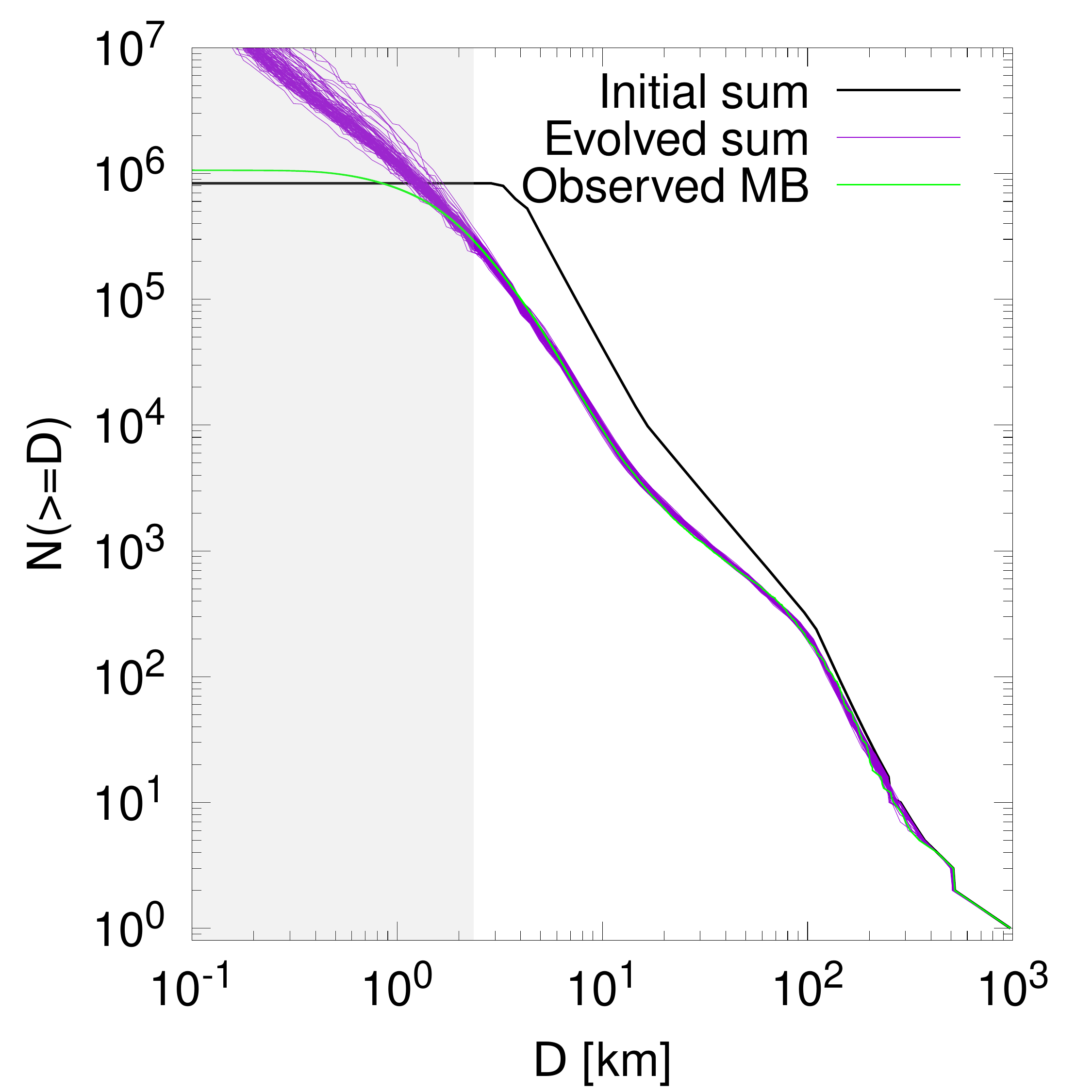} \\
\end{tabular}
\caption{Summed SFDs of all three parts from Fig.~\ref{sfd_4500_inner_43}.               
}
\label{sfd_4500_sum_43}
\end{figure}

\begin{figure}[hbt!]
\centering
\includegraphics[width=9.5cm]{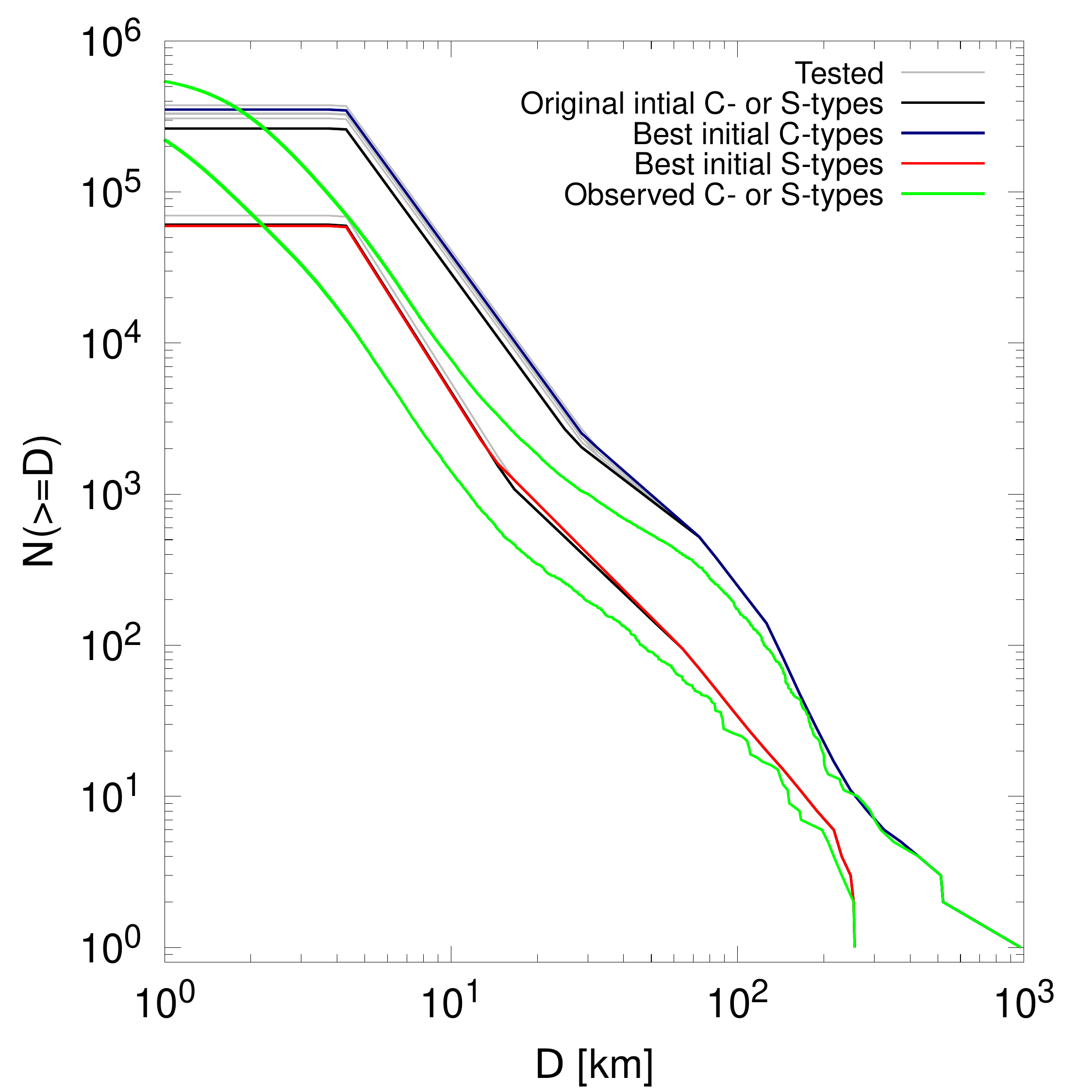}
\caption{
Six and three trial initial SFDs of C- and S-types (gray),
compared to the original ones (black). 
The pair of C- (blue) and S-types (red) initial SFDs,
which resulted in the lowest $\chi^2$, is also plotted.
}
\label{sfd_initial}
\end{figure}

\end{document}